\documentclass[aps,prb,singlecolumn,showpacs,preprint,superscriptaddress]{revtex4-1}

\usepackage{graphicx} 
\usepackage{amsmath, amsfonts, amssymb, bm}
\usepackage{float}
\usepackage{mathtools}
\usepackage[inline]{enumitem}
\usepackage{graphicx}
\usepackage{xcolor}
\usepackage{booktabs}
\usepackage[breaklinks,hypertexnames=false]{hyperref}
\usepackage{titlesec} 

\titleformat{\section}
  {\normalfont\Large\bfseries}{\thesection}{1em}{}

\begin{document}
\title{Theory of Tunneling Spectroscopy in Unconventional $p$-wave Magnet-Superconductor hybrid structures}
\author{Kazuki Maeda}
\affiliation{Department of Applied Physics, Nagoya University, Nagoya 464-8603, Japan}
\author{Bo Lu}
\affiliation{Department of Physics, Tianjin University, Tianjin 300072, China}
\author{Keiji Yada}
\affiliation{Department of Applied Physics, Nagoya University, Nagoya 464-8603, Japan}
\author{Yukio Tanaka}
\affiliation{Department of Applied Physics, Nagoya University, Nagoya 464-8603, Japan}

\date{\today}

\begin{abstract}
We theoretically study the tunneling conductance of a junction consisting of
a two-dimensional unconventional $p$-wave magnet (UPM) and a superconductor (SC) for
various pairing symmetries. The zero
bias conductance peaks arising from the dispersionless surface Andreev bound
states (SABSs) in $d_{xy}$-wave and $p_{x}$-wave superconductor junctions
are insensitive against varying the magnetic spin-splitting strength $\alpha _{y}$.
Moreover, for chiral $p$- or chiral $d$-wave SCs, zero bias conductance shows
a non-monotonic change as a function of $\alpha_{y}$ indicating the
existence of the dispersive SABSs. Our obtained results of tunneling
spectroscopy based on a UPM serve as an effective way for
the identification of the pairing symmetries of unconventional superconductors.
It is noted that our used Hamiltonian of UPM is also available for persistent spin helix systems. 
\end{abstract}

\maketitle

\section{Introduction}\label{sec:Intro}
Altermagnets (AMs) \cite{LiborSAv,Libor22,Hayami19,Hayami20,landscape22,MazinPNAS,MazinPRX22,Libor011028} 
or materials with momentum-dependent spin-splitting Fermi surfaces \cite{solovyev2006lattice,noda2016momentum,okugawa2018weakly,naka2019spin,naka2020anomalous,naka2021perovskite,naka2022anomalous,yuan2021strong} 
are an emerging class of magnetic materials as a third magnetic phase beyond ferromagnets (FMs) and antiferromagnets (AFMs). AMs are distinct from FMs in the sense that AMs have vanishing macroscopic magnetism. Instead, AMs possess alternating spin-polarized magnetic order in the momentum space in contrast with the spatially varying order in AFMs.  
Candidate AM materials include ${\mathrm{RuO}}_{2}$ \cite{Ahn19,Libor22,Fedchenko24}, $\mathrm{Mn}\mathrm{Te}$ \cite{MnTeLee,osumi2024,Krempaský2024},  ${\mathrm{Mn}_{5} \mathrm{Si}_{3}}$  \cite{Helena2021} as well as semiconductors/insulators like ${\mathrm{MnF}_{2}}$ and ${\mathrm{La}_{2}\mathrm{CuO}_{4}}$ \cite{Moreno16}.

In this context, the interplay between superconductor (SC) and AMs in heterostructures\cite{Beenakker23,Sun23,Papaj23,Ouassou23,Songbo23,Nagae2024,lu2024varphi}  is of particular interest since AMs could make it possible to fabricate superconducting spintronic devices 
discussed in ferromagnet / superconductor (FM/SC) junctions 
\cite{buzdin2005proximity,bergeret2005odd,linder2015superconducting,eschrig2015spin} 
with zero net magnetism. 
In FM/SC junctions, the spin-polarized field in ferromagnets makes the alignment of electron spins and generally suppresses the formation of Cooper pairs near the junction boundaries. 
Nevertheless, AMs may overcome the difficulty of the coexistence of exchange field and pair potential, e.g., with reduced stray field \cite{giil2023}.  
Moreover, several proposals have been formulated to realize topological superconductors in the proximitized altermagnet systems \cite{CanoArxiv,Zhongbo23}. 

Although antiferromagnet / superconductor (AFM/SC) junctions also exhibit anomalous features arising from e.g., so-called $Q$ reflection \cite{bobkova2004spin,andersen2005bound,jakobsen2020electrical,jakobsen2021electrically} with reduced stray field,  AM/SC junctions could show distinguished features due to spin-splitting in the momentum space.

On the other hand, to study Andreev reflection and charge transport in
SC junction has been a very fundamental problem in
superconductivity \cite{BTK82}. It is known that the Andreev reflection and
charge conductance depend significantly on both spin-splitting fields and
unconventional pairings. The former factor indicates that Andreev reflection is a spin-sensitive process and the latter usually gives rise to the surface Andreev bound states (SABSs) resulting in the enhanced Andreev probabilities  \cite%
{TK95,KT96}.
 Non-zero topological invariant obtained from bulk hamiltonian  
leads to 
  the zero-energy Majorana edge state and the corresponding  
flat-band zero energy SABS (ZESABS) or dispersive SABSs
\cite{Sato2011,Schnyder2011,Brydon2011,matsuura2013,Kobayashi14,Kobayashi16,Kobayashi18,tanaka2011symmetry,tanaka2024theory,Tamura2017}%
. In the case of spin-singlet $d$-wave SCs or spin-triplet $p$-wave SC with line nodes, a zero bias conductance peak (ZBCP) appears in normal metal/SC
junctions~\cite{TK95,KT96,Yamashiro98,Kashiwaya2000} due to the presence of
ZESABS \cite{Hu94,Kashiwaya2000}. The ZBCP was experimentally observed in
tunneling spectroscopy on high $T_{\mathrm{c}}$ cuprate ~\cite%
{Alff97,Wei98,Iguchi2000,Kashiwaya2000,Experiment3,Experiment5,Experiment6,Sharoni_2001,Millo2018,Bouscher_2020}. Also, the study of FM
/ unconventional SC junctions have been studied both for
spin-singlet $d$-wave \cite{KashiwayaPRB1999,Zutic1999,Ting2000} and
spin-triplet $p$-wave SCs \cite{HiraiPRB2003}. Basically, the
Andreev reflection and the height of ZBCP are suppressed by the exchange
field in FM /$d$-wave SC. On the other hand, in spin-triplet
junctions, the influence FM depends on the relative directions between the
exchange field and $d$-vector of spin-triplet pairing \cite{Hirai2001}.

Based on these backgrounds of the tunneling spectroscopy, studying  Andreev
reflection and charge transport in AM/SC junctions
for various pairing symmetry becomes an interesting topic. The transport
study of AM/SC junctions indicates a variety of new phenomena. In an AM/$s$%
-wave SC junction, studies show that Andreev reflection is sensitive to both
the crystal orientation and the strength of the spin-splitting field \cite%
{Sun23,Papaj23}, as compared to ferromagnetic materials which are orientationally
independent. The study also shows that the zero-biased conductance peak is
still prominent in the tunneling spectroscopy of an  AM/$d$-wave SC junction.
Interestingly, AM can display $0-\pi $ oscillations in the Josephson
junction without net magnetism\cite{Ouassou23,Songbo23,lu2024varphi}. It is also noted
that the symmetry of the AM is restricted to $d$-wave
in the previous studies of AM/SC junctions. Recent research has shown a variety
of other possibilities of symmetries of magnetism, such as  unconventional  $p$-wave magnet (UPM). 
 Unconventional $p$-wave magnetism  is different from Dirac spin-orbit coupling in that Dirac spin-orbit interaction is a weak interaction that appears in the order of $1/c^2$ expansion of the Dirac equation while  UPM  appears in the non-relativistic limit \cite{Hellenes23}. 
While it has been predicted that even-parity-wave collinear magnetism like $d$-wave AM can be realized in many kinds of materials, the realization of stable  UPM  was recently proposed in a non-coplanar magnetic structure \cite{Hellenes23}.  
Thus, it is worth studying Andreev reflection beside $d$-wave
AM and exploring extensive pairings in the transport property of AM/SC
junctions. 
Here, we study this problem based on an effective model of $p$-wave  magnet  which has been used as 
a model of persistent spin-helix \cite{
Ikegaya2021,Alidoust2021,Yang2017,Alidoust2020,Jacobsen2015,Lee2021,Liu2014}. \par
The organization of this paper is as follows. 
In section \ref{sec:M&F}, we explain the model and the 
theoretical formulation. 
We obtain the 
conductance formula analytically. 
In section \ref{sec:Results}, we show the tunneling conductance 
between   UPM/SC junctions by changing the 
pairing symmetry in SC. We choose 
$s$-wave, $d_{x^{2}-y^{2}}$-wave, $d_{xy}$-wave, 
$p_{x}$-wave, $p_{y}$-wave, chiral $p$-wave, and chiral $d$-wave pairings. The flat band ZESABS appears for  $d_{xy}$-wave and $p_{x}$-wave pairings 
and chiral $p$-wave and chrial $d$-wave pairings have
dispersive SABS. We also study the impact of ferromagnetic insulators  at the interface in the AM/SC junctions. 
In section \ref{sec:concl}, we summarized the obtained results.

\section{Model and Formulation}\label{sec:M&F}

\begin{figure}[tb]
\begin{center}
\includegraphics[width=0.95\columnwidth]{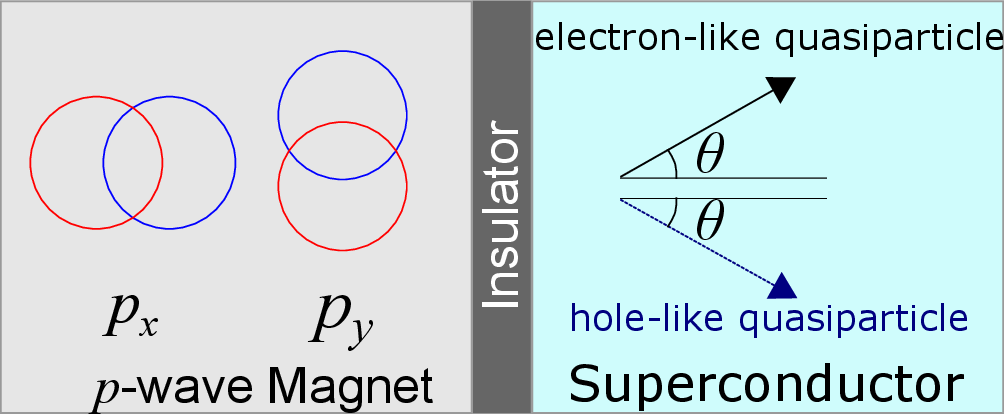}
\end{center}
\caption{(Color online) Schematic illustration of unconventional $p$-wave  magnet/Insulator/Superconductor junction. 
The angle $\theta$ corresponds to the direction of the electron-like quasiparticle in SC measured from the normal to the interface.}
\label{fig:Junction}
\end{figure}
In this section, we consider a  UPM  / Insulator (I) / SC junction as shown in Fig. \ref{fig:Junction}. 
The corresponding Bogoliugov-de Gennes (BdG) Hamiltonian in the system can be written 
by $4 \times 4$ matrix 
as follows: 
\begin{equation}
\check{\mathcal{H}}_{\mathrm{BdG}}=\left[\begin{array}{cc}
\hat{h}\left(\boldsymbol{k},x\right) & \hat{\Delta}\left(\hat{\boldsymbol{k}}\right)\Theta(x)\\
-\hat{\Delta}^{*}\left(-\hat{\boldsymbol{k}}\right)\Theta(x) & -\hat{h}^{*}\left(-\boldsymbol{k},x\right)
\end{array}\right]
\end{equation}
where $\hat{h}\left(\boldsymbol{k},x\right)$ is a 
single-particle Hamiltonian
\begin{equation}
    \label{eq:1particle_hamiltonian_AM}
    \hat{h}\left(\boldsymbol{k},x\right)=\mathrm{diag}(\xi_{+}(\boldsymbol{k},x),\xi_{-}(\boldsymbol{k},x))+\hat{U}_{0}\delta\left(x\right)
\end{equation}
\begin{equation}
\xi_{\pm}(\boldsymbol{k},x)=
\left(\boldsymbol{k}\pm\boldsymbol{\alpha}\Theta\left(-x\right)\right)^{\top}\frac{\hbar^{2}}{2m}\left(\boldsymbol{k}
\pm\boldsymbol{\alpha}\Theta\left(-x\right)\right)-\mu
\label{eq:pAM}
\end{equation}
with a momentum $\boldsymbol{k}=-i\nabla$, the effective mass of an electron 
$m$, and the chemical potential $\mu$. 
It is noted that we employ Eq. \eqref{eq:pAM} to describe UPM, exhibiting the same shape of Fermi surface and the sign of $S_z$ component given in Ref. \cite{Hellenes23}.
Here,  $2 \times 2$ matrix 
$\hat{U}_0$ given by 
\begin{equation}
\hat{U}_{0}=\mathrm{diag}\left(U_{\uparrow},U_{\downarrow}\right)
\end{equation}
denotes the flat insulating barrier at $x=0$. $\Theta(x)$ is the Heaviside step function.

\begin{figure}[htbp]
    \includegraphics[width=0.95\columnwidth]{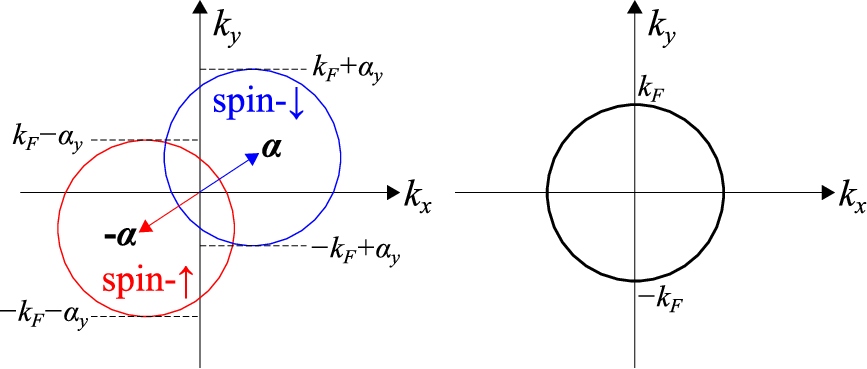}
    \caption{\label{fig:Fermisurfaces}
    (Color online) Schematic illustration of the Fermi surfaces in  unconventional $p$-wave  magnet  (UPM, left panel) and superconductor (SC, right panel). The magnitude and the direction of the splitting of the Fermi surface in  UPM  are described by $\boldsymbol{\alpha}$
vector.}
\end{figure}

Here, the Fermi surfaces for spin-$\uparrow$ and spin-$\downarrow$ 
electrons in  UPM  are split as shown in Fig. \ref{fig:Fermisurfaces} by  $\boldsymbol{\alpha}$ vector. It is noted that the present Hamiltonian
of  UPM  is essentially equivalent to the Hamiltonian of persistent helix
\cite{HelixZhang,HelixKohda,HelixIkegaya,
Ikegaya2021,Alidoust2021,Yang2017,Alidoust2020,Jacobsen2015,Lee2021,Liu2014}. 
We have assumed that the chemical potentials in UPM and SC have the same value $\mu$. Although we can introduce different values $\mu_\mathrm{UPM}$ and $\mu_\mathrm{SC}$ by changing Eq.~\ref{eq:pAM} as
\begin{equation}
    \xi_{\pm}(\boldsymbol{k},x)=
    \left(\boldsymbol{k}\pm\boldsymbol{\alpha}\Theta\left(-x\right)\right)^{\top}\frac{\hbar^{2}}{2m}\left(\boldsymbol{k}
    \pm\boldsymbol{\alpha}\Theta\left(-x\right)\right)
    -(\mu_\mathrm{UPM}\Theta(-x)+\mu_\mathrm{SC}\Theta(x)),
\end{equation}
which leads to different radii of the Fermi surfaces for UPM and SC,
it does not change the results qualitatively.
Thus, we hereafter concentrate on the cases with 
$\mu=\mu_\mathrm{UPM}=\mu_\mathrm{SC}$.

The Nambu spinor $\Psi\left(x,y\right)=\left[u_{\uparrow},u_{\downarrow},v_{\uparrow},v_{\downarrow}\right]^{\top}$ on the field operator basis $\left[\psi_{\uparrow},\psi_{\downarrow},\psi_{\uparrow}^{\dagger},\psi_{\downarrow}^{\dagger}\right]^{\top}$ follows the BdG equation
\begin{equation}
\check{\mathcal{H}}_{\mathrm{BdG}}\Psi=E\Psi
\end{equation}
In the SC region $x>0$, the pair potential is written as 
\begin{equation}
\hat{\Delta}\left(\pm\hat{\boldsymbol{k}}\right)=\left[\begin{array}{cc}
0 & \Delta\left(\theta\right)\\
-\Delta\left(\theta\right) & 0
\end{array}\right]
\end{equation}
\begin{equation}
    \hat{\boldsymbol{k}}=\frac{\boldsymbol{k}}{|\boldsymbol{k}|}
\end{equation}
\begin{equation}
\theta=\arctan\frac{k_{y}}{k_{x}}
\end{equation}
for spin-singlet SC, or 
\begin{equation}
\hat{\Delta}\left(\pm\hat{\boldsymbol{k}}\right)=\pm\left[\begin{array}{cc}
0 & \Delta\left(\theta\right)\\
\Delta\left(\theta\right) & 0
\end{array}\right]
\end{equation}
for spin-triplet SC with a $\boldsymbol{d}$ vector parallel to the $z$-axis, or the N\'{e}el vector of the  UPM, which corresponds to the Cooper pair with the $z$-component of spin $S_z=0$. 
The $y$-component of the wave vector $k_y$ is preserved and the wave function 
can be written as
\begin{equation}
\Psi_{\rho}\left(x,y\right)=\Psi_{\rho}\left(x,k_{y}\right)e^{ik_{y}y}.
\end{equation}
The boundary conditions of the wave function at $x=0$ are written as 
\begin{equation}
\Psi_{\rho} \left( x,k_{y}\right) \Bigg{|}_{x=0_{+}}=\Psi_{\rho} \left( x,k_{y}\right) \Bigg{|}_{x=0_{-}},
\label{eq:boundary_condition_1}
\end{equation}
\begin{equation}
\label{eq:boundary_condition_2}
\check{v}_{x}\Psi_{\rho} \left( x,k_{y}\right) \Bigg{|}_{x=0_{+}}-\check{v}_{x}\Psi_{\rho} \left(
x,k_{y}\right) \Bigg{|}_{x=0_{-}}=\frac{2}{i\hbar}\left(\hat{I}\otimes \hat{U}_{0}\right)\check{\tau}_{z}\Psi_{\rho}\left( 0,k_{y}\right), 
\end{equation}
where $4 \times 4$  matrix 
$\check{v}_{x}$ is a velocity operator given by 
\begin{equation}
\check{v}_{x}=\frac{1}{\hbar}\frac{\partial \check{\mathcal{H}}_{\mathrm{BdG}}}{\partial k_{x}}=\frac{\hbar}{m}\left(
\check{\tau}_{z}\frac{1}{i}\frac{\partial}{\partial x}+\check{\sigma}_{z}\alpha_{x}\Theta(-x)\right),
\label{eq:v_x}
\end{equation}
with
\begin{equation}
\hat{I}=\mathrm{diag}\left(1,1\right), \
\check{\tau}_{z}=\mathrm{diag}\left(1,1,-1,-1\right), \ 
\check{\sigma}_{z}=\mathrm{diag}\left(1,-1,1,-1\right).
\end{equation}

\begin{figure}
    \centering
    \includegraphics{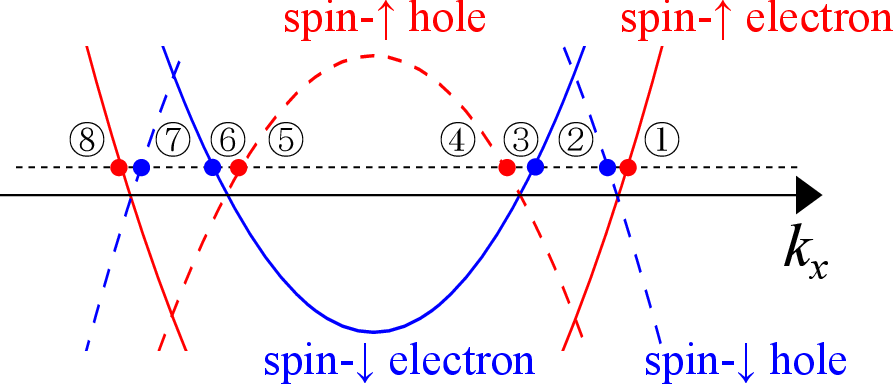}
    \caption{(Color online) Schematic illustration of the dispersion relation in $p_y$-wave  magnet  for $\alpha_{y}k_{y}<0$. The points \textcircled{\scriptsize 1}--\textcircled{\scriptsize 8} correspond to $k_{e\uparrow}^{+},k_{h\downarrow}^{-}, k_{e\downarrow}^{+},k_{h\uparrow}^{-},k_{h\uparrow}^{+},k_{e\downarrow}^{-},k_{h\downarrow}^{+},$ and $k_{e\uparrow}^{-}$, respectively.}
    \label{fig:dispersion_AM}
\end{figure}

In the SC region $x<0$, we derived the expression of the wave functions based on the standard theory of tunneling spectroscopy of unconventional superconductors \cite{Kashiwaya2000} (see Appendix~\ref{sec:appendixA}.)\\

In the  UPM  region $x<0$, the $x$-components of the possible wave vectors for fixed $E$ and $k_y$  are given by  
\begin{equation}
\label{eq:keup}
k_{e\uparrow}^{\pm}=-\alpha_{x}\pm\sqrt{\frac{2m}{\hbar^{2}}\left(E+\mu\right)-\left(k_{y}+\alpha_{y}\right)^{2}},
\end{equation}
\begin{equation}
k_{e\downarrow}^{\pm}=\alpha_{x}\pm\sqrt{\frac{2m}{\hbar^{2}}\left(E+\mu\right)-\left(k_{y}-\alpha_{y}\right)^{2}}, 
\label{eq:kedown}
\end{equation}
\begin{equation}
k_{h\uparrow}^{\pm}=\alpha_{x}\mp\sqrt{\frac{2m}{\hbar^{2}}\left(-E+\mu\right)-\left(k_{y}-\alpha_{y}\right)^{2}}, 
\label{eq:khup}
\end{equation}
\begin{equation}
\label{eq:khdown}
k_{h\downarrow}^{\pm}=-\alpha_{x}\mp\sqrt{\frac{2m}{\hbar^{2}}\left(-E+\mu\right)-\left(k_{y}+\alpha_{y}\right)^{2}}, 
\end{equation}
where the subscripts $e$ and $h$ correspond to an electron and a hole respectively, $\uparrow,\downarrow$ denote the spin, and the superscripts $\pm$ correspond to the sign of the eigenvalues of $\check{v}_{x}$. 
For spin-$\uparrow$ electron injection from $x<0$, the wave function in  UPM  can
be written as 
\begin{equation}
\Psi_{\uparrow}\left(x,k_{y}\right)=\left(\begin{array}{c}
1\\
0\\
0\\
0
\end{array}\right)e^{ik_{e\uparrow}^{+}x}
+r_{\uparrow}\left(\begin{array}{c}
1\\
0\\
0\\
0
\end{array}\right)e^{ik_{e\uparrow}^{-}x}
+r_{\uparrow}^{A}\left(\begin{array}{c}
0\\
0\\
0\\
1
\end{array}\right)e^{ik_{h\downarrow}^{-}x}
\label{eq:psiamup}
\end{equation}
and that in SC can be written as Eq. $\eqref{eq:psiscup}$. In the same way, for spin-$\downarrow$ electron injection, the corresponding 
wave function in  UPM  can
be written as 
\begin{equation}
\Psi_{\downarrow}\left(x,k_{y}\right)=\left(\begin{array}{c}
0\\
1\\
0\\
0
\end{array}\right)e^{ik_{e\downarrow}^{+}x}
+r_{\downarrow}\left(\begin{array}{c}
0\\
1\\
0\\
0
\end{array}\right)e^{ik_{e\downarrow}^{-}x}
+r_{\downarrow}^{A}\left(\begin{array}{c}
0\\
0\\
1\\
0
\end{array}\right)e^{ik_{h\uparrow}^{-}x}
\label{eq:psiamdn}
\end{equation}
and in SC as given in Eq. \eqref{eq:psiscdn}.
Fig. \ref{fig:dispersion_AM} shows the dispersion relation and the $x$-components of the wave vectors  in $p_y$-wave  magnet  with $\boldsymbol{\alpha}\parallel\hat{\boldsymbol{y}}$. 
We see that the absolute value of the $x$-component of the velocity is approximately the same for a spin-$\uparrow(\downarrow)$ electron and a spin-$\downarrow(\uparrow)$ hole for $|E|\ll\mu$.
It is noted that $\alpha_x$ only shifts the curves in Fig. \ref{fig:dispersion_AM} in the $k_{x}$-direction and does not affect the values of the $x$-components of the velocity. 

For spin-$\rho$ $\left(\rho=\uparrow,\downarrow\right)$ electron
injection with particular $k_{y}$, the particle flow 
\begin{equation}
j_{\rho}=\mathrm{Re}
\left(\Psi_{\rho}^{\dagger}\check{v}_{x}\check{\tau}_{z}\Psi_{\rho}\right)\label{eq:particleflow}
\end{equation}
is preserved. We can obtain the transparency  for spin-$\uparrow$ electron injection
\begin{equation}
\sigma_{\uparrow}^{S}\left(E,k_{y}\right)=1+\mathrm{Re}\left(\frac{q_{e\uparrow}^{-}}{q_{e\uparrow}^{+}}\right)\left|r_{\uparrow}\right|^{2}-\mathrm{Re}\left(\frac{q_{h\downarrow}^{-}}{q_{e\uparrow}^{+}}\right)\left|r_{\uparrow}^{A}\right|^{2}
\end{equation}
dividing Eq. $\eqref{eq:particleflow}$ by the flow of the injection
wave. In the same way, the transparency for spin-$\downarrow$ electron injection becomes 
\begin{equation}
\sigma_{\downarrow}^{S}\left(E,k_{y}\right)=1+\mathrm{Re}\left(\frac{q_{e\downarrow}^{-}}{q_{e\downarrow}^{+}}\right)\left|r_{\downarrow}\right|^{2}-\mathrm{Re}\left(\frac{q_{h\uparrow}^{-}}{q_{e\downarrow}^{+}}\right)\left|r_{\downarrow}^{A}\right|^{2}
\end{equation}
Here, 
\begin{equation}
q_{e\uparrow}^{\pm}=k_{e\uparrow}^{\pm}+\alpha_{x},
q_{e\downarrow}^{\pm}=k_{e\downarrow}^{\pm}-\alpha_{x},
q_{h\uparrow}^{\pm}=-k_{h\uparrow}^{\pm}+\alpha_{x},
q_{h\downarrow}^{\pm}=-k_{h\downarrow}^{\pm}-\alpha_{x}
\end{equation}
We can also obtain the transparency for normal metal state $\sigma_{\uparrow}^{N}\left(E,k_{y}\right),\sigma_{\downarrow}^{N}\left(E,k_{y}\right)$
by substituting $\Delta\left(\theta\right)=0$.

The differential conductance normalized by that for the normal metal state as a function of $eV$ with bias voltage $V$ can be written as,
\begin{equation}
\frac{G}{G_{0}}=\frac
{\int_{-k_{F}}^{k_{F}}dk_{y}\left[\sigma_{\uparrow}^{S}\left(E,k_{y}\right)+\sigma_{\downarrow}^{S}\left(E,k_{y}\right)\right]}
{\int_{-k_{F}}^{k_{F}}dk_{y}\left[\sigma_{\uparrow}^{N}\left(E,k_{y}\right)+\sigma_{\downarrow}^{N}\left(E,k_{y}\right)\right]},
\end{equation}
with $E=eV$. 
Note that when $q_{e\uparrow}^{+}$ $\left(q_{e\downarrow}^{+}\right)$
becomes a purely imaginary number, 
it is natural to assume  $\sigma_{\uparrow}^{S}\left(E,k_{y}\right)=\sigma_{\uparrow}^{N}\left(E,k_{y}\right)=0$ 
($\sigma_{\downarrow}^{S}\left(E,k_{y}\right)=\sigma_{\downarrow}^{N}\left(E,k_{y}\right)=0$) 
since there is no traveling wave.
It is noted that $\sigma_{\uparrow(\downarrow)}^{S}\left(E,k_{y}\right)$ and $\sigma_{\uparrow(\downarrow)}^{N}\left(E,k_{y}\right)$ are independent of $\alpha_x$ (see Appendix~\ref{sec:appendixB}).

Now, we assume that $E,\left|\Delta\left(\theta_{\pm}\right)\right|\ll\mu$ following quasiclassical approximation 
\cite{TK95,Kashiwaya2000}, 
\begin{equation}
q_{e\uparrow}^{+}\approx-q_{e\uparrow}^{-}\approx-q_{h\downarrow}^{-}\approx q_{\uparrow}^{\mathrm{AM}}=\sqrt{\frac{2m\mu}{\hbar^{2}}-\left(k_{y}+\alpha_{y}\right)^{2}},
\end{equation}
\begin{equation}
q_{e\downarrow}^{+}\approx-q_{e\downarrow}^{-}\approx-q_{h\uparrow}^{-}\approx q_{\downarrow}^{\mathrm{AM}}=\sqrt{\frac{2m\mu}{\hbar^{2}}-\left(k_{y}-\alpha_{y}\right)^{2}},
\end{equation}
\begin{equation}
k_{e}^{s}\approx k_{h}^{s}\approx k^{s}=\sqrt{\frac{2m\mu}{\hbar^{2}}-k_{y}^{2}},
\end{equation}
where $k_e^s$ and $k_h^s$ are wavevectors of transmitting wave for SC. The concrete form of $k_e^s$ and $k_h^s$ are given by Eqs.~\ref{eq:kes} and \ref{eq:khs}. Then, we obtain the following forms of the conductance in the normal state $\sigma_{\rho}^{N}$ and superconducting state $\sigma_{\rho}^{S}$, 
\begin{equation}
\sigma_{\rho}^{N}\left(E,k_{y}\right)=\frac{\tilde{q}_{\rho}^{\mathrm{AM}}\tilde{k}^{s}}{Z^{2}+\frac{1}{4}\left(\tilde{q}_{\rho}^{\mathrm{AM}}+\tilde{k}^{s}\right)^{2}}\label{eq:sigmaN}
\end{equation}
\begin{equation}
r_{\rho}=\frac{\frac{1}{2}\left(\tilde{q}_{\rho}^{\mathrm{AM}}-\tilde{k}^{s}\right)+iZ}{\frac{1}{2}\left(\tilde{q}_{\rho}^{\mathrm{AM}}+\tilde{k}^{s}\right)+iZ}\frac{\Gamma_{+}\Gamma_{-}}{1-\left[1-\sigma_{\rho}^{N}\left(E,k_{y}\right)\right]\Gamma_{+}\Gamma_{-}}
\end{equation}
\begin{equation}
r_{\rho}^{A}=\begin{cases}
\mathrm{sgn}\left(\rho\right)\frac{\Gamma_{+}\sigma_{\rho}^{N}\left(E,k_{y}\right)}{1-\left[1-\sigma_{\rho}^{N}\left(E,k_{y}\right)\right]\Gamma_{+}\Gamma_{-}} & \mathrm{singlet}\\
\frac{\Gamma_{+}\sigma_{\rho}^{N}\left(E,k_{y}\right)}{1-\left[1-\sigma_{\rho}^{N}\left(E,k_{y}\right)\right]\Gamma_{+}\Gamma_{-}} & \mathrm{triplet}
\end{cases}
\end{equation}
\begin{equation}
\frac{\sigma_{\rho}^{S}\left(E,k_{y}\right)}{\sigma_{\rho}^{N}\left(E,k_{y}\right)}=\frac{1+\sigma_{\rho}^{N}\left(E,k_{y}\right)\left|\Gamma_{+}\right|^{2}-\left[1-\sigma_{\rho}^{N}\left(E,k_{y}\right)\right]\left|\Gamma_{+}\Gamma_{-}\right|^{2}}{\left|1-\left[1-\sigma_{\rho}^{N}\left(E,k_{y}\right)\right]\Gamma_{+}\Gamma_{-}\right|^{2}}
\label{eq:sigmaS}
\end{equation}
with $\rho=\uparrow,\downarrow$. 
In the above, 
$\tilde{q}_{\rho}^{\mathrm{AM}}$, $\tilde{k}^{s}$, and $\Gamma_{\pm}$  
are given by 
\begin{equation}
\tilde{q}_{\rho}^{\mathrm{AM}}=
\frac{q_{\rho}^{\mathrm{AM}}}{k_{F}}, \ 
\tilde{k}^{s} = \frac{k^{s}}{k_{F}}, \ 
k_{F}=\frac{\sqrt{2m\mu}}{\hbar},
\end{equation}

\begin{equation}
\label{eq:Gamma}
\Gamma_{+}=\frac{\Delta^{*}\left(\theta_{+}\right)}{E+\Omega_{+}},\Gamma_{-}=\frac{\Delta\left(\theta_{-}\right)}{E+\Omega_{-}},
\end{equation}
\begin{equation}
\theta_{+}=\arcsin\frac{k_{y}}{k_{F}}, \theta_{-}=\pi-\theta_{+},
\end{equation}
\begin{equation}
\Omega_{\pm}=\lim_{\delta\to0_{+}}\sqrt{\left(E+i\delta\right)^{2}-\left|\Delta\left(\theta_{\pm}\right)\right|^{2}}.
\end{equation}

It is noted that Eq. \eqref{eq:sigmaS}
has the same structure 
as compared to the 
tunneling conductance formula in 
normal metal / unconventional 
superconductor junctions \cite{TK95}. 
Here, the values with a tilde represent that they are divided by the Fermi wavenumber in SC
and $Z$ is the dimensionless parameter 
\begin{equation}
Z=\frac{mU}{\hbar^{2} k_{F}}
\end{equation}
If we consider spin-triplet superconductors with arbitrary direction of $\boldsymbol{d}$-vector, 
the pair potential can be written as 
\begin{equation}
    \hat{\Delta}\left(\pm\hat{\boldsymbol{k}}\right)
    =\pm i\boldsymbol{d}(\theta)\cdot\hat{\boldsymbol{\sigma}}\hat{\sigma}_{y}
    =\pm\left[\begin{array}{cc}
    -d_x(\theta)+id_y(\theta) & d_z(\theta)\\
    d_z(\theta) & d_x(\theta)+id_y(\theta)
    \end{array}\right], 
\end{equation}
with Pauli matrices
\begin{equation}
     \hat{\boldsymbol{\sigma}}=
     \left(\hat{\sigma}_{x},\hat{\sigma}_{y},\hat{\sigma}_{z}\right),
     \end{equation}
     \begin{equation}
     \hat{\sigma}_x
    =\left[\begin{array}{cc}
    0 & 1\\
    1 & 0
    \end{array}\right],
    \hat{\sigma}_y
    =\left[\begin{array}{cc}
    0 & -i\\
    i & 0
    \end{array}\right],
    \hat{\sigma}_z
    =\left[\begin{array}{cc}
    1 & 0\\
   0 & -1
    \end{array}\right].
\end{equation}
In addition to the $\boldsymbol{d\parallel\hat{z}}$ case, we perform calculations for $\boldsymbol{d\parallel\hat{x}}$ case, where the pair potential can be written as 
\begin{equation}
\pm\hat{\Delta}\left(\pm\hat{\boldsymbol{k}}\right)=\left[\begin{array}{cc}
-\Delta\left(\theta\right) & 0\\
0 & \Delta\left(\theta\right)
\end{array}\right]
\end{equation}
with $\hat{\boldsymbol{x}}$ and $\hat{\boldsymbol{z}}$ being the unit vectors parallel to the $x$-axis and the $z$-axis, respectively.

In the presence of a magnetically active  barrier, 
the barrier potential should be 
expressed by $2 \times 2$ matrix with 
\begin{equation}
\hat{Z}=\mathrm{diag}(Z_{\uparrow},Z_{\downarrow})=\frac{m\hat{U}_{0}}{\hbar k_{F}^{2}}. 
\end{equation}

For a spin-$\uparrow(\downarrow)$ electron with a particular value of $k_{y}$ contributing to the conduction process, 
the incident electron and the transparent electron must form traveling waves. 
For $|E|,|\Delta(\theta_{\pm})|\ll\mu$, this condition can be approximately rewritten as 
\begin{equation}
    k_{\uparrow(\downarrow)}^{\mathrm{min}} < k_{y} <   k_{\uparrow(\downarrow)}^{\mathrm{max}}
    \label{eq:kycondition_1},
\end{equation}
where
\begin{equation}
    k_{\uparrow}^{\mathrm{min}} = \max(-k_F,-k_F-\alpha_{y}),  
    k_{\uparrow}^{\mathrm{max}} = \min(k_F,k_F-\alpha_{y}),
\end{equation}
\begin{equation}
    k_{\downarrow}^{\mathrm{min}} = \max(-k_F,-k_F+\alpha_{y}),  
    k_{\downarrow}^{\mathrm{max}} = \min(k_F,k_F+\alpha_{y}), 
\end{equation}
as we can see in Fig. \ref{fig:Fermisurfaces}.
Especially for $|E|<|\Delta(\theta_{\pm})|$, Andreev reflection is needed for the conduction process.
In the cases of spin-singlet pairing or spin-triplet pairing with $\boldsymbol{d}\parallel\hat{\boldsymbol{z}}$, the spin-$\uparrow(\downarrow)$ electron is reflected as a spin-$\downarrow(\uparrow)$ hole. 
The wave vector of the reflected hole is nearly equivalent to that of a spin-$\uparrow(\downarrow)$ electron for $|E|,|\Delta(\theta_{\pm})|\ll\mu$.
Hence, the condition Eq. \eqref{eq:kycondition_1} does not change.
On the other hand, in the case of spin-triplet pairing with $\boldsymbol{d}\parallel\hat{\boldsymbol{x}}$, the spin-$\uparrow(\downarrow)$ electron is reflected as a spin-$\uparrow(\downarrow)$ hole. 
The wave vector of the reflected hole is nearly equivalent to that of a spin-$\downarrow(\uparrow)$ electron for  $|E|,|\Delta(\theta_{\pm})|\ll\mu$.
This changes the condition Eq. \eqref{eq:kycondition_1} for  $|E|,|\Delta(\theta_{\pm})|\ll\mu$ into 
\begin{equation}
    k^{\mathrm{min}} < k_{y} <   k^{\mathrm{max}}
    \label{eq:kycondition_2}
\end{equation}
for both spin-$\uparrow$ and spin-$\downarrow$ electrons, where
\begin{equation}
    k^{\mathrm{min}} = -k_F+|\alpha_{y}|,  
    k^{\mathrm{max}} = k_F-|\alpha_{y}|.
    \label{eq:kmax_kmin}
\end{equation}
It is noted that the strength of  UPM  must satisfy $|\alpha_y|/k_F<2$ to satisfy $k_{\uparrow(\downarrow)}^{\mathrm{min}} <  k_{\uparrow(\downarrow)}^{\mathrm{max}}$ in Eq.~\eqref{eq:kycondition_1}.
\section{Results}\label{sec:Results}


\begin{figure}[htb]
    \includegraphics[scale=1]{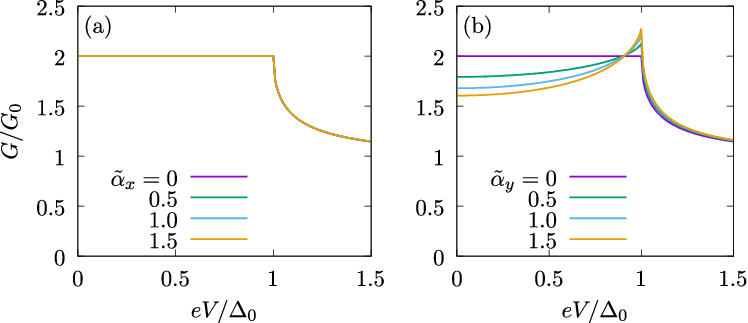}
    \caption{\label{fig:Result1}(Color online) Normalized conductance $G/G_{0}$ of $p$-wave  magnet/insulator/superconductor junctions with s-wave superconductor without the barrier potential ($Z=0$) for (a)$p_{x}$-wave  magnet  case where the Fermi surfaces shifts to the $x$-direction, and for (b)$p_{y}$-wave  magnet  case where the Fermi surfaces shift to the $y$-direction.}
\end{figure}

In Fig. \ref{fig:Result1}, the normalized conductance of UPM/I/SC junction $G/G_{0}$ as a function of the bias voltage $V$ is plotted for various values of $\tilde{\boldsymbol{\alpha}}=\left(\tilde{\alpha}_{x},\tilde{\alpha}_{y}\right)={\boldsymbol{\alpha}}/k_{F}$. 
Here, we choose conventional $s$-wave superconductor 
where pair potential is given by $\Delta_{0}$. 
Figure \ref{fig:Result1}(a) shows the results for $p_{x}$-wave  magnet  with $\boldsymbol{\alpha}$ parallel to the $x$-axis, $i.e.$, the normal to the interface. 
It is noted that $G$ does not depend on 
$\tilde{\alpha}_{x}$ as shown in Fig. \ref{fig:Result1}(a) 
since $\tilde{q}_{\rho}^{\mathrm{AM}}$ is independent of 
$\tilde{\alpha}_{x}$. 
On the other hand, 
as shown in Fig. \ref{fig:Result1}(b), 
$G$ depends on $\tilde{\alpha}_{y}$ since 
$\tilde{q}_{\rho}^{\mathrm{AM}}$ depends on 
$\tilde{\alpha}_{y}$. 
It is noted that when we change the sign of the bias voltage $V$, $G/G_{0}$ always becomes an even function of $eV$ for the results in this paper. 

\begin{figure}[tb]
    \includegraphics{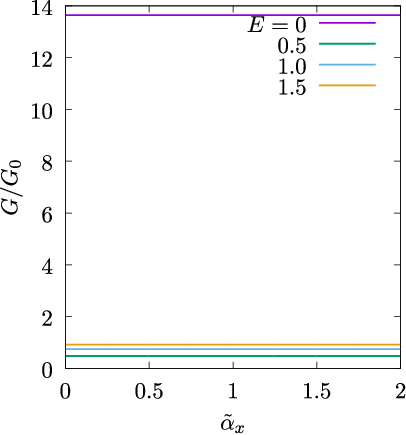}
    \caption{\label{fig:Result_alphax_independence}(Color online) Normalized conductance $G/G_{0}$ of unconventional $p$-wave  magnet/insulator/$d_{xy}$-wave superconductor junction. The barrier potential is set to $Z=2$ and $\tilde{\alpha}_y$ is fixed to $\tilde{\alpha}_y=0.5$.}
\end{figure}

In Fig. \ref{fig:Result_alphax_independence}, the normalized conductance of UPM/I/SC junction $G/G_{0}$ as a function of $\tilde{\alpha}_{x}$ is plotted with a fixed value of $\tilde{\alpha}_{y}$ for various values of $E=eV$. 
This shows that the conductance does not change by $\tilde{\alpha}_{x}$ even for $\tilde{\alpha}_{y}\neq 0$ as proved generally by Eqs. \eqref{eq:boundary_condition_1}, \eqref{eq:boundary_condition_2}, and \eqref{eq:vx_Psi_up_AM}--\eqref{eq:vx_Psi_down_SC}.
Thus we can assume $\tilde{\alpha}_x=0$ without loss of generality.
In the rest of this chapter, we show the results for $p_y$-wave   magnet  cases with $\boldsymbol{\alpha}\parallel\hat{\boldsymbol{y}}$.

\begin{figure}[tb]
  \includegraphics[scale=1]{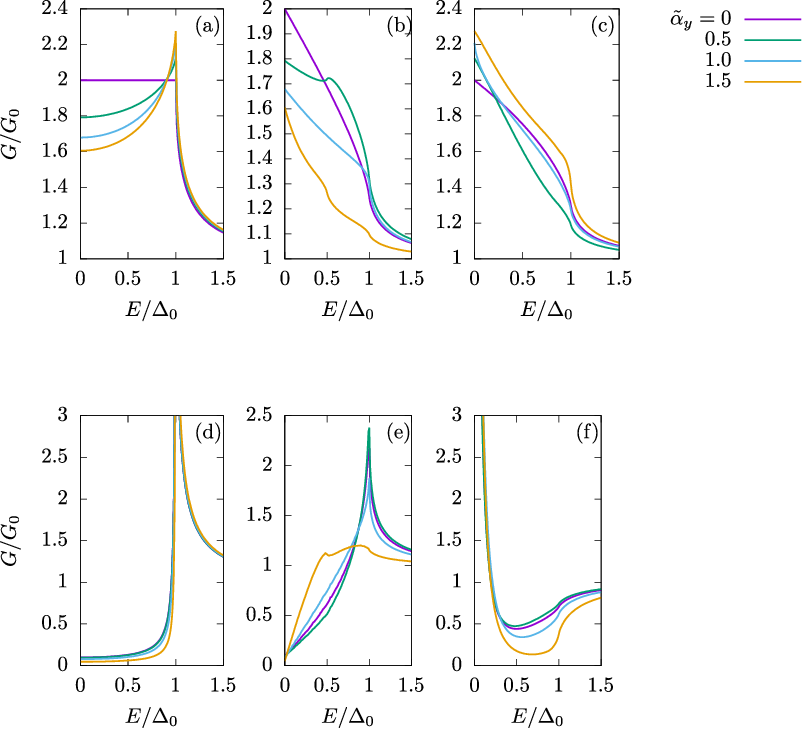}
  \caption{\label{fig:Result2}(Color online) Normalized conductance $G/G_{0}$ of $p_{y}$-wave magnet/insulator/superconductor junctions with spin-singlet superconductors. The barrier potential $Z=0$ for upper panels ((a),(b),(c)), and $Z=2$ for lower panels ((d),(e),(f)). The pairing symmetry of SCs are $s$-wave ((a),(d)), $d_{x^2-y^2}$-wave SC((b),(e)), and $d_{xy}$-wave((c),(f)).}
\end{figure}

\begin{figure}[tb]
  \includegraphics[scale=1]{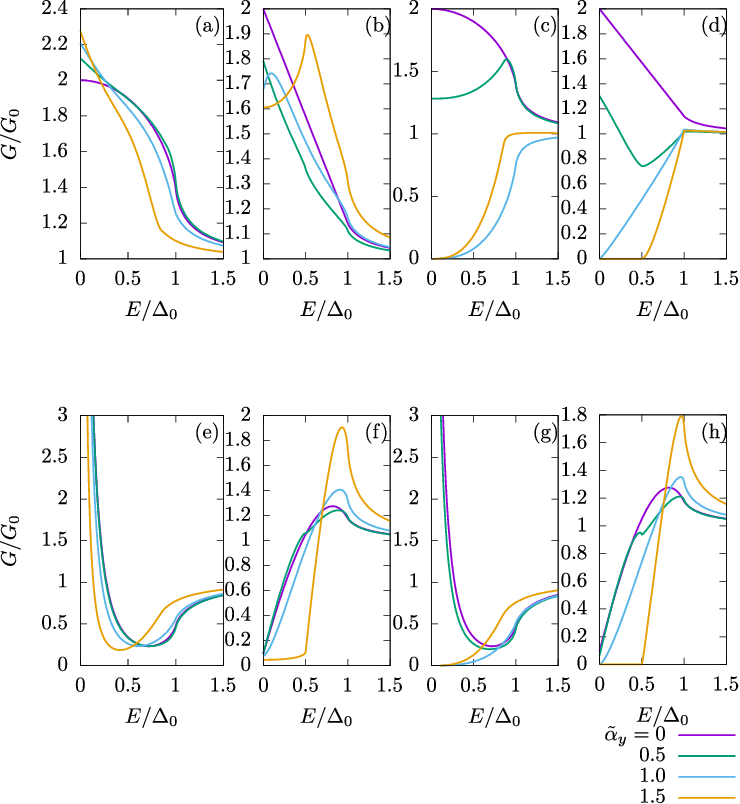}
  \caption{\label{fig:Result3}(Color online) Normalized conductance $G/G_{0}$ of $p_{y}$-wave magnet/insulator/superconductor junctions with spin-triplet superconductors.
  The barrier potential $Z=0$ for upper panels ((a), (b), (c), (d)), and $Z=2$ for lower panels ((e), (f), (g), (h)). The pairing symmetry of SCs are $p_{x}$-wave ((a),(e)), $p_{y}$-wave ((b),(f)) with $\boldsymbol{d} \parallel \hat{\boldsymbol{z}}$, and $p_{x}$-wave ((c),(g)), $p_{y}$-wave ((d),(h)) with $\boldsymbol{d} \parallel \hat{\boldsymbol{x}}$.}
\end{figure}

\begin{figure}[tb]
    \includegraphics{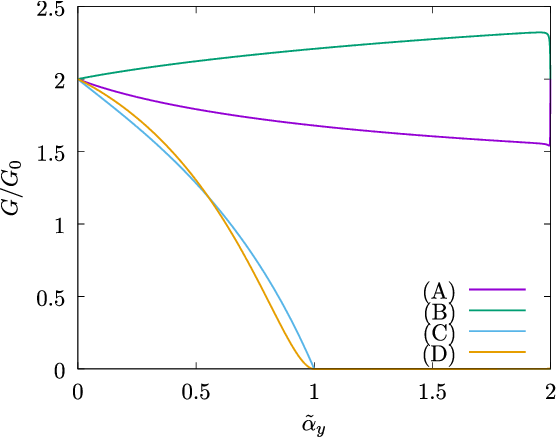}
    \caption{\label{fig:continuous_alpha_Z=0}(Color online) The $\tilde{\alpha}_{y}$ dependence of zero bias conductance at $Z=0$. Pairing symmetries of SCs are
    (A)$s$-wave, $p_y$-wave with $\boldsymbol{d}\parallel\boldsymbol{\hat{z}}$ , $d_{x^2-y^2}$-wave, 
    (B)$p_x$-wave with $\boldsymbol{d}\parallel\boldsymbol{\hat{z}}$ , $d_{xy}$-wave, 
    (C)$p_x$-wave with $\boldsymbol{d}\parallel\boldsymbol{\hat{x}}$, and 
    (D)$p_y$-wave with $\boldsymbol{d}\parallel\boldsymbol{\hat{x}}$
    }
\end{figure}

\begin{figure}[tb]
  \includegraphics{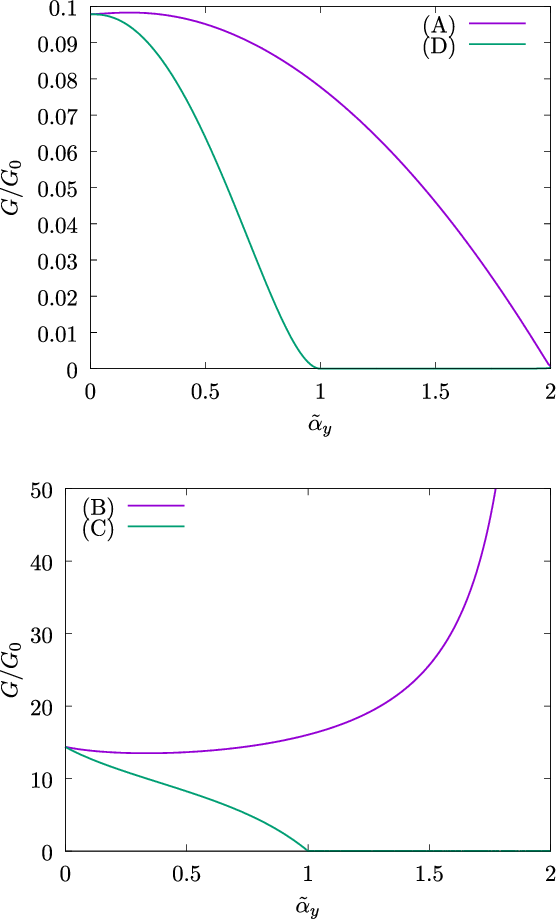}
  \caption{\label{fig:continuous_alpha_Z=2}(Color online) The $\tilde{\alpha}_{y}$ dependence of zero bias conductance at $Z=2$. The pairing symmetry of SCs are
    (A)$s$-wave, $p_y$-wave with $\boldsymbol{d}\parallel\boldsymbol{\hat{z}}$ , $d_{x^2-y^2}$-wave, 
    (B)$p_x$-wave with $\boldsymbol{d}\parallel\boldsymbol{\hat{z}}$ , $d_{xy}$-wave, 
    (C)$p_x$-wave with $\boldsymbol{d}\parallel\boldsymbol{\hat{x}}$, and
    (D)$p_y$-wave with $\boldsymbol{d}\parallel\boldsymbol{\hat{x}}$
    }
\end{figure}

\begin{figure}[tb]
  \includegraphics[scale=1]{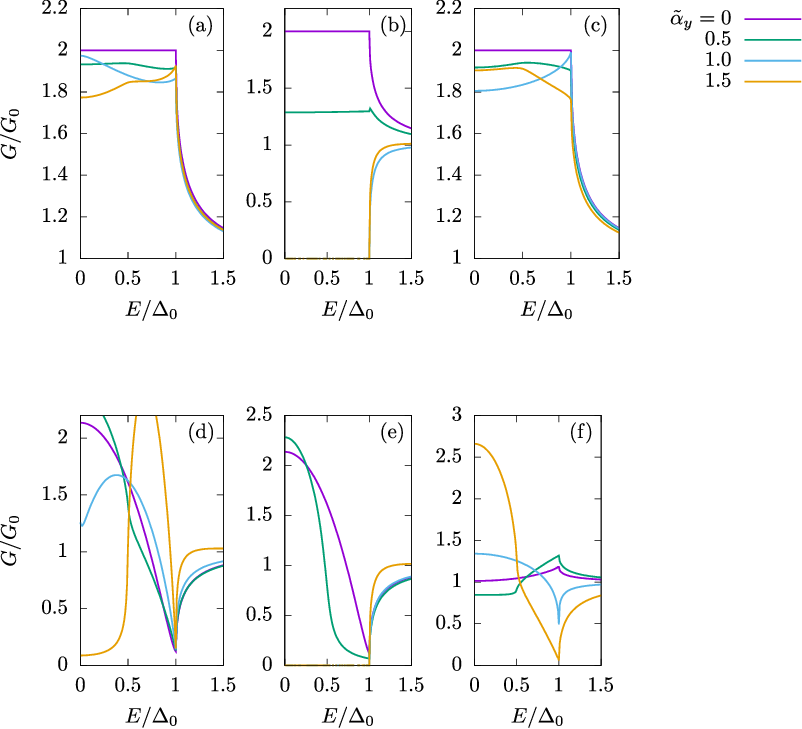}
  \caption{\label{fig:Result4}(Color online) Normalized conductance $G/G_{0}$ of $p_{y}$-wave magnet/insulator/superconductor junctions with chiral superconductors. The barrier potential $Z=0$ for upper panels ((a), (b), (c)), and $Z=2$ for lower panels ((d), (e), (f)).
   The pairing symmetry of SCs are chiral $p$-wave with $\boldsymbol{d} \parallel \hat{\boldsymbol{z}}$ ((a), (d)), chiral $p$-wave SC with $\boldsymbol{d} \parallel \hat{\boldsymbol{x}}$ ((b), (e)), and chiral $d$-wave ((c), (f)).}
\end{figure}

Figures \ref{fig:Result2}--\ref{fig:Result4} show the $\tilde{\alpha}_{y}$ dependence of $G/G_{0}$ for various types  of pairing symmetry of SC.
As shown in Figs. \ref{fig:Result2}(a), (b), (d), and (e) for $s$-wave SC with  $\Delta(\theta)=\Delta_{0}$ and $d_{x^{2}-y^{2}}$-wave SC with anisotropic pair potential $\Delta(\theta)=\Delta_{0} \cos 2\theta$, normalized conductance $G/G_{0}$ is suppressed with the increase of $|\tilde{\alpha}_{y}|$ around zero bias voltage either in high-transparency ($Z=0$) or low-transparency junction ($Z=2$). 
This result can be understood by the suppression of the magnitude of 
$\sigma_{\uparrow(\downarrow)}^{N}(E,k_y)$.
The discrepancy between two values of $\tilde{q}_{\uparrow(\downarrow)}^{AM}$ and $\tilde{k}^{S}$ makes $\sigma_{\uparrow(\downarrow)}^{N}(E,{k}_y)$ in Eq. \eqref{eq:sigmaN} smaller. 
This yields a smaller value of $\sigma_{\uparrow(\downarrow)}^{S}(E,{k}_y)/\sigma_{\uparrow(\downarrow)}^{N}(E,{k}_y)$ for $s$- and $d_{x^{2}-y^{2}}$-wave SC where $|\Gamma_{+}|=|\Gamma_{-}|=1,\Gamma_{+}\Gamma_{-}=-1$ are satisfied for any $k_y$ at $E=0$. 
In contrast, Figs. \ref{fig:Result2}(c) and (f) show that, for $d_{xy}$-wave SC with $\Delta(\theta)=\Delta_{0} \sin 2\theta$, $G/G_{0}$ is enhanced with the increase of $|\tilde{\alpha}_{y}|$ for almost all of $|\tilde{\alpha}_y|$ with $0\le|\tilde{\alpha}_y|\le2$. 
It can be understood that the decrease of $\sigma_{\uparrow(\downarrow)}^{N}(E,{k}_y)$ results in larger magnitude of 
$\sigma_{\uparrow(\downarrow)}^{S}(E,{k}_y)/\sigma_{\uparrow(\downarrow)}^{N}(E,{k}_y)$ for $d_{xy}$-wave SC where $|\Gamma_{+}|=|\Gamma_{-}|=1,\Gamma_{+}\Gamma_{-}=1$ are satisfied for any $k_y$ at $E=0$. 

It is noted that Eq. \eqref{eq:sigmaS}  shows that $\sigma_{\uparrow(\downarrow)}^{S}(E,{k}_y)=2$ is always satisfied for $d_{xy}$-wave SC case regardless of the values of $Z$ and $\tilde{\alpha}_y$. 
This means that an electron at $E=0$ injected from the UPM side is reflected as a hole by Andreev reflection with probability unity no matter how strong the barrier potential of the insulator is.
This is because the group velocities of an incident electron and the corresponding Andreev-reflected hole are the same value in the limit of $|E|/\mu\ll1$ in the case of UPM/I/SC junction, whereas they differ in the case of $d$-wave AM/I/SC junction for the general $k_y$ \cite{Sun23}. 

We perform a similar calculation for spin-triplet SCs as shown in Fig. \ref{fig:Result3}. 
Figures \ref{fig:Result3}(a), (b), (e), and (f) show that  in the cases with  $\boldsymbol{d}\parallel \hat{\boldsymbol{z}}$, 
$G/G_{0}$ is enhanced with the increase of $|\tilde{\alpha}_{y}|$ around the zero bias voltage for $p_{x}$-wave SC with $\Delta(\theta)=\Delta_{0} \cos \theta$, 
while $G/G_{0}$ is suppressed with the increase of $|\tilde{\alpha}_{y}|$ around the zero bias voltage for $p_{y}$-wave SC with $\Delta(\theta)=\Delta_{0} \sin \theta$. 
These features are consistent with Eqs. \eqref{eq:sigmaN}-\eqref{eq:sigmaS} again, 
which is available for both spin-singlet SC and spin-triplet SC with $\boldsymbol{d}\parallel \hat{\boldsymbol{z}}$.

By contrast, for $\boldsymbol{d}\parallel\hat{\boldsymbol{x}}$, 
as shown in Figs. \ref{fig:Result3}(c), (d), (g), and (h), 
$G/G_{0}$ is strongly suppressed by $|\tilde{\alpha}_{y}|$. 
Especially, the conductance becomes 0 at 
$eV=0$ for $|\tilde{\alpha}_{y}| \geq 1$. 
This result can be explained as follows. 
For $\boldsymbol{d}\parallel\hat{\boldsymbol{x}}$, an injected electron and the Andreev-reflected hole must have the same spin angular momentum. 
Andreev reflection only occurs for $k_{y}$ with which both $k_{e\uparrow(\downarrow)}^{+}$ and $k_{h\uparrow(\downarrow)}^{-}$ are real numbers. 
As shown in Eqs. \eqref{eq:keup} and \eqref{eq:khup} (\eqref{eq:kedown} and \eqref{eq:khdown}), the region of $k_{y}$ where both $k_{e\uparrow(\downarrow)}^{+}$ and $k_{h\uparrow(\downarrow)}^{-}$ are real becomes narrow with the increase of $|\tilde{\alpha}_y|$.
This results in a more strict condition for $k_{y}$ to make conduction possible than that in the normal metal case.
Especially, for $|\tilde{\alpha}_y|\geq 1$, at least one of $k_{e\uparrow(\downarrow)}^{+}$ and $k_{h\uparrow(\downarrow)}^{-}$ becomes imaginary number for all of $k_{y}$ with $|eV|\ll\mu$, which results in $G/G_{0}=0$.

To summarize the results so far, we calculate $G/G_{0}$ at $eV=0$ as a function of $\tilde{\alpha}_{y}$ as shown in Figs.~\ref{fig:continuous_alpha_Z=0} and \ref{fig:continuous_alpha_Z=2}.
These figures show that the zero-bias conductance of $p_y$-wave  magnet/I/SC junction is dramatically changed by the pairing symmetry of the superconductor and that the junction can be used to detect the pairing symmetry. 
We also anticipate that it is possible to determine the symmetries of unknown candidates of 
 magnets with momentum-dependent spin-splitting  
by measuring the zero-bias conductance of  magnet/SC junctions.
Actually, Figs.~\ref{fig:continuous_alpha_Z=0} and \ref{fig:continuous_alpha_Z=2} show that the conductance of the UPM/spin-triplet SC junction with $\boldsymbol{d}\parallel\hat{\boldsymbol{z}}$ is generally larger than that of the cases with the $\boldsymbol{d}\perp\hat{\boldsymbol{z}}$. 
On the other hand, Figs.~\ref{fig:dxyAM_continuous_alpha_Z=2} and \ref{fig:dx2-y2AM_continuous_alpha_Z=2} in Appendix~\ref{sec:appendixC} show that the conductance for the $\boldsymbol{d}\parallel\hat{\boldsymbol{z}}$ case is smaller than or equal to that for the $\boldsymbol{d}\perp\hat{\boldsymbol{z}}$ case. 
Qualitative change of unnormalized zero-bias conductance is summarized in Table~\ref{tab:table1}. 
The most significant difference between the cases of a  UPM  junction and a $d$-wave AM one is that the change of conductance under the $\pi/2$ rotation of the $c$-crystal axis. 
In the case of a  UPM  junction, the normalized exchange energy ($\tilde{\alpha}_x$ or $\tilde{\alpha}_y$) dependence of the conductance significantly changes because of the broken fourfold symmetry of the Fermi surface, while there is no change in the case of a $d$-wave AM junction. 
In addition, the suppression of the conductance by the exchange energy of the $d$-wave AM junction is relatively smaller than that of the  UPM  cases for $\boldsymbol{d}\perp\hat{\boldsymbol{z}}$.
\begin{table}[]

\caption{The qualitative change of conductance $G$ of magnet/I/SC junctions with the increase of the normalized strength of  magnets  $\tilde{\alpha}_x,\tilde{\alpha}_y,\tilde{\alpha}_1,$ or $\tilde{\alpha}_2$ where $\tilde{\alpha_x}$ and $\tilde{\alpha_y}$ denote the normalized strength of the exchange field in  unconventional $p$-wave magnet  and $\tilde{\alpha}_1$ and $\tilde{\alpha}_2$ denote those in $d$-wave altermagnet. It is noted that the values of conductance are not normalized by that in the normal state. 
These results occur for both the cases $Z=2$ and $Z=5$. The expressions ``decrease*'' and ``decrease**'' indicate that the conductance reaches 0 when the strength of  magnets  $\tilde{\alpha}_x,\tilde{\alpha}_y,\tilde{\alpha}_1,$ or $\tilde{\alpha}_2$ reaches 1 and 2, respectively. On the other hand, ``suppressed'' and ``enhanced'' denote that the change of the conductance is insignificant.}
\label{tab:table1}
\begin{tabular}{@{}ccccc@{}}
\hline
& $p_x$ UPM & $p_y$ UPM   & $d_{xy}$ AM & $d_{x^2-y^2}$ AM \\ 
\hline
$p_x$ SC $\boldsymbol{d}\parallel\hat{\boldsymbol{z}}$      & constant & decrease** & constant    & decrease*        \\
$p_x$ SC $\boldsymbol{d}\perp\hat{\boldsymbol{z}}$          & constant & decrease*  & constant    & suppressed         \\ 
\hline
chiral $p$ SC $\boldsymbol{d}\parallel\hat{\boldsymbol{z}}$ & constant & decrease** & enhanced    & decrease*        \\
chiral $p$ SC $\boldsymbol{d}\perp\hat{\boldsymbol{z}}$     & constant & decrease*  & enhanced    & suppressed         \\ 
\hline
\end{tabular}

\end{table}

As shown in Fig. \ref{fig:Result4}, the $\tilde{\alpha}_{y}$ dependence of $G/G_{0}$ becomes more complicated for chiral $p$-wave SC with the pair potential $\Delta(\theta)=\Delta_{0}e^{i\theta}$ and chiral $d$-wave one with $\Delta(\theta)=\Delta_{0}e^{2i\theta}$ as compared to  Figs. \ref{fig:Result2} and \ref{fig:Result3}.  

\begin{figure}[tb]
     \includegraphics[scale=1]{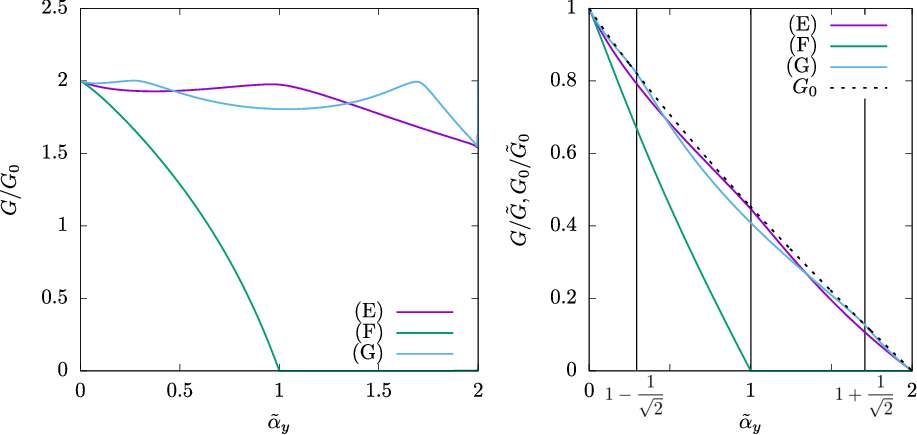}
    \caption{\label{fig:continuous_alpha_Z=0_chiral}(Color online) The $\tilde{\alpha}_{y}$ dependence of normalized zero-bias conductance $G/G_{0}$ (left panel)  and the unnormalized value $G$ divided by its maximum value $\tilde{G}$ (right panel) at $Z=0$. Pairing symmetries of SCs are
    (E)chiral $p$-wave with $\boldsymbol{d}\parallel\boldsymbol{\hat{z}}$, 
    (F)chiral $p$-wave with $\boldsymbol{d}\parallel\boldsymbol{\hat{x}}$, and 
    (G)chiral $d$-wave. 
    The dotted line corresponds to the conductance of the normal-metal state $G_0$. 
    }
\end{figure}

\begin{figure}[tb]
     \includegraphics[scale=1]{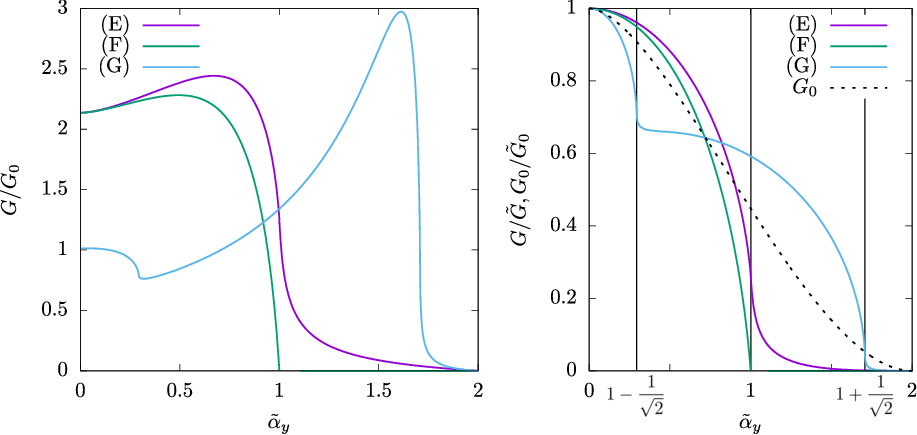}
    \caption{\label{fig:continuous_alpha_Z=2_chiral}(Color online) The $\tilde{\alpha}_{y}$ dependence of normalized zero-bias conductance $G/G_{0}$ (left panel)  and the unnormalized value $G$ divided by its maximum value $\tilde{G}$ (right panel) at $Z=2$. Pairing symmetries of SCs are
    (E)chiral $p$-wave with $\boldsymbol{d}\parallel\boldsymbol{\hat{z}}$, 
    (F)chiral $p$-wave with $\boldsymbol{d}\parallel\boldsymbol{\hat{x}}$, and 
    (G)chiral $d$-wave. 
    The dotted line corresponds to the conductance of the normal-metal state $G_0$. 
    }
\end{figure}

To clarify these features, we calculate $G/G_{0}$ for chiral SC cases at $eV=0$ as a function of $\tilde{\alpha}_{y}$ as shown in the left panels of Figs. \ref{fig:continuous_alpha_Z=0_chiral} and \ref{fig:continuous_alpha_Z=2_chiral}. 
$G/G_{0}$ has at most one local maximum in the chiral $p$-wave SC cases, while for chiral $d$-wave SC, $G/G_{0}$ has two local maxima, as shown in Figs. \ref{fig:continuous_alpha_Z=0_chiral} and \ref{fig:continuous_alpha_Z=2_chiral}. 
These features are significantly different from those of $p_{x}$-wave and $d_{xy}$-wave pairings. 
For the comparison, we show the value of  $G$ and $G_{0}$ for chiral SC cases at $eV=0$ as a function of $\tilde{\alpha}_{y}$ divided by each maximum value $\tilde{G}$ or $\tilde{G}_0$ in the right panels of Figs. \ref{fig:continuous_alpha_Z=0_chiral} and \ref{fig:continuous_alpha_Z=2_chiral}. 
Since both $G$ and $G_{0}$ are monotonic decreasing functions of $|\tilde{\alpha}_{y}|$ 
and $\partial G_{0}/\partial \tilde{\alpha}_y$ does not vary significantly by $\tilde{\alpha}_{y}$, 
$G/G_{0}$ is enhanced with the increase of $\tilde{\alpha}_y$ when $|\partial G/\partial \tilde{\alpha}_y|$ has a particularly small value.

\begin{figure}[tb]
  \includegraphics{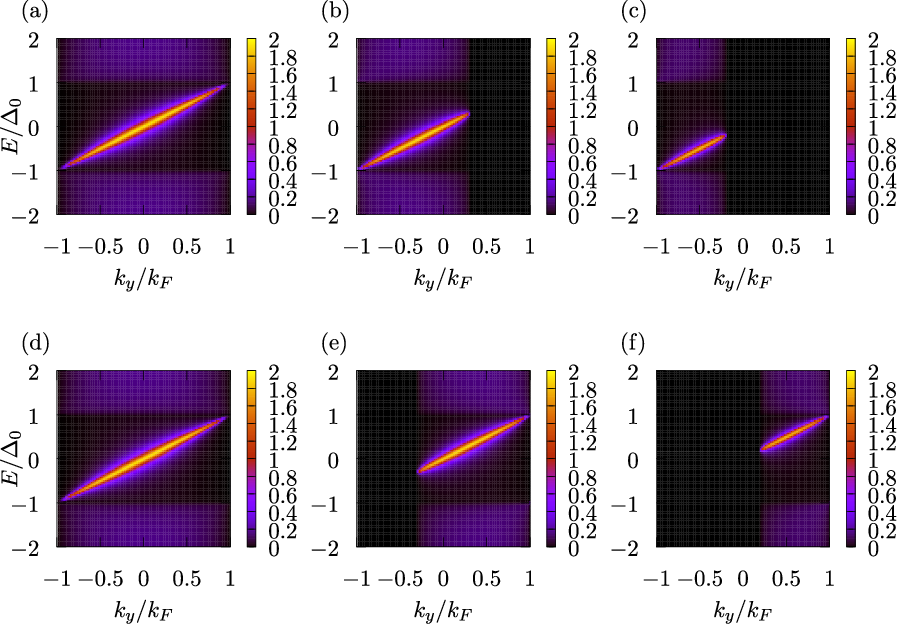}
  \caption{\label{fig:kyvsG_chiralpSC_dz}(Color online) Momentum resolved conductance $\sigma_{\uparrow}^{S}(E,k_y)$ ((a),(b),(c)) and $\sigma_{\downarrow}^{S}(E,k_y)$ ((d),(e),(f)) for $p_{y}$-wave magnet/Insulator/chiral $p$-wave superconductor with $d \parallel \hat{z}$ junction where $Z=2$. The strength of  $p$-wave  magnet  is set to $\tilde{\alpha}_{y}=0$((a),(d)), $0.7$((b),(e)), and $1.2$((c),(f)).
  The number of chiral-edge mode is $N=1$ and the values of $M$ and $k_1$ in Eq. \eqref{eq:k_M} becomes $M=1$ and $k_{1}/k_{F}=0$.}
\end{figure}
\begin{figure}[tb]
  \includegraphics{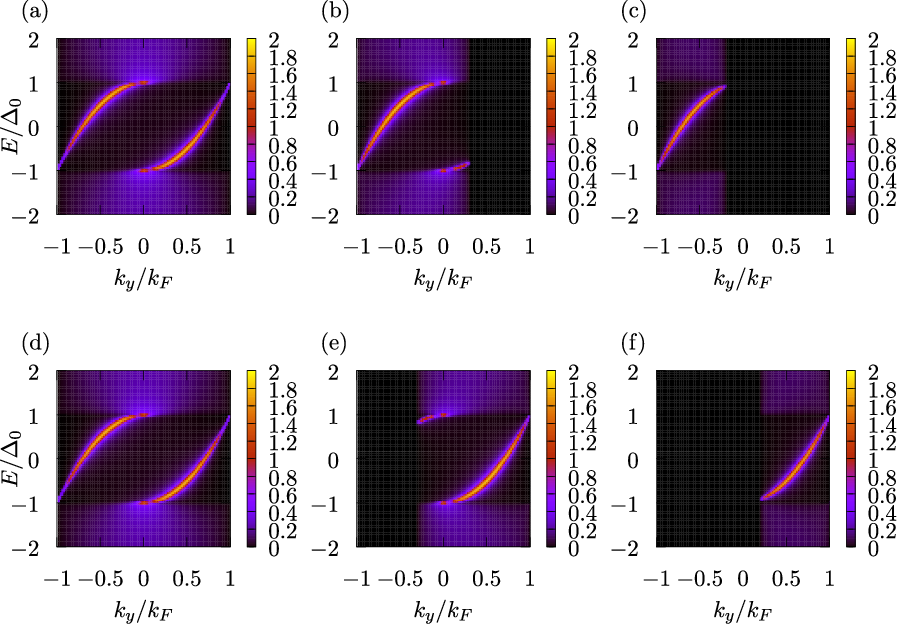}
  \caption{\label{fig:kyvsG_chiraldSC}(Color online) Momentum resolved conductance $\sigma_{\uparrow}^{S}(E,k_y)$ ((a),(b),(c)) and $\sigma_{\downarrow}^{S}(E,k_y)$ ((d),(e),(f)) for $p_{y}$-wave magnet/Insulator/chiral $d$-wave superconductor junction where $Z=2$. The strength of  $p$-wave  magnet  is set to $\tilde{\alpha}_{y}=0$((a),(d)), $0.7$((b),(e)), and $1.2$((c),(f)).
  The number of chiral-edge modes is $N=2$ and the values of $M$ and $k_1$ in Eq. \eqref{eq:k_M} becomes $M=1$ and $k_{1}/k_{F}=1/\sqrt{2}$.}
\end{figure}
\begin{figure}[tb]
  \includegraphics{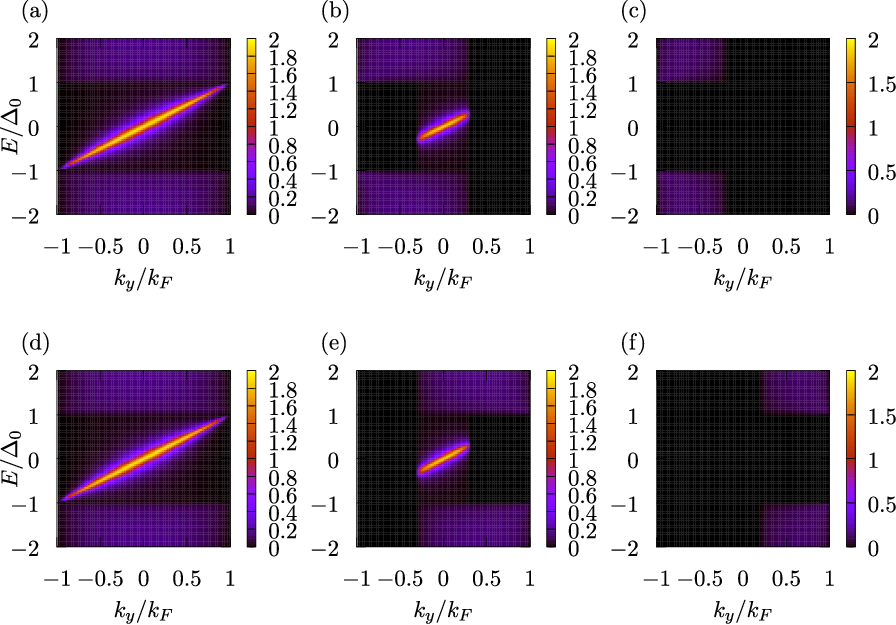}
  \caption{\label{fig:kyvsG_chiralpSC_dx}(Color online) Momentum resolved conductance $\sigma_{\uparrow}^{S}(E,k_y)$ ((a),(b),(c)) and $\sigma_{\downarrow}^{S}(E,k_y)$ ((d),(e),(f)) for $p_{y}$-wave magnet/Insulator/chiral $p$-wave superconductor with $d \parallel \hat{x}$ junction where $Z=2$. The strength of  $p$-wave  magnet  is set to $\tilde{\alpha}_{y}=0$((a),(d)), $0.7$((b),(e)), and $1.2$((c),(f)).
  The number of chiral-edge mode is $N=1$ and the values of $M$ and $k_1$ in Eq. \eqref{eq:k_M} becomes $M=1$ and $k_{1}/k_{F}=0$.}
\end{figure}

To understand this non-monotonic $\tilde{\alpha}_y$ dependence of $G/G_0$ for chiral SC cases, we show the $k_y$- and spin-resolved conductance of AM/I/SC junction 
$\sigma_{\uparrow}^{S}\left(E,{k}_{y}\right)$ and $\sigma_{\downarrow}^{S}\left(E,{k}_{y}\right)$
in Figs. \ref{fig:kyvsG_chiralpSC_dz}--\ref{fig:kyvsG_chiralpSC_dx}. 
As shown in Figs. \ref{fig:kyvsG_chiralpSC_dx}(a), (d), \ref{fig:kyvsG_chiraldSC}(a), (d), and \ref{fig:kyvsG_chiralpSC_dz}(a), (d), the energy dispersion corresponding to the maximum value of $\sigma_{\uparrow}^{S}\left(E,{k}_{y}\right)$ and $\sigma_{\downarrow}^{S}\left(E,{k}_{y}\right)$ at $\tilde{\alpha}_y=0$ corresponds to so-called chiral edge mode. As the value of $|\tilde{\alpha}_y|$ increases, the range of $k_y$ contributing to the conduction process becomes restricted.

Both in the cases of the spin-triplet chiral $p$-wave SC with $\boldsymbol{d} \parallel\hat{\boldsymbol{z}}$ and spin-singlet chiral $d$-wave SC, the range of $k_y$ contributing to the conduction process becomes restricted 
as shown in Figs. \ref{fig:kyvsG_chiralpSC_dz} and \ref{fig:kyvsG_chiraldSC}. 
This corresponds to Eq. \eqref{eq:kycondition_1}.
For the other values, $\sigma_{\uparrow(\downarrow)}^{S}\left(E,{k}_{y}\right)=0$.

For generality, we assume $\sigma_{\uparrow(\downarrow)}^{S}\left(E=0,{k}_{y}\right)$ 
is prominently enhanced around ${k}_{y} = \pm k_{1}$, $\pm k_{2}$, $\dots$, $\pm k_{M}$ with $0\le k_{1} < k_{2} < \dots < k_{M}$ 
corresponding to chiral edge mode with $E=0$ in spin-singlet SC or spin-triplet SC with $\boldsymbol{d}\parallel\hat{\boldsymbol{x}}$. 
Here, the integer $M$ is related to the number of chiral-edge modes $N$ as 
\begin{equation}
\left\{ \,
    \begin{aligned}
    & M=\frac{N+1}{2}, k_1=0 & \text{for odd } N,\\
    & M=\frac{N}{2}, k_1\neq 0 & \text{for even } N.
    \end{aligned}
    \label{eq:k_M}
\right.
\end{equation}
When $|\tilde{\alpha}_y|<1\pm k_{M}/k_{F}$ is satisfied, except for $|\tilde{\alpha}_y|\approx1\pm{k}_{i}/k_{F}$ with $i=1,2,\dots,M$, the increase of $|\tilde{\alpha}_y|$ does not substantially influence $G$ at $E=0$ 
since the values of $\sigma_{\uparrow(\downarrow)}^{S}\left(E=0,{k}_{y}\right)$ at 
$k_{\uparrow(\downarrow)}^{\mathrm{min}}, k_{\uparrow(\downarrow)}^{\mathrm{max}}$ are much smaller than that for ${k}_{y} \approx \pm k_{1}, \pm k_{2}, \dots, \pm k_{M}$. 
On the other hand, for $|\tilde{\alpha}_y|\approx1\pm k_{i}/k_{F}$ with $i=1,2,\dots,M$,
the increase of $|\tilde{\alpha}_y|$ significantly suppresses $G$ at $E=0$ since $|k_{\uparrow}^{\mathrm{max}}|=|k_{\downarrow}^{\mathrm{min}}|\approx k_{i}$ ($|k_{\downarrow}^{\mathrm{max}}|=|k_{\uparrow}^{\mathrm{min}}|\approx k_{i}$) and the values of $\sigma_{\uparrow(\downarrow)}^{S}\left(E=0,{k}_{y}\right)$  at $k_{\uparrow}^{\mathrm{max}}$ and $k_{\downarrow}^{\mathrm{min}}$ ($k_{\downarrow}^{\mathrm{max}}$ and $k_{\uparrow}^{\mathrm{min}}$) are not significantly smaller than the maximum value of $\sigma_{\uparrow(\downarrow)}^{S}\left(E=0,{k}_{y}\right)$.
When the value of $|\tilde{\alpha}_y|$ exceeds $1+{k}_{M}/k_{F}$, the maximum value of $\sigma_{\uparrow(\downarrow)}^{S}\left(E=0,{k}_{y}\right)$ for Eq. \eqref{eq:kycondition_1} is reduced  and $G/G_{0}$ is suppressed with the increase of $|\tilde{\alpha}_y|$. 
Hence, $G/G_0$ at $E=0$ as a function of $|\tilde{\alpha}_{y}|$ has $N$ local maxima.
As shown in Fig. \ref{fig:kyvsG_chiralpSC_dz}, $\sigma_{\uparrow(\downarrow)}^{S}\left(E=0,{k}_{y}\right)$ is prominently enhanced around ${k}_{y}=0$ for chiral $p$-wave SC with $\boldsymbol{d}\parallel\hat{\boldsymbol{z}}$.  This corresponds to $N=1$ and  $G/G_0$ at $E=0$ as a function of $|\tilde{\alpha}_{y}|$ has one local maximum. Since $N=2$ (in the present calculation, $k_1/k_F=1/\sqrt{2}$) for chiral $d$-wave SC as shown in Fig. \ref{fig:kyvsG_chiraldSC}, the corresponding $G/G_0$ at $E=0$ as a function of $|\tilde{\alpha}_{y}|$ has two local maxima.

In the case of the spin-triplet chiral $p$-wave SC with $\boldsymbol{d} \parallel\hat{\boldsymbol{x}}$, 
the range of $k_y$ contributing to the conduction process becomes restricted 
as shown in Fig. \ref{fig:kyvsG_chiralpSC_dx}. 
This corresponds to Eq. \eqref{eq:kycondition_2}.
For the other values, $\sigma_{\uparrow(\downarrow)}^{S}\left(E,{k}_{y}\right)=0$ with $|E|<\Delta_0$. 
As we see in Fig. \ref{fig:kyvsG_chiralpSC_dx}, $\sigma_{\uparrow(\downarrow)}^{S}(E=0,{k}_{y})$ is prominently enhanced around $k_{y}=0$ for the chiral $p$-wave pairing case with $\boldsymbol{d}\parallel\hat{\boldsymbol{x}}$. 
For sufficiently large $Z$ with strong barrier, when $|\tilde{\alpha}_y|<1$ is satisfied, except for $|\tilde{\alpha}_y|\approx1$, the increase of $|\tilde{\alpha}_y|$ does not substantially influence $G$ at $E=0$ since the values of $\sigma_{\uparrow(\downarrow)}^{S}\left(E=0,{k}_{y}\right)$ at 
$k_{y}=k^{\mathrm{min}}$ and $k_y=k^{\mathrm{max}}$ are much smaller than that for ${k}_y\approx0$. 
However, for $|\tilde{\alpha}_y|\approx$1, the increase of $|\tilde{\alpha}_y|$ significantly suppresses $G$ at $E=0$. 
In this case, $|k^{\mathrm{min}}|=|k^{\mathrm{max}}|\approx0$ is satisfied, and the values of $\sigma_{\uparrow(\downarrow)}^{S}\left(E=0,{k}_{y}\neq 0\right)$ at 
$k^{\mathrm{min}}$ and $k^{\mathrm{max}}$  are not significantly smaller as compared to the maximum value $\sigma_{\uparrow(\downarrow)}^{S}\left(E=0,{k}_{y}=0\right)$. 
This results in suppression of $G/G_{0}$ like the other spin-triplet SC cases shown in Figs. \ref{fig:Result3}(c), (d), (g), and (h).
It is noted that, for a weak barrier,  $\sigma_{\uparrow(\downarrow)}^{S}(E=0,{k}_{y})$ at $k_y$ apart from $k_y\approx0$ is a little bit smaller than $\sigma_{\uparrow(\downarrow)}^{S}(E=0,{k}_{y}=0)$, and $G$ decreases almost linearly as a function of $|\tilde{\alpha}_y|$ for $|\tilde{\alpha}_y|<1$ as shown in the right panel of Fig. \ref{fig:continuous_alpha_Z=0_chiral}.
This may lead to no local maxima of $G/G_{0}$ as shown in Fig. \ref{fig:continuous_alpha_Z=0_chiral}(G).\par
 
Besides the $\tilde\alpha_y$-dependence of $G/G_0$ at $eV = 0$, 
the voltage dependence of $G/G_0$ with a single value of $\tilde\alpha_y$ around $\tilde\alpha_y = 1.5$ 
and strong barrier 
can be an important feature to distinguish chiral $p$-wave SC and chiral $d$-wave ones. 
As shown in Figs.~\ref{fig:kyvsG_chiralpSC_dz}~(c), (f), for $p_y$-wave  magnet  / chiral $p$-wave SC $(\boldsymbol{d}\parallel\hat{\boldsymbol{x}})$ junctions with $\tilde{\alpha}_y>1$, $k_y\approx 0$ corresponding to the chiral edge mode with $E\approx 0$ cannot contribute to the conduction while $k_y\approx \pm k_F$ corresponding to the chiral edge mode with $E\approx \pm \Delta_0$ can.
In this case, as shown in Fig.~\ref{fig:Result4}~(d),  
$G/G_0$ as a function of $E=eV$ has a conspicuous concave around $E=0$.
On the other hand, Figs.~\ref{fig:kyvsG_chiraldSC}~(c), (f) shows that for $p_y$-wave  magnet  / chiral $d$-wave SC junctions with $1<\tilde{\alpha}_y<1+1/\sqrt2$, $k_y\approx \pm 1/\sqrt{2}$ corresponding to the chiral edge mode with $E\approx 0$ can contribute to the conduction of both spin-$\uparrow$ and $\downarrow$ electrons, while $k_y\approx 0$ which is one of the two values of $k_y$ corresponding to the chiral edge mode with a certain energy of $E\approx \pm \Delta_0$ cannot contribute to the conduction.
In this case, as shown in Fig.~\ref{fig:Result4}~(f),  
$G/G_0$ as a function of $E=eV$ has a conspicuous convex around $E=0$.
As shown in Fig.~\ref{fig:Result4}~(e), 
$G/G_0$ becomes zero for $|eV|<\Delta_0$ for $p_y$-wave  magnet  / chiral $p$-wave SC $(\boldsymbol{d}\parallel\hat{\boldsymbol{x}})$ junctions with $\tilde{\alpha}_y>1$. 
These features are so distinctive that it would be better to use UPM/SC junctions instead of normal-metal / SC ones to clarify the SCs with broken time-reversal symmetry. 
\par
It is noted that the non-monotonic changes of $G/G_{0}$ as a function of the strength of AM in AM/I/chiral SC junction like those shown in Figs. \ref{fig:continuous_alpha_Z=0_chiral} and \ref{fig:continuous_alpha_Z=2_chiral} cannot be seen in the $d$-wave AM case as shown in Appendix \ref{sec:appendixC}.

\begin{figure}[tb]
    \includegraphics[scale=1]{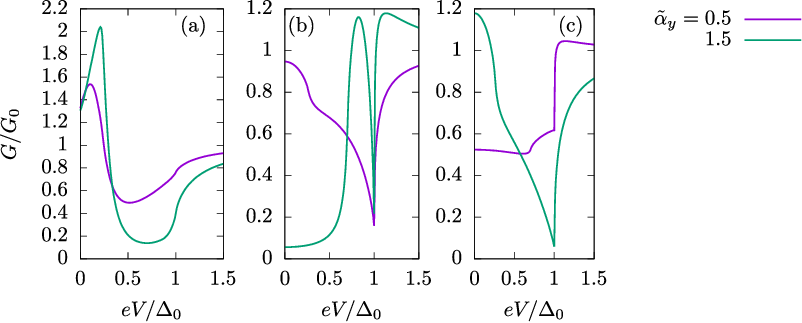}
    \caption{\label{fig:Result_FI}(Color online) Normalized conductance $G/G_{0}$ of $p_{y}$-wave magnet/Ferromagnetic Insulator/superconductor junctions. The barrier is set to $Z_{\uparrow}=1, Z_{\downarrow}=3$. (a)$d_{xy}$-wave SC. (b)chiral $p$-wave SC with $\boldsymbol{d} \parallel \hat{\boldsymbol{z}}$.  (c)chiral $d$-wave SC.}
\end{figure}

\begin{figure}[tb]
  \includegraphics[scale=1]{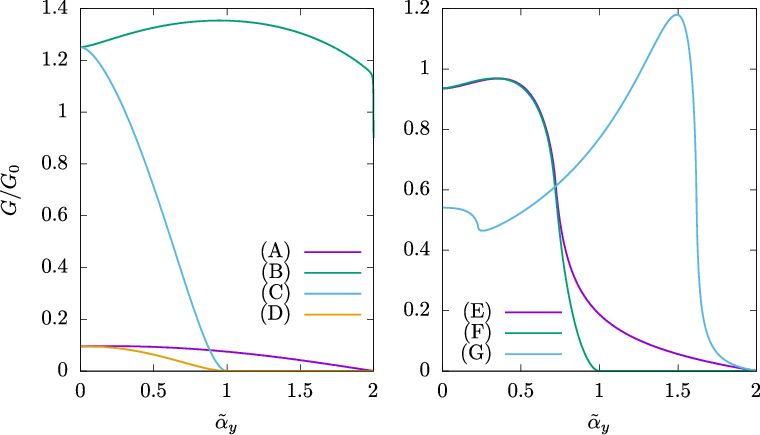}
  \caption{\label{fig:continuous_alpha_Z=1,3}(Color online) The $\alpha_{y}$ dependency of zero bias conductance at $Z_{\uparrow}=1,Z_{\downarrow}=3$.
    (A)$s$-wave, $p_y$-wave with $\boldsymbol{d}\parallel\boldsymbol{\hat{z}}$ , $d_{x^2-y^2}$-wave, 
    (B)$p_x$-wave with $\boldsymbol{d}\parallel\boldsymbol{\hat{z}}$ , $d_{xy}$-wave, 
    (C)$p_x$-wave with $\boldsymbol{d}\parallel\boldsymbol{\hat{x}}$,
    (D)$p_y$-wave with $\boldsymbol{d}\parallel\boldsymbol{\hat{x}}$, 
    (E)chiral $p$-wave with $\boldsymbol{d}\parallel\boldsymbol{\hat{z}}$, 
    (F)chiral $p$-wave with $\boldsymbol{d}\parallel\boldsymbol{\hat{x}}$, and 
    (G)chiral $d$-wave. }
\end{figure}

Since electrons in  UPM  are spin-polarized in each sublattice like those in conventional antiferromagnet \cite{Libor22},  UPM  strictly leads to time-reversal symmetry breaking which is not reflected in the Hamiltonian as shown in Eq. \eqref{eq:1particle_hamiltonian_AM}. 
The consequence of antiferromagnetic spin structure in real space may appear in the boundary condition of each spin. 
Since the distances of the sublattice from the boundary for $\uparrow$ and $\downarrow$ spin are different from each other, the boundary condition or the barrier potential is spin-dependent. 
To estimate the effect of this, we perform the calculation of the normalized conductance $G/G_0$ as a function of $eV$ with $Z_{\uparrow} \neq Z_{\downarrow}$.
For the $d_{xy}$-wave SC case with $Z_{\uparrow} = Z_{\downarrow}$, the ZBCP appears as shown in Fig. \ref{fig:Result2}(f). 
When we consider the case with $Z_{\uparrow} \neq Z_{\downarrow}$, the ZBCP splits into two as shown in Fig. \ref{fig:Result_FI}(a)
similar to the case of normal metal / ferromagnetic insulator /  $d_{xy}$-wave superconductor junctions \cite{KashiwayaPRB1999}. 
Fig. \ref{fig:Result_FI}(b) and (c) show The $\tilde{\alpha}_{y}$ 
 and $eV$ dependence of conductance  
similar to that for $Z_{\uparrow} = Z_{\downarrow} = 2$. 
For the $d_{xy}$-wave pairing, SC is protected by time-reversal symmetry. 
In this case, our calculating model of the UPM/I/SC junction breaks time-reversal symmetry only if $Z_{\uparrow}\neq Z_{\downarrow}$. 
This results in a qualitative change of conductance by assuming the spin-dependent barrier at the interface.
By contrast, conductance changes only quantitatively by assuming the spin-dependent barrier $Z_{\uparrow} \neq Z_{\downarrow}$ at the interface for the chiral $p$- and chiral $d$-wave pairing.
It is relevant to the fact that the time-reversal symmetry is broken
in these superconducting states themselves. 
We also calculate $G/G_{0}$ at $eV=0$ as a function of $\tilde{\alpha}_y$ with $Z_{\uparrow}\neq Z_{\downarrow}$ as shown in Fig. \ref{fig:continuous_alpha_Z=1,3}. 
In this figure, we see one local maximum for chiral $p$-wave SC and two local maxima for chiral $d$-wave SC even in the presence of the spin-dependent barrier. 
This shows that the charge conductance of the junction is available for the distinction between chiral $p$-wave SC and chiral $d$-wave one. 

\section{Conclusion}\label{sec:concl}
In this paper, we have
studied the tunneling conductance between two-dimensional 
 unconventional $p$-wave magnet (UPM) 
/ superconductor (SC) junctions.
We choose various types of pairing symmetries of superconductors such as $s$-wave, $d_{x^{2}-y^{2}}$-wave,
$d_{xy}$-wave, $p_{x}$-wave, $p_{y}$-wave, chiral $p$-wave, and chiral $d$-wave pairings.
The zero bias conductance peak due to
the zero energy surface Andreev bound states
(ZESABS) in $d_{xy}$-wave and $p_{x}$-wave superconductor junctions are insensitive against the change of $\alpha_{y}$
which is an indicator of the magnitude of the momentum-dependent band splitting.
Changing the orientation of  UPM  has the same effect as changing the strength of the UPM.
For chiral $p$- or chiral $d$-wave SCs,
zero bias conductance shows a non-monotonic change as a function of the strength of altermagnet since the surface Andreev bound states have a momentum dependence.
The tunneling spectroscopy based on a 
 unconventional $p$-wave magnet  
can be a useful way to detect the SABS with momentum dependence.
It is noted that our obtained conductance formula
is available for persistent spin-helix  / SC junctions
since  unconventional $p$-wave magnetism  
is essentially equivalent to the persistent
spin-helix system. \par
It is also noted that 
theoretical works about the superconducting diode effect 
in altermagnetic junctions \cite{Banerjee2024altermagnetic}
and orientationally dependence on Josephson current 
in spin-triplet superconductor junction
has been started \cite{Cheng2024}. 
As a future work, it is interesting to study the Josephson effect involving altermagnet and unconventional
superconductors because the presence of ZESABS
seriously influences on the
current phase relation and temperature dependence of the Josephson current \cite{TKJosephson96, TKJosephson97,Kashiwaya2000}.
Also, it is intriguing to clarify the impact of ferromagnetic insulator  
at the interface in these junctions. \cite{TANAKAPhyscaC1997,Tanaka2000}


\section{Acknowledgements}\label{sec:Ack}
 We thank  S. Ikegaya for the valuable information 
 and constructive discussion. We also thank 
 J. Linder, S. Kashiwaya, D. Hirai, and S. Onari,   
 for their discussions.   
 Y. T. acknowledges support from JSPS with 
 Grants-in-Aid for Scientific Research ( KAKENHI Grants Nos. 23K17668, 24K00583, and24K00556).

\appendix

\renewcommand{\thefigure}{\thesection\arabic{figure}}
\setcounter{figure}{0}

\section{Wave Functions in the Superconductor}\label{sec:appendixA}
In this section, we introduce the wave functions in spin-singlet SC or spin-triplet SC with a $\boldsymbol{d}$-vector parallel to the $z$-axis based on the standard theory of tunneling spectroscopy of unconventional superconductors \cite{Kashiwaya2000}.
The wave function in SC can be written as
\begin{equation}
\Psi_{\uparrow}\left(x,k_{y}\right)=t_{\uparrow}\left(\begin{array}{c}
1\\
0\\
0\\
\Gamma_{+}
\end{array}\right)e^{ik_{e}^{s}x}+t_{\uparrow}^{A}\left(\begin{array}{c}
\Gamma_{-}\\
0\\
0\\
1
\end{array}\right)e^{-ik_{h}^{s}x}\label{eq:psiscup}
\end{equation}
\begin{equation}
\Psi_{\downarrow}\left(x,k_{y}\right)=t_{\downarrow}\left(\begin{array}{c}
0\\
1\\
\mp\Gamma_{+}\\
0
\end{array}\right)e^{ik_{e}^{s}x}+t_{\downarrow}^{A}\left(\begin{array}{c}
0\\
\mp\Gamma_{-}\\
1\\
0
\end{array}\right)e^{-ik_{h}^{s}x}\label{eq:psiscdn}
\end{equation}
where $\rho=\uparrow,\downarrow$ denotes the spin index of an injected electron. In Eq. (\ref{eq:psiscdn}), the sign $\mp$ becomes $-$ for spin-singlet SC and $+$ for spin-triplet SC,  respectively. 
In the above, we have used the following relations
\begin{equation}\label{eq:kes}
k_{e}^{s}=\sqrt{\frac{2m}{\hbar^{2}}\left(\mu+\Omega_{+}\right)-k_{y}^{2}}, 
\end{equation}
\begin{equation}\label{eq:khs}
k_{h}^{s}=\sqrt{\frac{2m}{\hbar^{2}}\left(\mu-\Omega_{-}\right)-k_{y}^{2}}.
\end{equation}

\section{Independence of the Transparency by $\alpha_x$}\label{sec:appendixB}
The $k_y$-resolved transparencies of UPM/SC and UPM/normal metal junctions $\sigma_{\uparrow(\downarrow)}^{N}\left(E,k_{y}\right)$ and $\sigma_{\uparrow(\downarrow)}^{N}\left(E,k_{y}\right)$ are independent of $\alpha_x$ since
\begin{equation}
\check{v}_x\Psi_{\uparrow}\left(x=0_{-},k_{y}\right)=
\frac{\hbar{}k_F}{m}
\left(\begin{array}{c}
q_{e\uparrow}^{+}+r_{\uparrow}q_{e\uparrow}^{-}\\
0\\0\\
r_{\uparrow}^{A}q_{h\downarrow}^{-}
\end{array}\right)
\label{eq:vx_Psi_up_AM}
\end{equation}
\begin{equation}
\check{v}_x\Psi_{\downarrow}\left(x=0_{-},k_{y}\right)=
\frac{\hbar{}k_F}{m}
\left(\begin{array}{c}
0\\
q_{e\downarrow}^{+}+r_{\downarrow}q_{e\downarrow}^{-}\\
r_{\downarrow}^{A}q_{h\uparrow}^{-}\\
0
\end{array}\right)
\end{equation}
\begin{equation}
\check{v}_x\Psi_{\uparrow}\left(x=0_{+},k_{y}\right)=
\frac{\hbar{}k_F}{m}
\left(\begin{array}{c}
t_{\uparrow}k_{e}^{s}-t_{\uparrow}^{A}k_{h}^{s}\Gamma_{-}\\
0\\0\\
-t_{\uparrow}k_{e}^{s}\Gamma_{+}+t_{\uparrow}^{A}k_{h}^{s}
\end{array}\right)
\end{equation}
\begin{equation}
\check{v}_x\Psi_{\downarrow}\left(x=0_{+},k_{y}\right)=
\frac{\hbar{}k_F}{m}
\left(\begin{array}{c}
0\\
t_{\downarrow}k_{e}^{s}\pm t_{\downarrow}^{A}k_{h}^{s}\Gamma_{-}\\
\pm t_{\downarrow}k_{e}^{s}\Gamma_{+}+t_{\downarrow}^{A}k_{h}^{s}\\
0
\end{array}\right)
\label{eq:vx_Psi_down_SC}
\end{equation}
based on Eq. \eqref{eq:v_x} where $q_{e\uparrow(\downarrow)}^{\pm}$, $q_{h\uparrow(\downarrow)}^{\pm}$, $k_{e}^{s}$, and $k_{h}^{s}$ are independent of $\alpha_x$, with which Eqs. \eqref{eq:boundary_condition_1}  and \eqref{eq:boundary_condition_2} yields $r_{\uparrow(\downarrow)}$ and $r_{\uparrow(\downarrow)}^{A}$ independent of  $\alpha_x$.

\section{Comparison with $d$-wave Altermagnet}\label{sec:appendixC}
In this section, we consider a $d$-wave AM / Insulator (I) / SC junction for comparison with the  UPM case. The corresponding BdG Hamiltonian in
the system can be written by $4\times 4$ matrix as follows:
\begin{equation}
\check{\mathcal{H}}_{\mathrm{BdG}}=\left[
\begin{array}{cc}
\hat{h}\left( \boldsymbol{k},x\right)  & \hat{\Delta}\left( \hat{\boldsymbol{%
k}}\right) \Theta (x) \\
-\hat{\Delta}^{\ast }\left( -\hat{\boldsymbol{k}}\right) \Theta (x) & -\hat{h%
}^{\ast }\left( -\boldsymbol{k},x\right)
\end{array}%
\right]   \label{eq:d-AMH}
\end{equation}%
where the single-particle Hamiltonian $\hat{h}\left( \boldsymbol{k},x\right)
$ can be written as \cite{Sun23}
\begin{equation}
\hat{h}\left( \boldsymbol{k},x\right) =\mathrm{diag}(\xi _{+}(\boldsymbol{k}%
,x),\xi _{-}(\boldsymbol{k},x))+\hat{U}_{0}\delta \left( x\right)
\label{eq:1particle_hamiltonian_dAM}
\end{equation}%
\begin{equation}
\xi _{\pm }=\frac{\hbar ^{2}}{2m}\boldsymbol{k}^{2}\pm \left[ \alpha
_{1}k_{x}k_{y}+\frac{\alpha _{2}}{2}(k_{x}^{2}-k_{y}^{2})\right] \Theta
(x)-\mu .  \label{eq:dAM}
\end{equation}%
Here, $2\times 2$ matrix $\hat{U}_{0}$ given by
\begin{equation}
\hat{U}_{0}=U\hat{I},\hat{I}=\mathrm{diag}(1,1)
\end{equation}%
denotes the insulating barrier at $x=0$. Here, the shapes of the polarized
Fermi surfaces for spin-$\uparrow $ and spin-$\downarrow $ electron species
are changed differently by parameters $\alpha _{1}$ and $\alpha _{2}$. We
define a dimensionless parameter $Z=mU/\left( \hbar ^{2}k_{F}\right) $ with $%
k_{F}$ being the Fermi wave vector on the superconducting side. In the $d$%
-wave AM region $x<0$, the $x$-components of the possible wave vectors for
fixed $E$ and $k_{y}$ are given by
\begin{equation}
k_{e\uparrow }^{\pm }=k_{F}\left[ \pm \frac{1}{1+\tilde{\alpha}_{2}}\sqrt{%
\left( 1+\frac{E}{\mu }\right) \left( 1+\tilde{\alpha}_{2}\right) +\tilde{k}%
_{y}^{2}\left( \tilde{\alpha}_{1}^{2}+\tilde{\alpha}_{2}^{2}-1\right) }-%
\frac{\tilde{\alpha}_{1}}{1+\tilde{\alpha}_{2}}\tilde{k}_{y}\right] ,
\label{eq:keup_dAM}
\end{equation}%
\begin{equation}
k_{e\downarrow }^{\pm }=k_{F}\left[ \pm \frac{1}{1-\tilde{\alpha}_{2}}\sqrt{%
\left( 1+\frac{E}{\mu }\right) \left( 1-\tilde{\alpha}_{2}\right) +\tilde{k}%
_{y}^{2}\left( \tilde{\alpha}_{1}^{2}+\tilde{\alpha}_{2}^{2}-1\right) }+%
\frac{\tilde{\alpha}_{1}}{1-\tilde{\alpha}_{2}}\tilde{k}_{y}\right] ,
\end{equation}%
\begin{equation}
k_{h\uparrow }^{\mp }=k_{F}\left[ \mp \frac{1}{1+\tilde{\alpha}_{2}}\sqrt{%
\left( 1-\frac{E}{\mu }\right) \left( 1+\tilde{\alpha}_{2}\right) +\tilde{k}%
_{y}^{2}\left( \tilde{\alpha}_{1}^{2}+\tilde{\alpha}_{2}^{2}-1\right) }-%
\frac{\tilde{\alpha}_{1}}{1+\tilde{\alpha}_{2}}\tilde{k}_{y}\right] ,
\end{equation}%
\begin{equation}
k_{h\downarrow }^{\mp }=k_{F}\left[ \mp \frac{1}{1-\tilde{\alpha}_{2}}\sqrt{%
\left( 1-\frac{E}{\mu }\right) \left( 1-\tilde{\alpha}_{2}\right) +\tilde{k}%
_{y}^{2}\left( \tilde{\alpha}_{1}^{2}+\tilde{\alpha}_{2}^{2}-1\right) }+%
\frac{\tilde{\alpha}_{1}}{1-\tilde{\alpha}_{2}}\tilde{k}_{y}\right] ,
\label{eq:khdown_dAM}
\end{equation}%
with dimensionless parameters $\tilde{\alpha}_{1}=m\alpha _{1}/\hbar ^{2},%
\tilde{\alpha}_{2}=m\alpha _{2}/\hbar ^{2}$, and $\tilde{k}_{y}=k_{y}/k_{F}$%
. Here, like Eqs. \eqref{eq:keup}--\eqref{eq:khdown}, the subscripts $e$ and
$h$ correspond to an electron and a hole respectively, $\uparrow, \downarrow
$ denote the spin, and the superscripts $\pm $ correspond to the sign of the
eigenvalues of the velocity operator
\begin{equation}
\check{v}_{x}=\frac{1}{\hbar }\frac{\partial \check{\mathcal{H}}_{\mathrm{BdG%
}}}{\partial k_{x}}=\check{\tau}_{z}\left[ \frac{\hbar }{m}+\check{\sigma}%
_{z}\frac{\alpha _{2}}{\hbar }\Theta (-x)\right] \frac{1}{i}\frac{\partial
}{\partial x}+\check{\tau}_{z}\check{\sigma}_{z}\frac{\alpha _{1}}{\hbar }%
k_{y}\Theta (-x).  \label{eq:vx_dAM}
\end{equation}%
%
From the Hamiltonian Eq. \ref{eq:d-AMH}, we obtain wavefunctions $\Psi \left(
x,y\right) = \Psi \left(x,k_{y}\right) e^{ik_{y}y}$ which follows the boundary condition%
\begin{equation}
\Psi \left( x,k_{y}\right) \Bigg{|}_{x=0_{+}}=\Psi \left( x,k_{y}\right) \Bigg{|}_{x=0_{-}},
\end{equation}%
\begin{equation}
\check{v}_{x}\Psi \left( x,k_{y}\right) \Bigg{|}_{x=0_{+}}-\check{v}_{x}\Psi \left(
x,k_{y}\right) \Bigg{|}_{x=0_{-}}=\frac{2U}{i\hbar }\check{\tau} _{3}\Psi \left( 0,k_{y}\right) .
\end{equation}%
The scattering coefficients as well as the conductance can thus be solved following the same way in the main text.

It is noted that the strength of AM must satisfy $|\tilde{\alpha}_{1}^{2}+%
\tilde{\alpha}_{2}^{2}|<1$ to keep the Fermi surface $E=0$ in a restricted
domain in the momentum space. In the same way as Eqs. %
\eqref{eq:kycondition_1}--\eqref{eq:kmax_kmin}, we get the conditions for
injected electrons with a particular value of $k_{y}$ to contribute to the
conduction process with $|E|\ll \mu $ as follows. In the case of $%
|E|>|\Delta (\theta _{\pm })|$, the condition for a spin-$\uparrow
(\downarrow )$ electron can be approximately rewritten as
\begin{equation}
|k_{y}|<k_{\uparrow (\downarrow )}^{c},  \label{eq:kycondition_1_dAM}
\end{equation}%
with
\begin{equation}
\frac{k_{\uparrow }^{c}}{k_{F}}=\min \left( 1,\sqrt{\frac{1+\tilde{\alpha}%
_{2}}{1-\tilde{\alpha}_{1}^{2}-\tilde{\alpha}_{2}^{2}}}\right) ,\frac{%
k_{\downarrow }^{c}}{k_{F}}=\min \left( 1,\sqrt{\frac{1-\tilde{\alpha}_{2}}{%
1-\tilde{\alpha}_{1}^{2}-\tilde{\alpha}_{2}^{2}}}\right) .
\end{equation}%
For $|E|<|\Delta (\theta _{\pm })|$,
in the cases of spin-singlet pairing or spin-triplet pairing with $%
\boldsymbol{d}\parallel \hat{\boldsymbol{z}}$,
the wave vector of the reflected spin-$\downarrow (\uparrow )$ hole is
nearly equivalent to that of a spin-$\downarrow (\uparrow )$ electron and
the condition in Eq. \eqref{eq:kycondition_1_dAM} changes into
\begin{equation}
|k_{y}|<k^{c},  \label{eq:kycondition_2_dAM}
\end{equation}%
with
\begin{equation}
\frac{k^{c}}{k_{F}}=\min \left( 1,\sqrt{\frac{1+\tilde{\alpha}_{2}}{1-\tilde{%
\alpha}_{1}^{2}-\tilde{\alpha}_{2}^{2}}},\sqrt{\frac{1-\tilde{\alpha}_{2}}{1-%
\tilde{\alpha}_{1}^{2}-\tilde{\alpha}_{2}^{2}}}\right) ,
\end{equation}%
for both an injected spin-$\uparrow $ electron and a spin-$\downarrow $ one.
On the other hand, in the case of spin-triplet pairing with $\boldsymbol{d}%
\parallel \hat{\boldsymbol{x}}$, the wave vector of the reflected spin-$%
\uparrow (\downarrow )$ hole is nearly equivalent to that of a spin-$%
\uparrow (\downarrow )$ electron and the condition in Eq. %
\eqref{eq:kycondition_1_dAM} does not change. It is noted that, for the $%
d_{xy}$-wave AM case with $\tilde{\alpha}_{2}=0$, both the conditions in
Eqs. \eqref{eq:kycondition_1_dAM} and \eqref{eq:kycondition_2_dAM} can be
simply written as $|k_{y}|<k_{F}$.

\begin{figure}[tb]
  \includegraphics{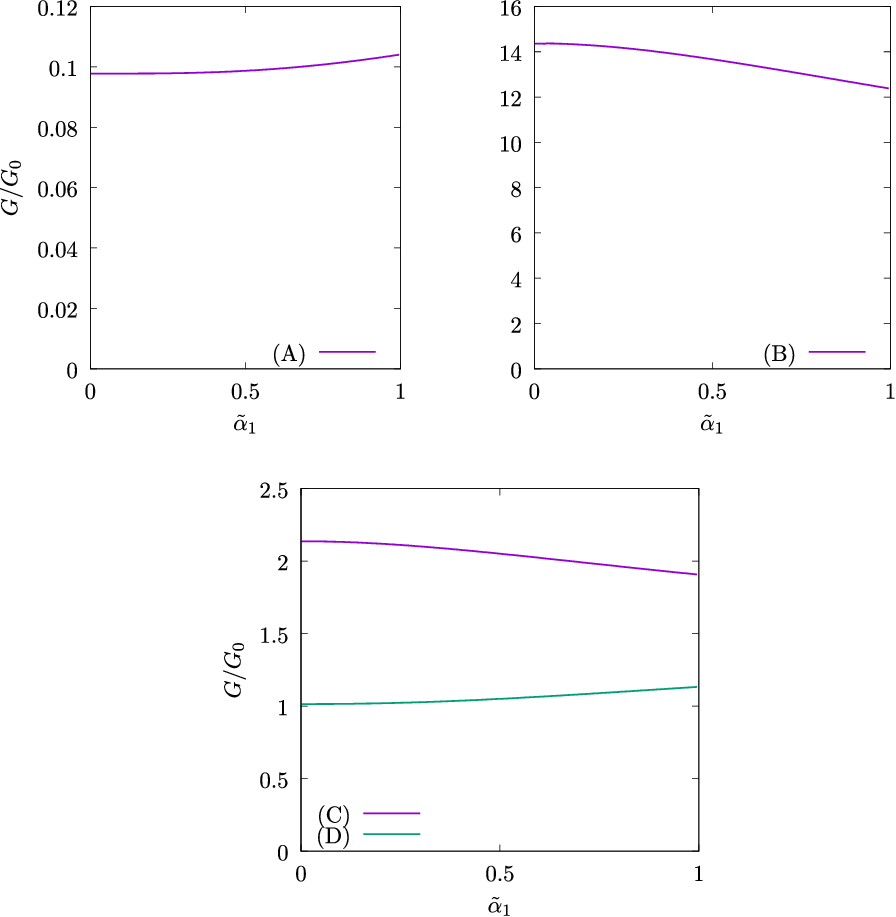}
  \caption{\label{fig:dxyAM_continuous_alpha_Z=2}(Color online) The $\tilde\alpha_{1}$ dependence of zero bias conductance of $d_{xy}$-wave AM/I/SC junctions at $Z=2$. The pairing symmetry of SCs are
    (A)$s$-wave, $p_y$-wave with $\boldsymbol{d}\parallel\boldsymbol{\hat{z}}$, $p_y$-wave with $\boldsymbol{d}\parallel\boldsymbol{\hat{x}}$, $d_{x^2-y^2}$-wave, 
    (B)$p_x$-wave with $\boldsymbol{d}\parallel\boldsymbol{\hat{z}}$,$p_x$-wave with $\boldsymbol{d}\parallel\boldsymbol{\hat{x}}$, $d_{xy}$-wave,
    (C)chiral $p$-wave with $\boldsymbol{d}\parallel\boldsymbol{\hat{z}}$, chiral $p$-wave with $\boldsymbol{d}\parallel\boldsymbol{\hat{x}}$, and
    (D)chiral $d$-wave. }
\end{figure}
\begin{figure}[tb]
  \includegraphics{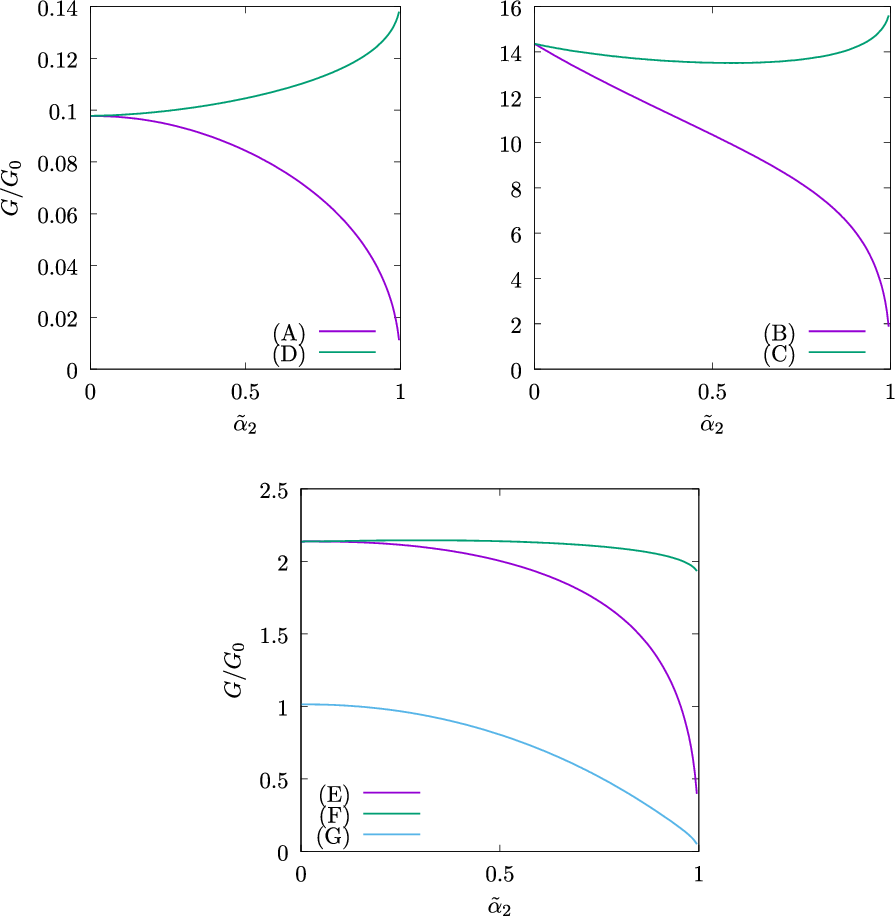}
  \caption{\label{fig:dx2-y2AM_continuous_alpha_Z=2}(Color online) The $\tilde\alpha_{2}$ dependence of zero bias conductance of $d_{x^2-y^2}$-wave AM/I/SC junctions at $Z=2$. The pairing symmetry of SCs are
    (A)$s$-wave, $p_y$-wave with $\boldsymbol{d}\parallel\boldsymbol{\hat{z}}$ , $d_{x^2-y^2}$-wave, 
    (B)$p_x$-wave with $\boldsymbol{d}\parallel\boldsymbol{\hat{z}}$, $d_{xy}$-wave, 
    (C)$p_x$-wave with $\boldsymbol{d}\parallel\boldsymbol{\hat{x}}$,
    (D)$p_y$-wave with $\boldsymbol{d}\parallel\boldsymbol{\hat{x}}$, 
    (E)chiral $p$-wave with $\boldsymbol{d}\parallel\boldsymbol{\hat{z}}$, 
    (F)chiral $p$-wave with $\boldsymbol{d}\parallel\boldsymbol{\hat{x}}$, and 
    (G)chiral $d$-wave. }
\end{figure}

To compare the behaviors of 
 unconventional $p$-wave magnet and $d$-wave altermagnet,
we calculate the normalized conductance $G/G_{0}$ of $d$-wave AM/I/SC junctions
at $eV=0$ as functions of $\tilde{\alpha}_{1}$ or $\tilde{\alpha}_{2}$ as
shown in Figs. \ref{fig:dxyAM_continuous_alpha_Z=2} and \ref%
{fig:dx2-y2AM_continuous_alpha_Z=2}. We set $\tilde{\alpha}_{2}=0$ for $%
d_{xy}$-wave AM and $\tilde{\alpha}_{1}=0$ for $d_{x^{2}-y^{2}}$-wave AM in
Figs. \ref{fig:dxyAM_continuous_alpha_Z=2} and \ref%
{fig:dx2-y2AM_continuous_alpha_Z=2}, respectively. As compared to the $p_{y}$%
-wave  magnet  where the normalized conductance $G/G_{0}$ develops a drastic
change with the strength of  magnetism, 
$G/G_{0}$ shows only a slight variation with
the increase of the strength of AM $|\tilde{\alpha}_{1}|$ for the $d_{xy}$%
-wave AM. Similarly, the variation of $G/G_{0}$ is small for spin-triplet SC
with $\boldsymbol{d}\parallel \hat{\boldsymbol{x}}$ in the $d_{x^{2}-y^{2}}$%
-wave AM case with the increase of $|\tilde{\alpha}_{2}|$. In these cases,
the condition of $k_{y}$ becomes the same for the normal state and the
superconducting state. In addition, the $x$-components of the group
velocities of electrons and holes in AM are even functions of $k_{y}$ as
well as those in SC, which we can derive from Eqs. \eqref{eq:keup_dAM}--%
\eqref{eq:vx_dAM}. As a consequence, the discrepancy among the group
velocities for electrons and holes in AM and SC becomes smaller than that in
the $p_{y}$-wave  magnet  case and the root of the boundary conditions in Eqs. %
\eqref{eq:boundary_condition_1} and \eqref{eq:boundary_condition_2} does not
change significantly with $\tilde{\alpha}_{1}$ or $\tilde{\alpha}_{2}$. By
contrast, for $d_{x^{2}-y^{2}}$-wave AM/I/SC junctions with spin-singlet SC
or spin-triplet one with $\boldsymbol{d}\parallel \hat{\boldsymbol{z}}$, $%
G/G_{0}$ is strongly suppressed with the increase of $|\tilde{\alpha}_{2}|$.
In this case, the range of $k_{y}$ contributing to the conduction process is
restricted similarly for the spin-$\uparrow $ electron injection and that of
spin-$\downarrow $ with $|E|<\Delta (\theta _{\pm })$ for the
superconducting state. By contrast, for the normal state, the range of $k_{y}$ is merely $|k_{y}|<k_{F}$
with $\tilde{\alpha}_{2}>0 (\tilde{\alpha}_{2}<0)$ for the
spin-$\uparrow (\downarrow)$ electron injection.

\begin{figure}[tb]
  \includegraphics{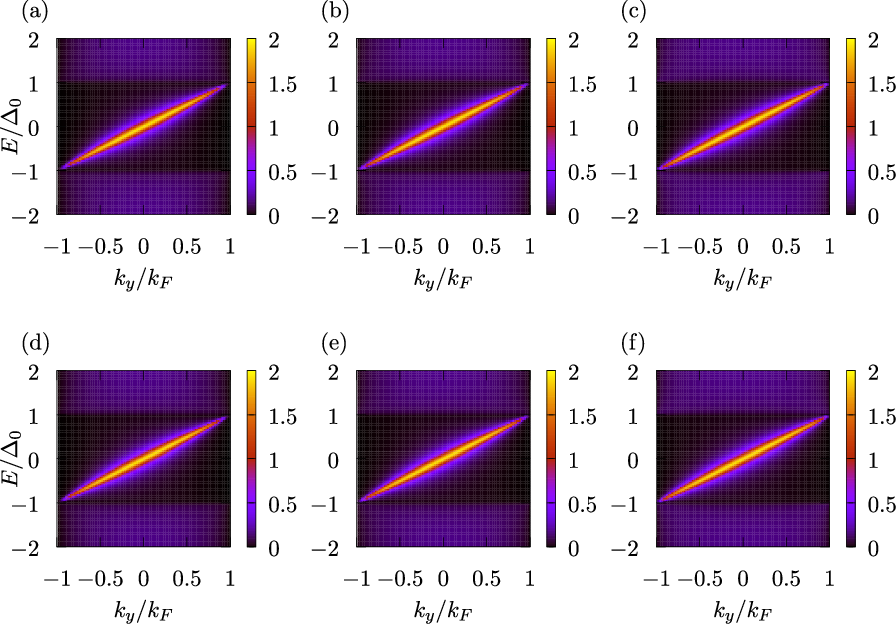}
  \caption{\label{fig:dxyAM_kyvsG_chiralpSC}(Color online) Momentum resolved conductance $\sigma_{\uparrow}^{S}(E,k_y)$ ((a),(b),(c)) and $\sigma_{\downarrow}^{S}(E,k_y)$ ((d),(e),(f)) for $d_{xy}$-wave altermagnet/Insulator/chiral $p$-wave superconductor with $\boldsymbol{d} \parallel \hat{\boldsymbol{z}}$ junction where $Z=2$. The strength of altermagnet is set to $\tilde{\alpha}_{1}=0$((a),(d)), $0.5$((b),(e)), and $0.9$((c),(f)).}
\end{figure}
\begin{figure}[tb]
  \includegraphics{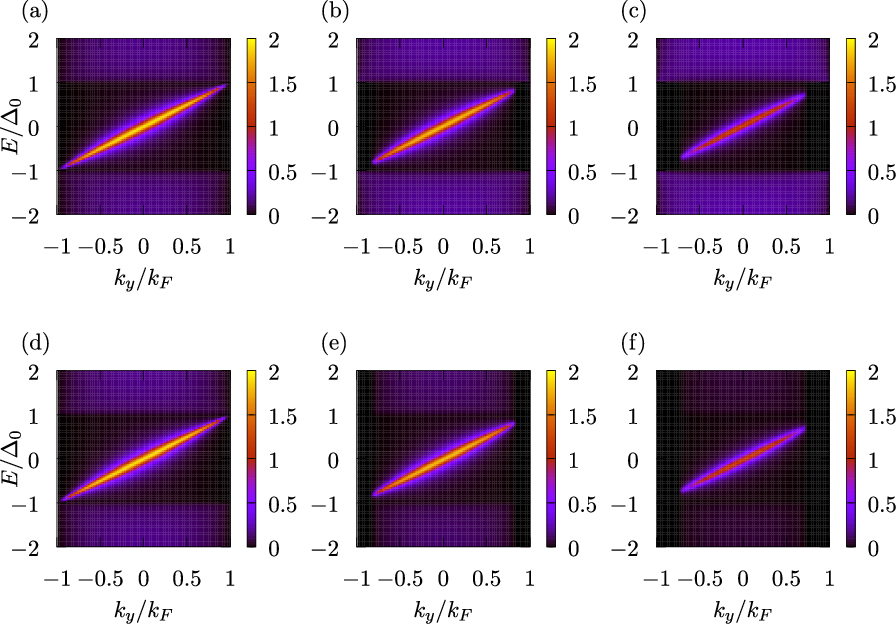}
  \caption{\label{fig:dx2-y2AM_kyvsG_chiralpSC}(Color online) Momentum resolved conductance $\sigma_{\uparrow}^{S}(E,k_y)$ ((a),(b),(c)) and $\sigma_{\downarrow}^{S}(E,k_y)$ ((d),(e),(f)) for $d_{x^2-y^2}$-wave altermagnet/Insulator/chiral $p$-wave superconductor with $\boldsymbol{d} \parallel \hat{\boldsymbol{z}}$ junction where $Z=2$. The strength of altermagnet is set to $\tilde{\alpha}_{2}=0$((a),(d)), $0.5$((b),(e)), and $0.9$((c),(f)).}
\end{figure}
\begin{figure}[tb]
  \includegraphics{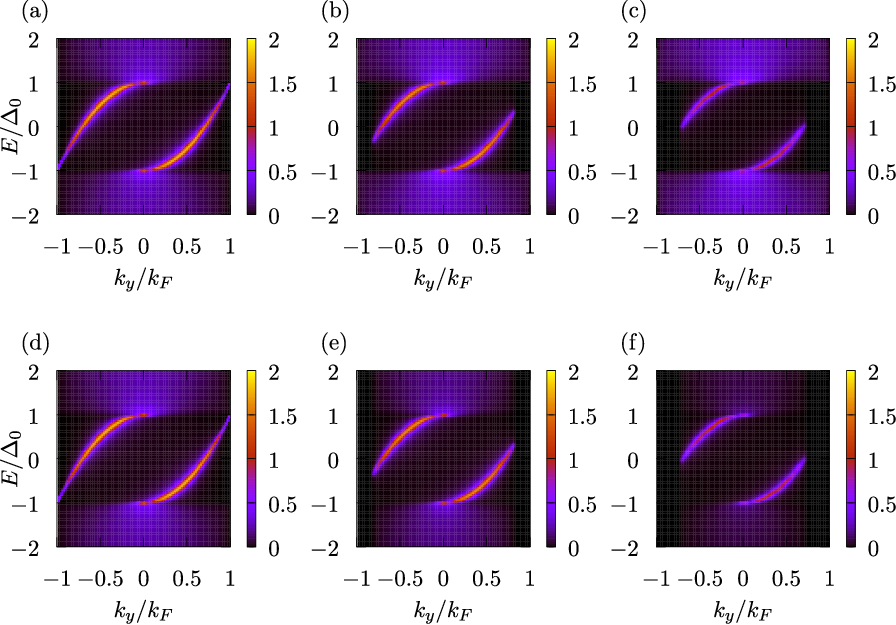}
  \caption{\label{fig:dx2-y2AM_kyvsG_chiraldSC}(Color online) Momentum resolved conductance $\sigma_{\uparrow}^{S}(E,k_y)$ ((a),(b),(c)) and $\sigma_{\downarrow}^{S}(E,k_y)$ ((d),(e),(f)) for $d_{x^2-y^2}$-wave altermagnet/Insulator/chiral $d$-wave superconductor junction where $Z=2$. The strength of altermagnet is set to $\tilde{\alpha}_{2}=0$((a),(d)), $0.5$((b),(e)), and $0.9$((c),(f)).}
\end{figure}
\begin{figure}[tb]
  \includegraphics{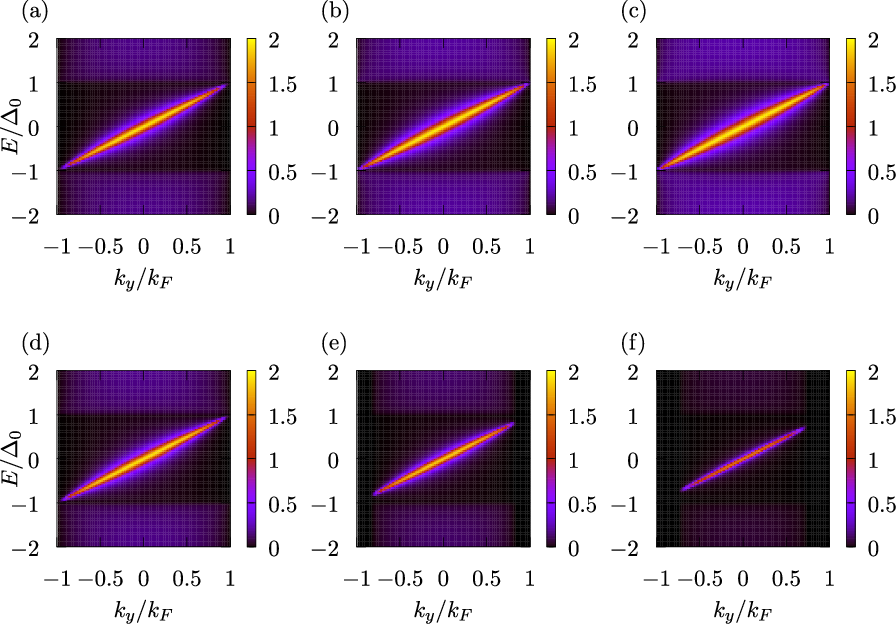}
  \caption{\label{fig:dx2-y2AM_kyvsG_chiralpSC_dx}(Color online) Momentum resolved conductance $\sigma_{\uparrow}^{S}(E,k_y)$ ((a),(b),(c)) and $\sigma_{\downarrow}^{S}(E,k_y)$ ((d),(e),(f)) for $d_{x^2-y^2}$-wave altermagnet/Insulator/chiral $p$-wave superconductor with $\boldsymbol{d} \parallel \hat{\boldsymbol{x}}$ junction where $Z=2$. The strength of altermagnet is set to $\tilde{\alpha}_{2}=0$((a),(d)), $0.5$((b),(e)), and $0.9$((c),(f)).}
\end{figure}

It is noted that $G/G_{0}$ for chiral SC as a function of $\tilde{\alpha}%
_{1},\tilde{\alpha}_{2}$ does not have conspicuous local maxima or local
minima like those in the  UPM cases shown in Figs. \ref%
{fig:continuous_alpha_Z=0_chiral}, \ref{fig:continuous_alpha_Z=2_chiral} and %
\ref{fig:continuous_alpha_Z=1,3}. To clarify this difference, we calculate
the $k_{y}$- and spin-resolved conductance $\sigma _{\uparrow }^{S}\left( E,{%
k}_{y}\right) $ and $\sigma _{\downarrow }^{S}\left( E,{k}_{y}\right) $ of $d
$-AM/I/chiral SC junctions as shown in Figs. \ref{fig:dxyAM_kyvsG_chiralpSC}--%
\ref{fig:dx2-y2AM_kyvsG_chiralpSC_dx}. Figure \ref{fig:dxyAM_kyvsG_chiralpSC}
is an example of the $d_{xy}$-AM cases and shows the change of $\sigma
_{\uparrow (\downarrow )}^{S}(E,k_{y})$ by $\tilde{\alpha}_{1}$ is small for
all $k_{y}$. For the $d_{x^{2}-y^{2}}$-AM cases, as shown in Figs. \ref%
{fig:dx2-y2AM_kyvsG_chiralpSC}--\ref{fig:dx2-y2AM_kyvsG_chiralpSC_dx}, the
range of $k_{y}$ contributing to the conduction process is changed by $%
\tilde{\alpha}_{2}$ following Eqs. \eqref{eq:kycondition_1_dAM} and %
\eqref{eq:kycondition_2_dAM}. This range cannot be narrower than $%
|k_{y}|/k_{F}<1/\sqrt{2}$ with $|\tilde{\alpha}_{2}|<1$ and the boundaries
of the momentum parallel to the interface $k^{c},k_{\uparrow (\downarrow
)}^{c}$ do not cross the chiral edge modes at $E=0$ for chiral $p$- and $d$%
-wave SCs. As a consequence, any prominent local maxima of $G/G_{0}$ as a
function of $\tilde{\alpha}_{2}$ cannot be seen in the $d_{x^{2}-y^{2}}$-AM
case. The discussion above indicates that the asymmetric restriction of $%
k_{y}$ and the boundary of $k_{y}$ reaching $k_{y}=0$ as shown in Figs. \ref%
{fig:kyvsG_chiralpSC_dz}--\ref{fig:kyvsG_chiralpSC_dx} are important
features of  UPM compared to $d$-wave AM.

\bibliography{71287}

\begin{thebibliography}{88}%
\makeatletter
\providecommand \@ifxundefined [1]{%
 \@ifx{#1\undefined}
}%
\providecommand \@ifnum [1]{%
 \ifnum #1\expandafter \@firstoftwo
 \else \expandafter \@secondoftwo
 \fi
}%
\providecommand \@ifx [1]{%
 \ifx #1\expandafter \@firstoftwo
 \else \expandafter \@secondoftwo
 \fi
}%
\providecommand \natexlab [1]{#1}%
\providecommand \enquote  [1]{``#1''}%
\providecommand \bibnamefont  [1]{#1}%
\providecommand \bibfnamefont [1]{#1}%
\providecommand \citenamefont [1]{#1}%
\providecommand \href@noop [0]{\@secondoftwo}%
\providecommand \href [0]{\begingroup \@sanitize@url \@href}%
\providecommand \@href[1]{\@@startlink{#1}\@@href}%
\providecommand \@@href[1]{\endgroup#1\@@endlink}%
\providecommand \@sanitize@url [0]{\catcode `\\12\catcode `\$12\catcode `\&12\catcode `\#12\catcode `\^12\catcode `\_12\catcode `\%12\relax}%
\providecommand \@@startlink[1]{}%
\providecommand \@@endlink[0]{}%
\providecommand \url  [0]{\begingroup\@sanitize@url \@url }%
\providecommand \@url [1]{\endgroup\@href {#1}{\urlprefix }}%
\providecommand \urlprefix  [0]{URL }%
\providecommand \Eprint [0]{\href }%
\providecommand \doibase [0]{http://dx.doi.org/}%
\providecommand \selectlanguage [0]{\@gobble}%
\providecommand \bibinfo  [0]{\@secondoftwo}%
\providecommand \bibfield  [0]{\@secondoftwo}%
\providecommand \translation [1]{[#1]}%
\providecommand \BibitemOpen [0]{}%
\providecommand \bibitemStop [0]{}%
\providecommand \bibitemNoStop [0]{.\EOS\space}%
\providecommand \EOS [0]{\spacefactor3000\relax}%
\providecommand \BibitemShut  [1]{\csname bibitem#1\endcsname}%
\let\auto@bib@innerbib\@empty
\bibitem [{\citenamefont {Šmejkal}\ \emph {et~al.}(2020)\citenamefont {Šmejkal}, \citenamefont {González-Hernández}, \citenamefont {Jungwirth},\ and\ \citenamefont {Sinova}}]{LiborSAv}%
  \BibitemOpen
  \bibfield  {author} {\bibinfo {author} {\bibfnamefont {L.}~\bibnamefont {Šmejkal}}, \bibinfo {author} {\bibfnamefont {R.}~\bibnamefont {González-Hernández}}, \bibinfo {author} {\bibfnamefont {T.}~\bibnamefont {Jungwirth}}, \ and\ \bibinfo {author} {\bibfnamefont {J.}~\bibnamefont {Sinova}},\ }\href {\doibase 10.1126/sciadv.aaz8809} {\bibfield  {journal} {\bibinfo  {journal} {Sci. Adv.}\ }\textbf {\bibinfo {volume} {6}},\ \bibinfo {pages} {eaaz8809} (\bibinfo {year} {2020})}\BibitemShut {NoStop}%
\bibitem [{\citenamefont {\ifmmode~\check{S}\else \v{S}\fi{}mejkal}\ \emph {et~al.}(2022{\natexlab{a}})\citenamefont {\ifmmode~\check{S}\else \v{S}\fi{}mejkal}, \citenamefont {Sinova},\ and\ \citenamefont {Jungwirth}}]{Libor22}%
  \BibitemOpen
  \bibfield  {author} {\bibinfo {author} {\bibfnamefont {L.}~\bibnamefont {\ifmmode~\check{S}\else \v{S}\fi{}mejkal}}, \bibinfo {author} {\bibfnamefont {J.}~\bibnamefont {Sinova}}, \ and\ \bibinfo {author} {\bibfnamefont {T.}~\bibnamefont {Jungwirth}},\ }\href {\doibase 10.1103/PhysRevX.12.031042} {\bibfield  {journal} {\bibinfo  {journal} {Phys. Rev. X}\ }\textbf {\bibinfo {volume} {12}},\ \bibinfo {pages} {031042} (\bibinfo {year} {2022}{\natexlab{a}})}\BibitemShut {NoStop}%
\bibitem [{\citenamefont {Hayami}\ \emph {et~al.}(2019)\citenamefont {Hayami}, \citenamefont {Yanagi},\ and\ \citenamefont {Kusunose}}]{Hayami19}%
  \BibitemOpen
  \bibfield  {author} {\bibinfo {author} {\bibfnamefont {S.}~\bibnamefont {Hayami}}, \bibinfo {author} {\bibfnamefont {Y.}~\bibnamefont {Yanagi}}, \ and\ \bibinfo {author} {\bibfnamefont {H.}~\bibnamefont {Kusunose}},\ }\href {\doibase 10.7566/JPSJ.88.123702} {\bibfield  {journal} {\bibinfo  {journal} {J. Phys. Soc. Jpn.}\ }\textbf {\bibinfo {volume} {88}},\ \bibinfo {pages} {123702} (\bibinfo {year} {2019})}\BibitemShut {NoStop}%
\bibitem [{\citenamefont {Hayami}\ \emph {et~al.}(2020)\citenamefont {Hayami}, \citenamefont {Yanagi},\ and\ \citenamefont {Kusunose}}]{Hayami20}%
  \BibitemOpen
  \bibfield  {author} {\bibinfo {author} {\bibfnamefont {S.}~\bibnamefont {Hayami}}, \bibinfo {author} {\bibfnamefont {Y.}~\bibnamefont {Yanagi}}, \ and\ \bibinfo {author} {\bibfnamefont {H.}~\bibnamefont {Kusunose}},\ }\href {\doibase 10.1103/PhysRevB.102.144441} {\bibfield  {journal} {\bibinfo  {journal} {Phys. Rev. B}\ }\textbf {\bibinfo {volume} {102}},\ \bibinfo {pages} {144441} (\bibinfo {year} {2020})}\BibitemShut {NoStop}%
\bibitem [{\citenamefont {\ifmmode~\check{S}\else \v{S}\fi{}mejkal}\ \emph {et~al.}(2022{\natexlab{b}})\citenamefont {\ifmmode~\check{S}\else \v{S}\fi{}mejkal}, \citenamefont {Sinova},\ and\ \citenamefont {Jungwirth}}]{landscape22}%
  \BibitemOpen
  \bibfield  {author} {\bibinfo {author} {\bibfnamefont {L.}~\bibnamefont {\ifmmode~\check{S}\else \v{S}\fi{}mejkal}}, \bibinfo {author} {\bibfnamefont {J.}~\bibnamefont {Sinova}}, \ and\ \bibinfo {author} {\bibfnamefont {T.}~\bibnamefont {Jungwirth}},\ }\href {\doibase 10.1103/PhysRevX.12.040501} {\bibfield  {journal} {\bibinfo  {journal} {Phys. Rev. X}\ }\textbf {\bibinfo {volume} {12}},\ \bibinfo {pages} {040501} (\bibinfo {year} {2022}{\natexlab{b}})}\BibitemShut {NoStop}%
\bibitem [{\citenamefont {Mazin}\ \emph {et~al.}(2021)\citenamefont {Mazin}, \citenamefont {Koepernik}, \citenamefont {Johannes}, \citenamefont {González-Hernández},\ and\ \citenamefont {Šmejkal}}]{MazinPNAS}%
  \BibitemOpen
  \bibfield  {author} {\bibinfo {author} {\bibfnamefont {I.~I.}\ \bibnamefont {Mazin}}, \bibinfo {author} {\bibfnamefont {K.}~\bibnamefont {Koepernik}}, \bibinfo {author} {\bibfnamefont {M.~D.}\ \bibnamefont {Johannes}}, \bibinfo {author} {\bibfnamefont {R.}~\bibnamefont {González-Hernández}}, \ and\ \bibinfo {author} {\bibfnamefont {L.}~\bibnamefont {Šmejkal}},\ }\href {\doibase 10.1073/pnas.2108924118} {\bibfield  {journal} {\bibinfo  {journal} {Proc. Natl. Acad. Sci. U.S.A.}\ }\textbf {\bibinfo {volume} {118}},\ \bibinfo {pages} {e2108924118} (\bibinfo {year} {2021})}\BibitemShut {NoStop}%
\bibitem [{\citenamefont {Mazin}(2022)}]{MazinPRX22}%
  \BibitemOpen
  \bibfield  {author} {\bibinfo {author} {\bibfnamefont {I.}~\bibnamefont {Mazin}} (\bibinfo {collaboration} {The PRX Editors}),\ }\href {\doibase 10.1103/PhysRevX.12.040002} {\bibfield  {journal} {\bibinfo  {journal} {Phys. Rev. X}\ }\textbf {\bibinfo {volume} {12}},\ \bibinfo {pages} {040002} (\bibinfo {year} {2022})}\BibitemShut {NoStop}%
\bibitem [{\citenamefont {\ifmmode~\check{S}\else \v{S}\fi{}mejkal}\ \emph {et~al.}(2022{\natexlab{c}})\citenamefont {\ifmmode~\check{S}\else \v{S}\fi{}mejkal}, \citenamefont {Hellenes}, \citenamefont {Gonz\'alez-Hern\'andez}, \citenamefont {Sinova},\ and\ \citenamefont {Jungwirth}}]{Libor011028}%
  \BibitemOpen
  \bibfield  {author} {\bibinfo {author} {\bibfnamefont {L.}~\bibnamefont {\ifmmode~\check{S}\else \v{S}\fi{}mejkal}}, \bibinfo {author} {\bibfnamefont {A.~B.}\ \bibnamefont {Hellenes}}, \bibinfo {author} {\bibfnamefont {R.}~\bibnamefont {Gonz\'alez-Hern\'andez}}, \bibinfo {author} {\bibfnamefont {J.}~\bibnamefont {Sinova}}, \ and\ \bibinfo {author} {\bibfnamefont {T.}~\bibnamefont {Jungwirth}},\ }\href {\doibase 10.1103/PhysRevX.12.011028} {\bibfield  {journal} {\bibinfo  {journal} {Phys. Rev. X}\ }\textbf {\bibinfo {volume} {12}},\ \bibinfo {pages} {011028} (\bibinfo {year} {2022}{\natexlab{c}})}\BibitemShut {NoStop}%
\bibitem [{\citenamefont {Solovyev}(2006)}]{solovyev2006lattice}%
  \BibitemOpen
  \bibfield  {author} {\bibinfo {author} {\bibfnamefont {I.}~\bibnamefont {Solovyev}},\ }\href@noop {} {\bibfield  {journal} {\bibinfo  {journal} {Physical Review B—Condensed Matter and Materials Physics}\ }\textbf {\bibinfo {volume} {74}},\ \bibinfo {pages} {054412} (\bibinfo {year} {2006})}\BibitemShut {NoStop}%
\bibitem [{\citenamefont {Noda}\ \emph {et~al.}(2016)\citenamefont {Noda}, \citenamefont {Ohno},\ and\ \citenamefont {Nakamura}}]{noda2016momentum}%
  \BibitemOpen
  \bibfield  {author} {\bibinfo {author} {\bibfnamefont {Y.}~\bibnamefont {Noda}}, \bibinfo {author} {\bibfnamefont {K.}~\bibnamefont {Ohno}}, \ and\ \bibinfo {author} {\bibfnamefont {S.}~\bibnamefont {Nakamura}},\ }\href@noop {} {\bibfield  {journal} {\bibinfo  {journal} {Physical Chemistry Chemical Physics}\ }\textbf {\bibinfo {volume} {18}},\ \bibinfo {pages} {13294} (\bibinfo {year} {2016})}\BibitemShut {NoStop}%
\bibitem [{\citenamefont {Okugawa}\ \emph {et~al.}(2018)\citenamefont {Okugawa}, \citenamefont {Ohno}, \citenamefont {Noda},\ and\ \citenamefont {Nakamura}}]{okugawa2018weakly}%
  \BibitemOpen
  \bibfield  {author} {\bibinfo {author} {\bibfnamefont {T.}~\bibnamefont {Okugawa}}, \bibinfo {author} {\bibfnamefont {K.}~\bibnamefont {Ohno}}, \bibinfo {author} {\bibfnamefont {Y.}~\bibnamefont {Noda}}, \ and\ \bibinfo {author} {\bibfnamefont {S.}~\bibnamefont {Nakamura}},\ }\href@noop {} {\bibfield  {journal} {\bibinfo  {journal} {Journal of Physics: Condensed Matter}\ }\textbf {\bibinfo {volume} {30}},\ \bibinfo {pages} {075502} (\bibinfo {year} {2018})}\BibitemShut {NoStop}%
\bibitem [{\citenamefont {Naka}\ \emph {et~al.}(2019)\citenamefont {Naka}, \citenamefont {Hayami}, \citenamefont {Kusunose}, \citenamefont {Yanagi}, \citenamefont {Motome},\ and\ \citenamefont {Seo}}]{naka2019spin}%
  \BibitemOpen
  \bibfield  {author} {\bibinfo {author} {\bibfnamefont {M.}~\bibnamefont {Naka}}, \bibinfo {author} {\bibfnamefont {S.}~\bibnamefont {Hayami}}, \bibinfo {author} {\bibfnamefont {H.}~\bibnamefont {Kusunose}}, \bibinfo {author} {\bibfnamefont {Y.}~\bibnamefont {Yanagi}}, \bibinfo {author} {\bibfnamefont {Y.}~\bibnamefont {Motome}}, \ and\ \bibinfo {author} {\bibfnamefont {H.}~\bibnamefont {Seo}},\ }\href@noop {} {\bibfield  {journal} {\bibinfo  {journal} {Nature communications}\ }\textbf {\bibinfo {volume} {10}},\ \bibinfo {pages} {4305} (\bibinfo {year} {2019})}\BibitemShut {NoStop}%
\bibitem [{\citenamefont {Naka}\ \emph {et~al.}(2020)\citenamefont {Naka}, \citenamefont {Hayami}, \citenamefont {Kusunose}, \citenamefont {Yanagi}, \citenamefont {Motome},\ and\ \citenamefont {Seo}}]{naka2020anomalous}%
  \BibitemOpen
  \bibfield  {author} {\bibinfo {author} {\bibfnamefont {M.}~\bibnamefont {Naka}}, \bibinfo {author} {\bibfnamefont {S.}~\bibnamefont {Hayami}}, \bibinfo {author} {\bibfnamefont {H.}~\bibnamefont {Kusunose}}, \bibinfo {author} {\bibfnamefont {Y.}~\bibnamefont {Yanagi}}, \bibinfo {author} {\bibfnamefont {Y.}~\bibnamefont {Motome}}, \ and\ \bibinfo {author} {\bibfnamefont {H.}~\bibnamefont {Seo}},\ }\href@noop {} {\bibfield  {journal} {\bibinfo  {journal} {Physical Review B}\ }\textbf {\bibinfo {volume} {102}},\ \bibinfo {pages} {075112} (\bibinfo {year} {2020})}\BibitemShut {NoStop}%
\bibitem [{\citenamefont {Naka}\ \emph {et~al.}(2021)\citenamefont {Naka}, \citenamefont {Motome},\ and\ \citenamefont {Seo}}]{naka2021perovskite}%
  \BibitemOpen
  \bibfield  {author} {\bibinfo {author} {\bibfnamefont {M.}~\bibnamefont {Naka}}, \bibinfo {author} {\bibfnamefont {Y.}~\bibnamefont {Motome}}, \ and\ \bibinfo {author} {\bibfnamefont {H.}~\bibnamefont {Seo}},\ }\href@noop {} {\bibfield  {journal} {\bibinfo  {journal} {Physical Review B}\ }\textbf {\bibinfo {volume} {103}},\ \bibinfo {pages} {125114} (\bibinfo {year} {2021})}\BibitemShut {NoStop}%
\bibitem [{\citenamefont {Naka}\ \emph {et~al.}(2022)\citenamefont {Naka}, \citenamefont {Motome},\ and\ \citenamefont {Seo}}]{naka2022anomalous}%
  \BibitemOpen
  \bibfield  {author} {\bibinfo {author} {\bibfnamefont {M.}~\bibnamefont {Naka}}, \bibinfo {author} {\bibfnamefont {Y.}~\bibnamefont {Motome}}, \ and\ \bibinfo {author} {\bibfnamefont {H.}~\bibnamefont {Seo}},\ }\href@noop {} {\bibfield  {journal} {\bibinfo  {journal} {Physical Review B}\ }\textbf {\bibinfo {volume} {106}},\ \bibinfo {pages} {195149} (\bibinfo {year} {2022})}\BibitemShut {NoStop}%
\bibitem [{\citenamefont {Yuan}\ \emph {et~al.}(2021)\citenamefont {Yuan}, \citenamefont {Wang}, \citenamefont {Luo},\ and\ \citenamefont {Zunger}}]{yuan2021strong}%
  \BibitemOpen
  \bibfield  {author} {\bibinfo {author} {\bibfnamefont {L.-D.}\ \bibnamefont {Yuan}}, \bibinfo {author} {\bibfnamefont {Z.}~\bibnamefont {Wang}}, \bibinfo {author} {\bibfnamefont {J.-W.}\ \bibnamefont {Luo}}, \ and\ \bibinfo {author} {\bibfnamefont {A.}~\bibnamefont {Zunger}},\ }\href@noop {} {\bibfield  {journal} {\bibinfo  {journal} {Physical Review B}\ }\textbf {\bibinfo {volume} {103}},\ \bibinfo {pages} {224410} (\bibinfo {year} {2021})}\BibitemShut {NoStop}%
\bibitem [{\citenamefont {Ahn}\ \emph {et~al.}(2019)\citenamefont {Ahn}, \citenamefont {Hariki}, \citenamefont {Lee},\ and\ \citenamefont {Kune\ifmmode~\check{s}\else \v{s}\fi{}}}]{Ahn19}%
  \BibitemOpen
  \bibfield  {author} {\bibinfo {author} {\bibfnamefont {K.-H.}\ \bibnamefont {Ahn}}, \bibinfo {author} {\bibfnamefont {A.}~\bibnamefont {Hariki}}, \bibinfo {author} {\bibfnamefont {K.-W.}\ \bibnamefont {Lee}}, \ and\ \bibinfo {author} {\bibfnamefont {J.}~\bibnamefont {Kune\ifmmode~\check{s}\else \v{s}\fi{}}},\ }\href {\doibase 10.1103/PhysRevB.99.184432} {\bibfield  {journal} {\bibinfo  {journal} {Phys. Rev. B}\ }\textbf {\bibinfo {volume} {99}},\ \bibinfo {pages} {184432} (\bibinfo {year} {2019})}\BibitemShut {NoStop}%
\bibitem [{\citenamefont {Fedchenko}\ \emph {et~al.}(2024)\citenamefont {Fedchenko}, \citenamefont {Minár}, \citenamefont {Akashdeep}, \citenamefont {D’Souza}, \citenamefont {Vasilyev}, \citenamefont {Tkach}, \citenamefont {Odenbreit}, \citenamefont {Nguyen}, \citenamefont {Kutnyakhov}, \citenamefont {Wind}, \citenamefont {Wenthaus}, \citenamefont {Scholz}, \citenamefont {Rossnagel}, \citenamefont {Hoesch}, \citenamefont {Aeschlimann}, \citenamefont {Stadtmüller}, \citenamefont {Kläui}, \citenamefont {Schönhense}, \citenamefont {Jungwirth}, \citenamefont {Hellenes}, \citenamefont {Jakob}, \citenamefont {Šmejkal}, \citenamefont {Sinova},\ and\ \citenamefont {Elmers}}]{Fedchenko24}%
  \BibitemOpen
  \bibfield  {author} {\bibinfo {author} {\bibfnamefont {O.}~\bibnamefont {Fedchenko}}, \bibinfo {author} {\bibfnamefont {J.}~\bibnamefont {Minár}}, \bibinfo {author} {\bibfnamefont {A.}~\bibnamefont {Akashdeep}}, \bibinfo {author} {\bibfnamefont {S.~W.}\ \bibnamefont {D’Souza}}, \bibinfo {author} {\bibfnamefont {D.}~\bibnamefont {Vasilyev}}, \bibinfo {author} {\bibfnamefont {O.}~\bibnamefont {Tkach}}, \bibinfo {author} {\bibfnamefont {L.}~\bibnamefont {Odenbreit}}, \bibinfo {author} {\bibfnamefont {Q.}~\bibnamefont {Nguyen}}, \bibinfo {author} {\bibfnamefont {D.}~\bibnamefont {Kutnyakhov}}, \bibinfo {author} {\bibfnamefont {N.}~\bibnamefont {Wind}}, \bibinfo {author} {\bibfnamefont {L.}~\bibnamefont {Wenthaus}}, \bibinfo {author} {\bibfnamefont {M.}~\bibnamefont {Scholz}}, \bibinfo {author} {\bibfnamefont {K.}~\bibnamefont {Rossnagel}}, \bibinfo {author} {\bibfnamefont {M.}~\bibnamefont {Hoesch}}, \bibinfo {author} {\bibfnamefont {M.}~\bibnamefont {Aeschlimann}}, \bibinfo {author} {\bibfnamefont
  {B.}~\bibnamefont {Stadtmüller}}, \bibinfo {author} {\bibfnamefont {M.}~\bibnamefont {Kläui}}, \bibinfo {author} {\bibfnamefont {G.}~\bibnamefont {Schönhense}}, \bibinfo {author} {\bibfnamefont {T.}~\bibnamefont {Jungwirth}}, \bibinfo {author} {\bibfnamefont {A.~B.}\ \bibnamefont {Hellenes}}, \bibinfo {author} {\bibfnamefont {G.}~\bibnamefont {Jakob}}, \bibinfo {author} {\bibfnamefont {L.}~\bibnamefont {Šmejkal}}, \bibinfo {author} {\bibfnamefont {J.}~\bibnamefont {Sinova}}, \ and\ \bibinfo {author} {\bibfnamefont {H.-J.}\ \bibnamefont {Elmers}},\ }\href {\doibase 10.1126/sciadv.adj4883} {\bibfield  {journal} {\bibinfo  {journal} {Sci. Adv.}\ }\textbf {\bibinfo {volume} {10}},\ \bibinfo {pages} {eadj4883} (\bibinfo {year} {2024})}\BibitemShut {NoStop}%
\bibitem [{\citenamefont {Lee}\ \emph {et~al.}(2024)\citenamefont {Lee}, \citenamefont {Lee}, \citenamefont {Jung}, \citenamefont {Jung}, \citenamefont {Kim}, \citenamefont {Lee}, \citenamefont {Seok}, \citenamefont {Kim}, \citenamefont {Park}, \citenamefont {\ifmmode~\check{S}\else \v{S}\fi{}mejkal}, \citenamefont {Kang},\ and\ \citenamefont {Kim}}]{MnTeLee}%
  \BibitemOpen
  \bibfield  {author} {\bibinfo {author} {\bibfnamefont {S.}~\bibnamefont {Lee}}, \bibinfo {author} {\bibfnamefont {S.}~\bibnamefont {Lee}}, \bibinfo {author} {\bibfnamefont {S.}~\bibnamefont {Jung}}, \bibinfo {author} {\bibfnamefont {J.}~\bibnamefont {Jung}}, \bibinfo {author} {\bibfnamefont {D.}~\bibnamefont {Kim}}, \bibinfo {author} {\bibfnamefont {Y.}~\bibnamefont {Lee}}, \bibinfo {author} {\bibfnamefont {B.}~\bibnamefont {Seok}}, \bibinfo {author} {\bibfnamefont {J.}~\bibnamefont {Kim}}, \bibinfo {author} {\bibfnamefont {B.~G.}\ \bibnamefont {Park}}, \bibinfo {author} {\bibfnamefont {L.}~\bibnamefont {\ifmmode~\check{S}\else \v{S}\fi{}mejkal}}, \bibinfo {author} {\bibfnamefont {C.-J.}\ \bibnamefont {Kang}}, \ and\ \bibinfo {author} {\bibfnamefont {C.}~\bibnamefont {Kim}},\ }\href {https://link.aps.org/doi/10.1103/PhysRevLett.132.036702} {\bibfield  {journal} {\bibinfo  {journal} {Phys. Rev. Lett.}\ }\textbf {\bibinfo {volume} {132}},\ \bibinfo {pages} {036702} (\bibinfo {year} {2024})}\BibitemShut
  {NoStop}%
\bibitem [{\citenamefont {Osumi}\ \emph {et~al.}(2024)\citenamefont {Osumi}, \citenamefont {Souma}, \citenamefont {Aoyama}, \citenamefont {Yamauchi}, \citenamefont {Honma}, \citenamefont {Nakayama}, \citenamefont {Takahashi}, \citenamefont {Ohgushi},\ and\ \citenamefont {Sato}}]{osumi2024}%
  \BibitemOpen
  \bibfield  {author} {\bibinfo {author} {\bibfnamefont {T.}~\bibnamefont {Osumi}}, \bibinfo {author} {\bibfnamefont {S.}~\bibnamefont {Souma}}, \bibinfo {author} {\bibfnamefont {T.}~\bibnamefont {Aoyama}}, \bibinfo {author} {\bibfnamefont {K.}~\bibnamefont {Yamauchi}}, \bibinfo {author} {\bibfnamefont {A.}~\bibnamefont {Honma}}, \bibinfo {author} {\bibfnamefont {K.}~\bibnamefont {Nakayama}}, \bibinfo {author} {\bibfnamefont {T.}~\bibnamefont {Takahashi}}, \bibinfo {author} {\bibfnamefont {K.}~\bibnamefont {Ohgushi}}, \ and\ \bibinfo {author} {\bibfnamefont {T.}~\bibnamefont {Sato}},\ }\href {\doibase 10.1103/PhysRevB.109.115102} {\bibfield  {journal} {\bibinfo  {journal} {Phys. Rev. B}\ }\textbf {\bibinfo {volume} {109}},\ \bibinfo {pages} {115102} (\bibinfo {year} {2024})}\BibitemShut {NoStop}%
\bibitem [{\citenamefont {Krempask{\'y}}\ \emph {et~al.}(2024)\citenamefont {Krempask{\'y}}, \citenamefont {{\v{S}}mejkal}, \citenamefont {D'Souza}, \citenamefont {Hajlaoui}, \citenamefont {Springholz}, \citenamefont {Uhl{\'i}{\v{r}}ov{\'a}}, \citenamefont {Alarab}, \citenamefont {Constantinou}, \citenamefont {Strocov}, \citenamefont {Usanov}, \citenamefont {Pudelko}, \citenamefont {Gonz{\'a}lez-Hern{\'a}ndez}, \citenamefont {Birk~Hellenes}, \citenamefont {Jansa}, \citenamefont {Reichlov{\'a}}, \citenamefont {{\v{S}}ob{\'a}{\v{n}}}, \citenamefont {Gonzalez~Betancourt}, \citenamefont {Wadley}, \citenamefont {Sinova}, \citenamefont {Kriegner}, \citenamefont {Min{\'a}r}, \citenamefont {Dil},\ and\ \citenamefont {Jungwirth}}]{Krempaský2024}%
  \BibitemOpen
  \bibfield  {author} {\bibinfo {author} {\bibfnamefont {J.}~\bibnamefont {Krempask{\'y}}}, \bibinfo {author} {\bibfnamefont {L.}~\bibnamefont {{\v{S}}mejkal}}, \bibinfo {author} {\bibfnamefont {S.~W.}\ \bibnamefont {D'Souza}}, \bibinfo {author} {\bibfnamefont {M.}~\bibnamefont {Hajlaoui}}, \bibinfo {author} {\bibfnamefont {G.}~\bibnamefont {Springholz}}, \bibinfo {author} {\bibfnamefont {K.}~\bibnamefont {Uhl{\'i}{\v{r}}ov{\'a}}}, \bibinfo {author} {\bibfnamefont {F.}~\bibnamefont {Alarab}}, \bibinfo {author} {\bibfnamefont {P.~C.}\ \bibnamefont {Constantinou}}, \bibinfo {author} {\bibfnamefont {V.}~\bibnamefont {Strocov}}, \bibinfo {author} {\bibfnamefont {D.}~\bibnamefont {Usanov}}, \bibinfo {author} {\bibfnamefont {W.~R.}\ \bibnamefont {Pudelko}}, \bibinfo {author} {\bibfnamefont {R.}~\bibnamefont {Gonz{\'a}lez-Hern{\'a}ndez}}, \bibinfo {author} {\bibfnamefont {A.}~\bibnamefont {Birk~Hellenes}}, \bibinfo {author} {\bibfnamefont {Z.}~\bibnamefont {Jansa}}, \bibinfo {author} {\bibfnamefont {H.}~\bibnamefont
  {Reichlov{\'a}}}, \bibinfo {author} {\bibfnamefont {Z.}~\bibnamefont {{\v{S}}ob{\'a}{\v{n}}}}, \bibinfo {author} {\bibfnamefont {R.~D.}\ \bibnamefont {Gonzalez~Betancourt}}, \bibinfo {author} {\bibfnamefont {P.}~\bibnamefont {Wadley}}, \bibinfo {author} {\bibfnamefont {J.}~\bibnamefont {Sinova}}, \bibinfo {author} {\bibfnamefont {D.}~\bibnamefont {Kriegner}}, \bibinfo {author} {\bibfnamefont {J.}~\bibnamefont {Min{\'a}r}}, \bibinfo {author} {\bibfnamefont {J.~H.}\ \bibnamefont {Dil}}, \ and\ \bibinfo {author} {\bibfnamefont {T.}~\bibnamefont {Jungwirth}},\ }\href {\doibase 10.1038/s41586-023-06907-7} {\bibfield  {journal} {\bibinfo  {journal} {Nature}\ }\textbf {\bibinfo {volume} {626}},\ \bibinfo {pages} {517} (\bibinfo {year} {2024})}\BibitemShut {NoStop}%
\bibitem [{\citenamefont {Reichlová}\ \emph {et~al.}(2021)\citenamefont {Reichlová}, \citenamefont {Seeger}, \citenamefont {González-Hernández}, \citenamefont {Kounta}, \citenamefont {Schlitz}, \citenamefont {Kriegner}, \citenamefont {Ritzinger}, \citenamefont {Lammel}, \citenamefont {Leiviskä}, \citenamefont {Petříček}, \citenamefont {Doležal}, \citenamefont {Schmoranzerová}, \citenamefont {Bad'ura}, \citenamefont {Thomas}, \citenamefont {Baltz}, \citenamefont {Michez}, \citenamefont {Sinova}, \citenamefont {Goennenwein}, \citenamefont {Jungwirth},\ and\ \citenamefont {Šmejkal}}]{Helena2021}%
  \BibitemOpen
  \bibfield  {author} {\bibinfo {author} {\bibfnamefont {H.}~\bibnamefont {Reichlová}}, \bibinfo {author} {\bibfnamefont {R.~L.}\ \bibnamefont {Seeger}}, \bibinfo {author} {\bibfnamefont {R.}~\bibnamefont {González-Hernández}}, \bibinfo {author} {\bibfnamefont {I.}~\bibnamefont {Kounta}}, \bibinfo {author} {\bibfnamefont {R.}~\bibnamefont {Schlitz}}, \bibinfo {author} {\bibfnamefont {D.}~\bibnamefont {Kriegner}}, \bibinfo {author} {\bibfnamefont {P.}~\bibnamefont {Ritzinger}}, \bibinfo {author} {\bibfnamefont {M.}~\bibnamefont {Lammel}}, \bibinfo {author} {\bibfnamefont {M.}~\bibnamefont {Leiviskä}}, \bibinfo {author} {\bibfnamefont {V.}~\bibnamefont {Petříček}}, \bibinfo {author} {\bibfnamefont {P.}~\bibnamefont {Doležal}}, \bibinfo {author} {\bibfnamefont {E.}~\bibnamefont {Schmoranzerová}}, \bibinfo {author} {\bibfnamefont {A.}~\bibnamefont {Bad'ura}}, \bibinfo {author} {\bibfnamefont {A.}~\bibnamefont {Thomas}}, \bibinfo {author} {\bibfnamefont {V.}~\bibnamefont {Baltz}}, \bibinfo {author}
  {\bibfnamefont {L.}~\bibnamefont {Michez}}, \bibinfo {author} {\bibfnamefont {J.}~\bibnamefont {Sinova}}, \bibinfo {author} {\bibfnamefont {S.~T.~B.}\ \bibnamefont {Goennenwein}}, \bibinfo {author} {\bibfnamefont {T.}~\bibnamefont {Jungwirth}}, \ and\ \bibinfo {author} {\bibfnamefont {L.}~\bibnamefont {Šmejkal}},\ }\href@noop {} {} (\bibinfo {year} {2021}),\ \Eprint {http://arxiv.org/abs/2012.15651} {arXiv:2012.15651} \BibitemShut {NoStop}%
\bibitem [{\citenamefont {López-Moreno}\ \emph {et~al.}(2016)\citenamefont {López-Moreno}, \citenamefont {Romero}, \citenamefont {Mejía-López},\ and\ \citenamefont {Muñoz}}]{Moreno16}%
  \BibitemOpen
  \bibfield  {author} {\bibinfo {author} {\bibfnamefont {S.}~\bibnamefont {López-Moreno}}, \bibinfo {author} {\bibfnamefont {A.~H.}\ \bibnamefont {Romero}}, \bibinfo {author} {\bibfnamefont {J.}~\bibnamefont {Mejía-López}}, \ and\ \bibinfo {author} {\bibfnamefont {A.}~\bibnamefont {Muñoz}},\ }\href {\doibase 10.1039/C6CP05467F} {\bibfield  {journal} {\bibinfo  {journal} {Phys. Chem. Chem. Phys.}\ }\textbf {\bibinfo {volume} {18}},\ \bibinfo {pages} {33250} (\bibinfo {year} {2016})}\BibitemShut {NoStop}%
\bibitem [{\citenamefont {Beenakker}\ and\ \citenamefont {Vakhtel}(2023)}]{Beenakker23}%
  \BibitemOpen
  \bibfield  {author} {\bibinfo {author} {\bibfnamefont {C.~W.~J.}\ \bibnamefont {Beenakker}}\ and\ \bibinfo {author} {\bibfnamefont {T.}~\bibnamefont {Vakhtel}},\ }\href {\doibase 10.1103/PhysRevB.108.075425} {\bibfield  {journal} {\bibinfo  {journal} {Phys. Rev. B}\ }\textbf {\bibinfo {volume} {108}},\ \bibinfo {pages} {075425} (\bibinfo {year} {2023})}\BibitemShut {NoStop}%
\bibitem [{\citenamefont {Sun}\ \emph {et~al.}(2023)\citenamefont {Sun}, \citenamefont {Brataas},\ and\ \citenamefont {Linder}}]{Sun23}%
  \BibitemOpen
  \bibfield  {author} {\bibinfo {author} {\bibfnamefont {C.}~\bibnamefont {Sun}}, \bibinfo {author} {\bibfnamefont {A.}~\bibnamefont {Brataas}}, \ and\ \bibinfo {author} {\bibfnamefont {J.}~\bibnamefont {Linder}},\ }\href {\doibase 10.1103/PhysRevB.108.054511} {\bibfield  {journal} {\bibinfo  {journal} {Phys. Rev. B}\ }\textbf {\bibinfo {volume} {108}},\ \bibinfo {pages} {054511} (\bibinfo {year} {2023})}\BibitemShut {NoStop}%
\bibitem [{\citenamefont {Papaj}(2023)}]{Papaj23}%
  \BibitemOpen
  \bibfield  {author} {\bibinfo {author} {\bibfnamefont {M.}~\bibnamefont {Papaj}},\ }\href {\doibase 10.1103/PhysRevB.108.L060508} {\bibfield  {journal} {\bibinfo  {journal} {Phys. Rev. B}\ }\textbf {\bibinfo {volume} {108}},\ \bibinfo {pages} {L060508} (\bibinfo {year} {2023})}\BibitemShut {NoStop}%
\bibitem [{\citenamefont {Ouassou}\ \emph {et~al.}(2023)\citenamefont {Ouassou}, \citenamefont {Brataas},\ and\ \citenamefont {Linder}}]{Ouassou23}%
  \BibitemOpen
  \bibfield  {author} {\bibinfo {author} {\bibfnamefont {J.~A.}\ \bibnamefont {Ouassou}}, \bibinfo {author} {\bibfnamefont {A.}~\bibnamefont {Brataas}}, \ and\ \bibinfo {author} {\bibfnamefont {J.}~\bibnamefont {Linder}},\ }\href {\doibase 10.1103/PhysRevLett.131.076003} {\bibfield  {journal} {\bibinfo  {journal} {Phys. Rev. Lett.}\ }\textbf {\bibinfo {volume} {131}},\ \bibinfo {pages} {076003} (\bibinfo {year} {2023})}\BibitemShut {NoStop}%
\bibitem [{\citenamefont {Zhang}\ \emph {et~al.}(2024)\citenamefont {Zhang}, \citenamefont {Hu},\ and\ \citenamefont {Neupert}}]{Songbo23}%
  \BibitemOpen
  \bibfield  {author} {\bibinfo {author} {\bibfnamefont {S.-B.}\ \bibnamefont {Zhang}}, \bibinfo {author} {\bibfnamefont {L.-H.}\ \bibnamefont {Hu}}, \ and\ \bibinfo {author} {\bibfnamefont {T.}~\bibnamefont {Neupert}},\ }\href@noop {} {\enquote {\bibinfo {title} {Finite-momentum cooper pairing in proximitized altermagnets},}\ } (\bibinfo {year} {2024})\BibitemShut {NoStop}%
\bibitem [{\citenamefont {Nagae}\ \emph {et~al.}(2024)\citenamefont {Nagae}, \citenamefont {Schnyder},\ and\ \citenamefont {Ikegaya}}]{Nagae2024}%
  \BibitemOpen
  \bibfield  {author} {\bibinfo {author} {\bibfnamefont {Y.}~\bibnamefont {Nagae}}, \bibinfo {author} {\bibfnamefont {A.~P.}\ \bibnamefont {Schnyder}}, \ and\ \bibinfo {author} {\bibfnamefont {S.}~\bibnamefont {Ikegaya}},\ }\href@noop {} {} (\bibinfo {year} {2024}),\ \Eprint {http://arxiv.org/abs/2403.07117} {arXiv:2403.07117} \BibitemShut {NoStop}%
\bibitem [{\citenamefont {Lu}\ \emph {et~al.}(2024)\citenamefont {Lu}, \citenamefont {Maeda}, \citenamefont {Ito}, \citenamefont {Yada},\ and\ \citenamefont {Tanaka}}]{lu2024varphi}%
  \BibitemOpen
  \bibfield  {author} {\bibinfo {author} {\bibfnamefont {B.}~\bibnamefont {Lu}}, \bibinfo {author} {\bibfnamefont {K.}~\bibnamefont {Maeda}}, \bibinfo {author} {\bibfnamefont {H.}~\bibnamefont {Ito}}, \bibinfo {author} {\bibfnamefont {K.}~\bibnamefont {Yada}}, \ and\ \bibinfo {author} {\bibfnamefont {Y.}~\bibnamefont {Tanaka}},\ }\href {\doibase 10.1103/PhysRevLett.133.226002} {\bibfield  {journal} {\bibinfo  {journal} {Phys. Rev. Lett.}\ }\textbf {\bibinfo {volume} {133}},\ \bibinfo {pages} {226002} (\bibinfo {year} {2024})}\BibitemShut {NoStop}%
\bibitem [{\citenamefont {Buzdin}(2005)}]{buzdin2005proximity}%
  \BibitemOpen
  \bibfield  {author} {\bibinfo {author} {\bibfnamefont {A.~I.}\ \bibnamefont {Buzdin}},\ }\href@noop {} {\bibfield  {journal} {\bibinfo  {journal} {Reviews of modern physics}\ }\textbf {\bibinfo {volume} {77}},\ \bibinfo {pages} {935} (\bibinfo {year} {2005})}\BibitemShut {NoStop}%
\bibitem [{\citenamefont {Bergeret}\ \emph {et~al.}(2005)\citenamefont {Bergeret}, \citenamefont {Volkov},\ and\ \citenamefont {Efetov}}]{bergeret2005odd}%
  \BibitemOpen
  \bibfield  {author} {\bibinfo {author} {\bibfnamefont {F.}~\bibnamefont {Bergeret}}, \bibinfo {author} {\bibfnamefont {A.~F.}\ \bibnamefont {Volkov}}, \ and\ \bibinfo {author} {\bibfnamefont {K.~B.}\ \bibnamefont {Efetov}},\ }\href@noop {} {\bibfield  {journal} {\bibinfo  {journal} {Reviews of modern physics}\ }\textbf {\bibinfo {volume} {77}},\ \bibinfo {pages} {1321} (\bibinfo {year} {2005})}\BibitemShut {NoStop}%
\bibitem [{\citenamefont {Linder}\ and\ \citenamefont {Robinson}(2015)}]{linder2015superconducting}%
  \BibitemOpen
  \bibfield  {author} {\bibinfo {author} {\bibfnamefont {J.}~\bibnamefont {Linder}}\ and\ \bibinfo {author} {\bibfnamefont {J.~W.}\ \bibnamefont {Robinson}},\ }\href@noop {} {\bibfield  {journal} {\bibinfo  {journal} {Nature Physics}\ }\textbf {\bibinfo {volume} {11}},\ \bibinfo {pages} {307} (\bibinfo {year} {2015})}\BibitemShut {NoStop}%
\bibitem [{\citenamefont {Eschrig}(2015)}]{eschrig2015spin}%
  \BibitemOpen
  \bibfield  {author} {\bibinfo {author} {\bibfnamefont {M.}~\bibnamefont {Eschrig}},\ }\href@noop {} {\bibfield  {journal} {\bibinfo  {journal} {Reports on Progress in Physics}\ }\textbf {\bibinfo {volume} {78}},\ \bibinfo {pages} {104501} (\bibinfo {year} {2015})}\BibitemShut {NoStop}%
\bibitem [{\citenamefont {Giil}\ and\ \citenamefont {Linder}(2024)}]{giil2023}%
  \BibitemOpen
  \bibfield  {author} {\bibinfo {author} {\bibfnamefont {H.~G.}\ \bibnamefont {Giil}}\ and\ \bibinfo {author} {\bibfnamefont {J.}~\bibnamefont {Linder}},\ }\href {\doibase 10.1103/PhysRevB.109.134511} {\bibfield  {journal} {\bibinfo  {journal} {Phys. Rev. B}\ }\textbf {\bibinfo {volume} {109}},\ \bibinfo {pages} {134511} (\bibinfo {year} {2024})}\BibitemShut {NoStop}%
\bibitem [{\citenamefont {Ghorashi}\ \emph {et~al.}(2023)\citenamefont {Ghorashi}, \citenamefont {Hughes},\ and\ \citenamefont {Cano}}]{CanoArxiv}%
  \BibitemOpen
  \bibfield  {author} {\bibinfo {author} {\bibfnamefont {S.~A.~A.}\ \bibnamefont {Ghorashi}}, \bibinfo {author} {\bibfnamefont {T.~L.}\ \bibnamefont {Hughes}}, \ and\ \bibinfo {author} {\bibfnamefont {J.}~\bibnamefont {Cano}},\ }\href@noop {} {} (\bibinfo {year} {2023}),\ \Eprint {http://arxiv.org/abs/2306.09413} {arXiv:2306.09413} \BibitemShut {NoStop}%
\bibitem [{\citenamefont {Zhu}\ \emph {et~al.}(2023)\citenamefont {Zhu}, \citenamefont {Zhuang}, \citenamefont {Wu},\ and\ \citenamefont {Yan}}]{Zhongbo23}%
  \BibitemOpen
  \bibfield  {author} {\bibinfo {author} {\bibfnamefont {D.}~\bibnamefont {Zhu}}, \bibinfo {author} {\bibfnamefont {Z.-Y.}\ \bibnamefont {Zhuang}}, \bibinfo {author} {\bibfnamefont {Z.}~\bibnamefont {Wu}}, \ and\ \bibinfo {author} {\bibfnamefont {Z.}~\bibnamefont {Yan}},\ }\href {\doibase 10.1103/PhysRevB.108.184505} {\bibfield  {journal} {\bibinfo  {journal} {Phys. Rev. B}\ }\textbf {\bibinfo {volume} {108}},\ \bibinfo {pages} {184505} (\bibinfo {year} {2023})}\BibitemShut {NoStop}%
\bibitem [{\citenamefont {Bobkova}\ \emph {et~al.}(2004)\citenamefont {Bobkova}, \citenamefont {Hirschfeld},\ and\ \citenamefont {Barash}}]{bobkova2004spin}%
  \BibitemOpen
  \bibfield  {author} {\bibinfo {author} {\bibfnamefont {I.}~\bibnamefont {Bobkova}}, \bibinfo {author} {\bibfnamefont {P.}~\bibnamefont {Hirschfeld}}, \ and\ \bibinfo {author} {\bibfnamefont {Y.~S.}\ \bibnamefont {Barash}},\ }\href@noop {} {\bibfield  {journal} {\bibinfo  {journal} {arXiv preprint cond-mat/0408032}\ } (\bibinfo {year} {2004})}\BibitemShut {NoStop}%
\bibitem [{\citenamefont {Andersen}\ \emph {et~al.}(2005)\citenamefont {Andersen}, \citenamefont {Bobkova}, \citenamefont {Hirschfeld},\ and\ \citenamefont {Barash}}]{andersen2005bound}%
  \BibitemOpen
  \bibfield  {author} {\bibinfo {author} {\bibfnamefont {B.~M.}\ \bibnamefont {Andersen}}, \bibinfo {author} {\bibfnamefont {I.}~\bibnamefont {Bobkova}}, \bibinfo {author} {\bibfnamefont {P.}~\bibnamefont {Hirschfeld}}, \ and\ \bibinfo {author} {\bibfnamefont {Y.~S.}\ \bibnamefont {Barash}},\ }\href@noop {} {\bibfield  {journal} {\bibinfo  {journal} {Physical Review B}\ }\textbf {\bibinfo {volume} {72}},\ \bibinfo {pages} {184510} (\bibinfo {year} {2005})}\BibitemShut {NoStop}%
\bibitem [{\citenamefont {Jakobsen}\ \emph {et~al.}(2020)\citenamefont {Jakobsen}, \citenamefont {Naess}, \citenamefont {Dutta}, \citenamefont {Brataas},\ and\ \citenamefont {Qaiumzadeh}}]{jakobsen2020electrical}%
  \BibitemOpen
  \bibfield  {author} {\bibinfo {author} {\bibfnamefont {M.~F.}\ \bibnamefont {Jakobsen}}, \bibinfo {author} {\bibfnamefont {K.~B.}\ \bibnamefont {Naess}}, \bibinfo {author} {\bibfnamefont {P.}~\bibnamefont {Dutta}}, \bibinfo {author} {\bibfnamefont {A.}~\bibnamefont {Brataas}}, \ and\ \bibinfo {author} {\bibfnamefont {A.}~\bibnamefont {Qaiumzadeh}},\ }\href@noop {} {\bibfield  {journal} {\bibinfo  {journal} {Physical Review B}\ }\textbf {\bibinfo {volume} {102}},\ \bibinfo {pages} {140504} (\bibinfo {year} {2020})}\BibitemShut {NoStop}%
\bibitem [{\citenamefont {Jakobsen}\ \emph {et~al.}(2021)\citenamefont {Jakobsen}, \citenamefont {Brataas},\ and\ \citenamefont {Qaiumzadeh}}]{jakobsen2021electrically}%
  \BibitemOpen
  \bibfield  {author} {\bibinfo {author} {\bibfnamefont {M.~F.}\ \bibnamefont {Jakobsen}}, \bibinfo {author} {\bibfnamefont {A.}~\bibnamefont {Brataas}}, \ and\ \bibinfo {author} {\bibfnamefont {A.}~\bibnamefont {Qaiumzadeh}},\ }\href@noop {} {\bibfield  {journal} {\bibinfo  {journal} {Physical Review Letters}\ }\textbf {\bibinfo {volume} {127}},\ \bibinfo {pages} {017701} (\bibinfo {year} {2021})}\BibitemShut {NoStop}%
\bibitem [{\citenamefont {Blonder}\ \emph {et~al.}(1982)\citenamefont {Blonder}, \citenamefont {Tinkham},\ and\ \citenamefont {Klapwijk}}]{BTK82}%
  \BibitemOpen
  \bibfield  {author} {\bibinfo {author} {\bibfnamefont {G.~E.}\ \bibnamefont {Blonder}}, \bibinfo {author} {\bibfnamefont {M.}~\bibnamefont {Tinkham}}, \ and\ \bibinfo {author} {\bibfnamefont {T.}~\bibnamefont {Klapwijk}},\ }\href {\doibase 10.1103/PhysRevB.25.4515} {\bibfield  {journal} {\bibinfo  {journal} {Phys. Rev. B}\ }\textbf {\bibinfo {volume} {25}},\ \bibinfo {pages} {4515} (\bibinfo {year} {1982})}\BibitemShut {NoStop}%
\bibitem [{\citenamefont {Tanaka}\ and\ \citenamefont {Kashiwaya}(1995)}]{TK95}%
  \BibitemOpen
  \bibfield  {author} {\bibinfo {author} {\bibfnamefont {Y.}~\bibnamefont {Tanaka}}\ and\ \bibinfo {author} {\bibfnamefont {S.}~\bibnamefont {Kashiwaya}},\ }\href {\doibase 10.1103/PhysRevLett.74.3451} {\bibfield  {journal} {\bibinfo  {journal} {Phys. Rev. Lett.}\ }\textbf {\bibinfo {volume} {74}},\ \bibinfo {pages} {3451} (\bibinfo {year} {1995})}\BibitemShut {NoStop}%
\bibitem [{\citenamefont {Kashiwaya}\ \emph {et~al.}(1996)\citenamefont {Kashiwaya}, \citenamefont {Tanaka}, \citenamefont {Koyanagi},\ and\ \citenamefont {Kajimura}}]{KT96}%
  \BibitemOpen
  \bibfield  {author} {\bibinfo {author} {\bibfnamefont {S.}~\bibnamefont {Kashiwaya}}, \bibinfo {author} {\bibfnamefont {Y.}~\bibnamefont {Tanaka}}, \bibinfo {author} {\bibfnamefont {M.}~\bibnamefont {Koyanagi}}, \ and\ \bibinfo {author} {\bibfnamefont {K.}~\bibnamefont {Kajimura}},\ }\href {\doibase 10.1103/PhysRevB.53.2667} {\bibfield  {journal} {\bibinfo  {journal} {Phys. Rev. B}\ }\textbf {\bibinfo {volume} {53}},\ \bibinfo {pages} {2667} (\bibinfo {year} {1996})}\BibitemShut {NoStop}%
\bibitem [{\citenamefont {Sato}\ \emph {et~al.}(2011)\citenamefont {Sato}, \citenamefont {Tanaka}, \citenamefont {Yada},\ and\ \citenamefont {Yokoyama}}]{Sato2011}%
  \BibitemOpen
  \bibfield  {author} {\bibinfo {author} {\bibfnamefont {M.}~\bibnamefont {Sato}}, \bibinfo {author} {\bibfnamefont {Y.}~\bibnamefont {Tanaka}}, \bibinfo {author} {\bibfnamefont {K.}~\bibnamefont {Yada}}, \ and\ \bibinfo {author} {\bibfnamefont {T.}~\bibnamefont {Yokoyama}},\ }\href {\doibase 10.1103/PhysRevB.83.224511} {\bibfield  {journal} {\bibinfo  {journal} {Phys. Rev. B}\ }\textbf {\bibinfo {volume} {83}},\ \bibinfo {pages} {224511} (\bibinfo {year} {2011})}\BibitemShut {NoStop}%
\bibitem [{\citenamefont {Schnyder}\ and\ \citenamefont {Ryu}(2011)}]{Schnyder2011}%
  \BibitemOpen
  \bibfield  {author} {\bibinfo {author} {\bibfnamefont {A.~P.}\ \bibnamefont {Schnyder}}\ and\ \bibinfo {author} {\bibfnamefont {S.}~\bibnamefont {Ryu}},\ }\href {\doibase 10.1103/PhysRevB.84.060504} {\bibfield  {journal} {\bibinfo  {journal} {Phys. Rev. B}\ }\textbf {\bibinfo {volume} {84}},\ \bibinfo {pages} {060504} (\bibinfo {year} {2011})}\BibitemShut {NoStop}%
\bibitem [{\citenamefont {Brydon}\ \emph {et~al.}(2011)\citenamefont {Brydon}, \citenamefont {Schnyder},\ and\ \citenamefont {Timm}}]{Brydon2011}%
  \BibitemOpen
  \bibfield  {author} {\bibinfo {author} {\bibfnamefont {P.~M.~R.}\ \bibnamefont {Brydon}}, \bibinfo {author} {\bibfnamefont {A.~P.}\ \bibnamefont {Schnyder}}, \ and\ \bibinfo {author} {\bibfnamefont {C.}~\bibnamefont {Timm}},\ }\href {\doibase 10.1103/PhysRevB.84.020501} {\bibfield  {journal} {\bibinfo  {journal} {Phys. Rev. B}\ }\textbf {\bibinfo {volume} {84}},\ \bibinfo {pages} {020501} (\bibinfo {year} {2011})}\BibitemShut {NoStop}%
\bibitem [{\citenamefont {Matsuura}\ \emph {et~al.}(2013)\citenamefont {Matsuura}, \citenamefont {Chang}, \citenamefont {Schnyder},\ and\ \citenamefont {Ryu}}]{matsuura2013}%
  \BibitemOpen
  \bibfield  {author} {\bibinfo {author} {\bibfnamefont {S.}~\bibnamefont {Matsuura}}, \bibinfo {author} {\bibfnamefont {P.-Y.}\ \bibnamefont {Chang}}, \bibinfo {author} {\bibfnamefont {A.~P.}\ \bibnamefont {Schnyder}}, \ and\ \bibinfo {author} {\bibfnamefont {S.}~\bibnamefont {Ryu}},\ }\href {\doibase 10.1088/1367-2630/15/6/065001} {\bibfield  {journal} {\bibinfo  {journal} {New Journal of Physics}\ }\textbf {\bibinfo {volume} {15}},\ \bibinfo {pages} {065001} (\bibinfo {year} {2013})}\BibitemShut {NoStop}%
\bibitem [{\citenamefont {Kobayashi}\ \emph {et~al.}(2014)\citenamefont {Kobayashi}, \citenamefont {Shiozaki}, \citenamefont {Tanaka},\ and\ \citenamefont {Sato}}]{Kobayashi14}%
  \BibitemOpen
  \bibfield  {author} {\bibinfo {author} {\bibfnamefont {S.}~\bibnamefont {Kobayashi}}, \bibinfo {author} {\bibfnamefont {K.}~\bibnamefont {Shiozaki}}, \bibinfo {author} {\bibfnamefont {Y.}~\bibnamefont {Tanaka}}, \ and\ \bibinfo {author} {\bibfnamefont {M.}~\bibnamefont {Sato}},\ }\href {\doibase 10.1103/PhysRevB.90.024516} {\bibfield  {journal} {\bibinfo  {journal} {Phys. Rev. B}\ }\textbf {\bibinfo {volume} {90}},\ \bibinfo {pages} {024516} (\bibinfo {year} {2014})}\BibitemShut {NoStop}%
\bibitem [{\citenamefont {Kobayashi}\ \emph {et~al.}(2016)\citenamefont {Kobayashi}, \citenamefont {Yanase},\ and\ \citenamefont {Sato}}]{Kobayashi16}%
  \BibitemOpen
  \bibfield  {author} {\bibinfo {author} {\bibfnamefont {S.}~\bibnamefont {Kobayashi}}, \bibinfo {author} {\bibfnamefont {Y.}~\bibnamefont {Yanase}}, \ and\ \bibinfo {author} {\bibfnamefont {M.}~\bibnamefont {Sato}},\ }\href {\doibase 10.1103/PhysRevB.94.134512} {\bibfield  {journal} {\bibinfo  {journal} {Phys. Rev. B}\ }\textbf {\bibinfo {volume} {94}},\ \bibinfo {pages} {134512} (\bibinfo {year} {2016})}\BibitemShut {NoStop}%
\bibitem [{\citenamefont {Kobayashi}\ \emph {et~al.}(2018)\citenamefont {Kobayashi}, \citenamefont {Sumita}, \citenamefont {Yanase},\ and\ \citenamefont {Sato}}]{Kobayashi18}%
  \BibitemOpen
  \bibfield  {author} {\bibinfo {author} {\bibfnamefont {S.}~\bibnamefont {Kobayashi}}, \bibinfo {author} {\bibfnamefont {S.}~\bibnamefont {Sumita}}, \bibinfo {author} {\bibfnamefont {Y.}~\bibnamefont {Yanase}}, \ and\ \bibinfo {author} {\bibfnamefont {M.}~\bibnamefont {Sato}},\ }\href {\doibase 10.1103/PhysRevB.97.180504} {\bibfield  {journal} {\bibinfo  {journal} {Phys. Rev. B}\ }\textbf {\bibinfo {volume} {97}},\ \bibinfo {pages} {180504(R)} (\bibinfo {year} {2018})}\BibitemShut {NoStop}%
\bibitem [{\citenamefont {Tanaka}\ \emph {et~al.}(2011)\citenamefont {Tanaka}, \citenamefont {Sato},\ and\ \citenamefont {Nagaosa}}]{tanaka2011symmetry}%
  \BibitemOpen
  \bibfield  {author} {\bibinfo {author} {\bibfnamefont {Y.}~\bibnamefont {Tanaka}}, \bibinfo {author} {\bibfnamefont {M.}~\bibnamefont {Sato}}, \ and\ \bibinfo {author} {\bibfnamefont {N.}~\bibnamefont {Nagaosa}},\ }\href@noop {} {\bibfield  {journal} {\bibinfo  {journal} {Journal of the Physical Society of Japan}\ }\textbf {\bibinfo {volume} {81}},\ \bibinfo {pages} {011013} (\bibinfo {year} {2011})}\BibitemShut {NoStop}%
\bibitem [{\citenamefont {Tanaka}\ \emph {et~al.}(2024)\citenamefont {Tanaka}, \citenamefont {Tamura},\ and\ \citenamefont {Cayao}}]{tanaka2024theory}%
  \BibitemOpen
  \bibfield  {author} {\bibinfo {author} {\bibfnamefont {Y.}~\bibnamefont {Tanaka}}, \bibinfo {author} {\bibfnamefont {S.}~\bibnamefont {Tamura}}, \ and\ \bibinfo {author} {\bibfnamefont {J.}~\bibnamefont {Cayao}},\ }\href@noop {} {\bibfield  {journal} {\bibinfo  {journal} {Progress of Theoretical and Experimental Physics}\ ,\ \bibinfo {pages} {ptae065}} (\bibinfo {year} {2024})}\BibitemShut {NoStop}%
\bibitem [{\citenamefont {Tamura}\ \emph {et~al.}(2017)\citenamefont {Tamura}, \citenamefont {Kobayashi}, \citenamefont {Bo},\ and\ \citenamefont {Tanaka}}]{Tamura2017}%
  \BibitemOpen
  \bibfield  {author} {\bibinfo {author} {\bibfnamefont {S.}~\bibnamefont {Tamura}}, \bibinfo {author} {\bibfnamefont {S.}~\bibnamefont {Kobayashi}}, \bibinfo {author} {\bibfnamefont {L.}~\bibnamefont {Bo}}, \ and\ \bibinfo {author} {\bibfnamefont {Y.}~\bibnamefont {Tanaka}},\ }\href {\doibase 10.1103/PhysRevB.95.104511} {\bibfield  {journal} {\bibinfo  {journal} {Phys. Rev. B}\ }\textbf {\bibinfo {volume} {95}},\ \bibinfo {pages} {104511} (\bibinfo {year} {2017})}\BibitemShut {NoStop}%
\bibitem [{\citenamefont {Yamashiro}\ \emph {et~al.}(1998)\citenamefont {Yamashiro}, \citenamefont {Tanaka}, \citenamefont {Tanuma},\ and\ \citenamefont {Kashiwaya}}]{Yamashiro98}%
  \BibitemOpen
  \bibfield  {author} {\bibinfo {author} {\bibfnamefont {M.}~\bibnamefont {Yamashiro}}, \bibinfo {author} {\bibfnamefont {Y.}~\bibnamefont {Tanaka}}, \bibinfo {author} {\bibfnamefont {Y.}~\bibnamefont {Tanuma}}, \ and\ \bibinfo {author} {\bibfnamefont {S.}~\bibnamefont {Kashiwaya}},\ }\href {\doibase 10.1143/JPSJ.67.3224} {\bibfield  {journal} {\bibinfo  {journal} {J. Phys. Soc. Jpn.}\ }\textbf {\bibinfo {volume} {67}},\ \bibinfo {pages} {3224} (\bibinfo {year} {1998})}\BibitemShut {NoStop}%
\bibitem [{\citenamefont {Kashiwaya}\ and\ \citenamefont {Tanaka}(2000)}]{Kashiwaya2000}%
  \BibitemOpen
  \bibfield  {author} {\bibinfo {author} {\bibfnamefont {S.}~\bibnamefont {Kashiwaya}}\ and\ \bibinfo {author} {\bibfnamefont {Y.}~\bibnamefont {Tanaka}},\ }\href {\doibase 10.1088/0034-4885/63/10/202} {\bibfield  {journal} {\bibinfo  {journal} {Rep. Prog. Phys.}\ }\textbf {\bibinfo {volume} {63}},\ \bibinfo {pages} {1641} (\bibinfo {year} {2000})}\BibitemShut {NoStop}%
\bibitem [{\citenamefont {Hu}(1994)}]{Hu94}%
  \BibitemOpen
  \bibfield  {author} {\bibinfo {author} {\bibfnamefont {C.-R.}\ \bibnamefont {Hu}},\ }\href {\doibase 10.1103/PhysRevLett.72.1526} {\bibfield  {journal} {\bibinfo  {journal} {Phys. Rev. Lett.}\ }\textbf {\bibinfo {volume} {72}},\ \bibinfo {pages} {1526} (\bibinfo {year} {1994})}\BibitemShut {NoStop}%
\bibitem [{\citenamefont {Alff}\ \emph {et~al.}(1997)\citenamefont {Alff}, \citenamefont {Takashima}, \citenamefont {Kashiwaya}, \citenamefont {Terada}, \citenamefont {Ihara}, \citenamefont {Tanaka}, \citenamefont {Koyanagi},\ and\ \citenamefont {Kajimura}}]{Alff97}%
  \BibitemOpen
  \bibfield  {author} {\bibinfo {author} {\bibfnamefont {L.}~\bibnamefont {Alff}}, \bibinfo {author} {\bibfnamefont {H.}~\bibnamefont {Takashima}}, \bibinfo {author} {\bibfnamefont {S.}~\bibnamefont {Kashiwaya}}, \bibinfo {author} {\bibfnamefont {N.}~\bibnamefont {Terada}}, \bibinfo {author} {\bibfnamefont {H.}~\bibnamefont {Ihara}}, \bibinfo {author} {\bibfnamefont {Y.}~\bibnamefont {Tanaka}}, \bibinfo {author} {\bibfnamefont {M.}~\bibnamefont {Koyanagi}}, \ and\ \bibinfo {author} {\bibfnamefont {K.}~\bibnamefont {Kajimura}},\ }\href {\doibase 10.1103/PhysRevB.55.R14757} {\bibfield  {journal} {\bibinfo  {journal} {Phys. Rev. B}\ }\textbf {\bibinfo {volume} {55}},\ \bibinfo {pages} {R14757} (\bibinfo {year} {1997})}\BibitemShut {NoStop}%
\bibitem [{\citenamefont {Wei}\ \emph {et~al.}(1998{\natexlab{a}})\citenamefont {Wei}, \citenamefont {Yeh}, \citenamefont {Garrigus},\ and\ \citenamefont {Strasik}}]{Wei98}%
  \BibitemOpen
  \bibfield  {author} {\bibinfo {author} {\bibfnamefont {J.~Y.~T.}\ \bibnamefont {Wei}}, \bibinfo {author} {\bibfnamefont {N.-C.}\ \bibnamefont {Yeh}}, \bibinfo {author} {\bibfnamefont {D.~F.}\ \bibnamefont {Garrigus}}, \ and\ \bibinfo {author} {\bibfnamefont {M.}~\bibnamefont {Strasik}},\ }\href {\doibase 10.1103/PhysRevLett.81.2542} {\bibfield  {journal} {\bibinfo  {journal} {Phys. Rev. Lett.}\ }\textbf {\bibinfo {volume} {81}},\ \bibinfo {pages} {2542} (\bibinfo {year} {1998}{\natexlab{a}})}\BibitemShut {NoStop}%
\bibitem [{\citenamefont {Iguchi}\ \emph {et~al.}(2000)\citenamefont {Iguchi}, \citenamefont {Wang}, \citenamefont {Yamazaki}, \citenamefont {Tanaka},\ and\ \citenamefont {Kashiwaya}}]{Iguchi2000}%
  \BibitemOpen
  \bibfield  {author} {\bibinfo {author} {\bibfnamefont {I.}~\bibnamefont {Iguchi}}, \bibinfo {author} {\bibfnamefont {W.}~\bibnamefont {Wang}}, \bibinfo {author} {\bibfnamefont {M.}~\bibnamefont {Yamazaki}}, \bibinfo {author} {\bibfnamefont {Y.}~\bibnamefont {Tanaka}}, \ and\ \bibinfo {author} {\bibfnamefont {S.}~\bibnamefont {Kashiwaya}},\ }\href {\doibase 10.1103/PhysRevB.62.R6131} {\bibfield  {journal} {\bibinfo  {journal} {Phys. Rev. B}\ }\textbf {\bibinfo {volume} {62}},\ \bibinfo {pages} {R6131} (\bibinfo {year} {2000})}\BibitemShut {NoStop}%
\bibitem [{\citenamefont {Covington}\ \emph {et~al.}(1997)\citenamefont {Covington}, \citenamefont {Aprili}, \citenamefont {Paraoanu}, \citenamefont {Greene}, \citenamefont {Xu}, \citenamefont {Zhu},\ and\ \citenamefont {Mirkin}}]{Experiment3}%
  \BibitemOpen
  \bibfield  {author} {\bibinfo {author} {\bibfnamefont {M.}~\bibnamefont {Covington}}, \bibinfo {author} {\bibfnamefont {M.}~\bibnamefont {Aprili}}, \bibinfo {author} {\bibfnamefont {E.}~\bibnamefont {Paraoanu}}, \bibinfo {author} {\bibfnamefont {L.~H.}\ \bibnamefont {Greene}}, \bibinfo {author} {\bibfnamefont {F.}~\bibnamefont {Xu}}, \bibinfo {author} {\bibfnamefont {J.}~\bibnamefont {Zhu}}, \ and\ \bibinfo {author} {\bibfnamefont {C.~A.}\ \bibnamefont {Mirkin}},\ }\href@noop {} {\bibfield  {journal} {\bibinfo  {journal} {Phys. Rev. Lett.}\ }\textbf {\bibinfo {volume} {79}},\ \bibinfo {pages} {277} (\bibinfo {year} {1997})}\BibitemShut {NoStop}%
\bibitem [{\citenamefont {Wei}\ \emph {et~al.}(1998{\natexlab{b}})\citenamefont {Wei}, \citenamefont {Yeh}, \citenamefont {Garrigus},\ and\ \citenamefont {Strasik}}]{Experiment5}%
  \BibitemOpen
  \bibfield  {author} {\bibinfo {author} {\bibfnamefont {J.~Y.~T.}\ \bibnamefont {Wei}}, \bibinfo {author} {\bibfnamefont {N.-C.}\ \bibnamefont {Yeh}}, \bibinfo {author} {\bibfnamefont {D.~F.}\ \bibnamefont {Garrigus}}, \ and\ \bibinfo {author} {\bibfnamefont {M.}~\bibnamefont {Strasik}},\ }\href@noop {} {\bibfield  {journal} {\bibinfo  {journal} {Phys. Rev. Lett.}\ }\textbf {\bibinfo {volume} {81}},\ \bibinfo {pages} {2542} (\bibinfo {year} {1998}{\natexlab{b}})}\BibitemShut {NoStop}%
\bibitem [{\citenamefont {Biswas}\ \emph {et~al.}(2002)\citenamefont {Biswas}, \citenamefont {Fournier}, \citenamefont {Qazilbash}, \citenamefont {Smolyaninova}, \citenamefont {Balci},\ and\ \citenamefont {Greene}}]{Experiment6}%
  \BibitemOpen
  \bibfield  {author} {\bibinfo {author} {\bibfnamefont {A.}~\bibnamefont {Biswas}}, \bibinfo {author} {\bibfnamefont {P.}~\bibnamefont {Fournier}}, \bibinfo {author} {\bibfnamefont {M.~M.}\ \bibnamefont {Qazilbash}}, \bibinfo {author} {\bibfnamefont {V.~N.}\ \bibnamefont {Smolyaninova}}, \bibinfo {author} {\bibfnamefont {H.}~\bibnamefont {Balci}}, \ and\ \bibinfo {author} {\bibfnamefont {R.~L.}\ \bibnamefont {Greene}},\ }\href@noop {} {\bibfield  {journal} {\bibinfo  {journal} {Phys. Rev. Lett.}\ }\textbf {\bibinfo {volume} {88}},\ \bibinfo {pages} {207004} (\bibinfo {year} {2002})}\BibitemShut {NoStop}%
\bibitem [{\citenamefont {Sharoni}\ \emph {et~al.}(2001)\citenamefont {Sharoni}, \citenamefont {Koren},\ and\ \citenamefont {Millo}}]{Sharoni_2001}%
  \BibitemOpen
  \bibfield  {author} {\bibinfo {author} {\bibfnamefont {A.}~\bibnamefont {Sharoni}}, \bibinfo {author} {\bibfnamefont {G.}~\bibnamefont {Koren}}, \ and\ \bibinfo {author} {\bibfnamefont {O.}~\bibnamefont {Millo}},\ }\href {\doibase 10.1209/epl/i2001-00368-1} {\bibfield  {journal} {\bibinfo  {journal} {Europhysics Letters}\ }\textbf {\bibinfo {volume} {54}},\ \bibinfo {pages} {675} (\bibinfo {year} {2001})}\BibitemShut {NoStop}%
\bibitem [{\citenamefont {Millo}\ and\ \citenamefont {Koren}(2018)}]{Millo2018}%
  \BibitemOpen
  \bibfield  {author} {\bibinfo {author} {\bibfnamefont {O.}~\bibnamefont {Millo}}\ and\ \bibinfo {author} {\bibfnamefont {G.}~\bibnamefont {Koren}},\ }\href@noop {} {\bibfield  {journal} {\bibinfo  {journal} {Philosophical Transactions of the Royal Society A: Mathematical, Physical and Engineering Sciences}\ }\textbf {\bibinfo {volume} {376}},\ \bibinfo {pages} {20140143} (\bibinfo {year} {2018})}\BibitemShut {NoStop}%
\bibitem [{\citenamefont {Bouscher}\ \emph {et~al.}(2020)\citenamefont {Bouscher}, \citenamefont {Kang}, \citenamefont {Balasubramanian}, \citenamefont {Panna}, \citenamefont {Yu}, \citenamefont {Chen},\ and\ \citenamefont {Hayat}}]{Bouscher_2020}%
  \BibitemOpen
  \bibfield  {author} {\bibinfo {author} {\bibfnamefont {S.}~\bibnamefont {Bouscher}}, \bibinfo {author} {\bibfnamefont {Z.}~\bibnamefont {Kang}}, \bibinfo {author} {\bibfnamefont {K.}~\bibnamefont {Balasubramanian}}, \bibinfo {author} {\bibfnamefont {D.}~\bibnamefont {Panna}}, \bibinfo {author} {\bibfnamefont {P.}~\bibnamefont {Yu}}, \bibinfo {author} {\bibfnamefont {X.}~\bibnamefont {Chen}}, \ and\ \bibinfo {author} {\bibfnamefont {A.}~\bibnamefont {Hayat}},\ }\href {\doibase 10.1088/1361-648X/abae18} {\bibfield  {journal} {\bibinfo  {journal} {Journal of Physics: Condensed Matter}\ }\textbf {\bibinfo {volume} {32}},\ \bibinfo {pages} {475502} (\bibinfo {year} {2020})}\BibitemShut {NoStop}%
\bibitem [{\citenamefont {Kashiwaya}\ \emph {et~al.}(1999)\citenamefont {Kashiwaya}, \citenamefont {Tanaka}, \citenamefont {Yoshida},\ and\ \citenamefont {Beasley}}]{KashiwayaPRB1999}%
  \BibitemOpen
  \bibfield  {author} {\bibinfo {author} {\bibfnamefont {S.}~\bibnamefont {Kashiwaya}}, \bibinfo {author} {\bibfnamefont {Y.}~\bibnamefont {Tanaka}}, \bibinfo {author} {\bibfnamefont {N.}~\bibnamefont {Yoshida}}, \ and\ \bibinfo {author} {\bibfnamefont {M.~R.}\ \bibnamefont {Beasley}},\ }\href {\doibase 10.1103/PhysRevB.60.3572} {\bibfield  {journal} {\bibinfo  {journal} {Phys. Rev. B}\ }\textbf {\bibinfo {volume} {60}},\ \bibinfo {pages} {3572} (\bibinfo {year} {1999})}\BibitemShut {NoStop}%
\bibitem [{\citenamefont {\ifmmode \check{Z}\else \v{Z}\fi{}uti\ifmmode~\acute{c}\else \'{c}\fi{}}\ and\ \citenamefont {Valls}(1999)}]{Zutic1999}%
  \BibitemOpen
  \bibfield  {author} {\bibinfo {author} {\bibfnamefont {I.}~\bibnamefont {\ifmmode \check{Z}\else \v{Z}\fi{}uti\ifmmode~\acute{c}\else \'{c}\fi{}}}\ and\ \bibinfo {author} {\bibfnamefont {O.~T.}\ \bibnamefont {Valls}},\ }\href {\doibase 10.1103/PhysRevB.60.6320} {\bibfield  {journal} {\bibinfo  {journal} {Phys. Rev. B}\ }\textbf {\bibinfo {volume} {60}},\ \bibinfo {pages} {6320} (\bibinfo {year} {1999})}\BibitemShut {NoStop}%
\bibitem [{\citenamefont {Zhu}\ and\ \citenamefont {Ting}(2000)}]{Ting2000}%
  \BibitemOpen
  \bibfield  {author} {\bibinfo {author} {\bibfnamefont {J.-X.}\ \bibnamefont {Zhu}}\ and\ \bibinfo {author} {\bibfnamefont {C.~S.}\ \bibnamefont {Ting}},\ }\href {\doibase 10.1103/PhysRevB.61.1456} {\bibfield  {journal} {\bibinfo  {journal} {Phys. Rev. B}\ }\textbf {\bibinfo {volume} {61}},\ \bibinfo {pages} {1456} (\bibinfo {year} {2000})}\BibitemShut {NoStop}%
\bibitem [{\citenamefont {Hirai}\ \emph {et~al.}(2003)\citenamefont {Hirai}, \citenamefont {Tanaka}, \citenamefont {Yoshida}, \citenamefont {Asano}, \citenamefont {Inoue},\ and\ \citenamefont {Kashiwaya}}]{HiraiPRB2003}%
  \BibitemOpen
  \bibfield  {author} {\bibinfo {author} {\bibfnamefont {T.}~\bibnamefont {Hirai}}, \bibinfo {author} {\bibfnamefont {Y.}~\bibnamefont {Tanaka}}, \bibinfo {author} {\bibfnamefont {N.}~\bibnamefont {Yoshida}}, \bibinfo {author} {\bibfnamefont {Y.}~\bibnamefont {Asano}}, \bibinfo {author} {\bibfnamefont {J.}~\bibnamefont {Inoue}}, \ and\ \bibinfo {author} {\bibfnamefont {S.}~\bibnamefont {Kashiwaya}},\ }\href {\doibase 10.1103/PhysRevB.67.174501} {\bibfield  {journal} {\bibinfo  {journal} {Phys. Rev. B}\ }\textbf {\bibinfo {volume} {67}},\ \bibinfo {pages} {174501} (\bibinfo {year} {2003})}\BibitemShut {NoStop}%
\bibitem [{\citenamefont {Hirai}\ \emph {et~al.}(2001)\citenamefont {Hirai}, \citenamefont {Yoshida}, \citenamefont {Tanaka}, \citenamefont {Inoue},\ and\ \citenamefont {Kashiwaya}}]{Hirai2001}%
  \BibitemOpen
  \bibfield  {author} {\bibinfo {author} {\bibfnamefont {T.}~\bibnamefont {Hirai}}, \bibinfo {author} {\bibfnamefont {N.}~\bibnamefont {Yoshida}}, \bibinfo {author} {\bibfnamefont {Y.}~\bibnamefont {Tanaka}}, \bibinfo {author} {\bibfnamefont {J.-i.}\ \bibnamefont {Inoue}}, \ and\ \bibinfo {author} {\bibfnamefont {S.}~\bibnamefont {Kashiwaya}},\ }\href {\doibase 10.1143/JPSJ.70.1885} {\bibfield  {journal} {\bibinfo  {journal} {J. Phys. Soc. Jpn.}\ }\textbf {\bibinfo {volume} {70}},\ \bibinfo {pages} {1885} (\bibinfo {year} {2001})}\BibitemShut {NoStop}%
\bibitem [{\citenamefont {Hellenes}\ \emph {et~al.}(2023)\citenamefont {Hellenes}, \citenamefont {Jungwirth}, \citenamefont {Sinova},\ and\ \citenamefont {Šmejkal}}]{Hellenes23}%
  \BibitemOpen
  \bibfield  {author} {\bibinfo {author} {\bibfnamefont {A.}~\bibnamefont {Hellenes}}, \bibinfo {author} {\bibfnamefont {T.}~\bibnamefont {Jungwirth}}, \bibinfo {author} {\bibfnamefont {J.}~\bibnamefont {Sinova}}, \ and\ \bibinfo {author} {\bibfnamefont {L.}~\bibnamefont {Šmejkal}},\ }\href@noop {} {} (\bibinfo {year} {2023}),\ \Eprint {http://arxiv.org/abs/2309.01607} {arXiv:2309.01607} \BibitemShut {NoStop}%
\bibitem [{\citenamefont {Ikegaya}\ \emph {et~al.}(2021)\citenamefont {Ikegaya}, \citenamefont {Lee}, \citenamefont {Schnyder},\ and\ \citenamefont {Asano}}]{Ikegaya2021}%
  \BibitemOpen
  \bibfield  {author} {\bibinfo {author} {\bibfnamefont {S.}~\bibnamefont {Ikegaya}}, \bibinfo {author} {\bibfnamefont {J.}~\bibnamefont {Lee}}, \bibinfo {author} {\bibfnamefont {A.~P.}\ \bibnamefont {Schnyder}}, \ and\ \bibinfo {author} {\bibfnamefont {Y.}~\bibnamefont {Asano}},\ }\href {\doibase 10.1103/PhysRevB.104.L020502} {\bibfield  {journal} {\bibinfo  {journal} {Phys. Rev. B}\ }\textbf {\bibinfo {volume} {104}},\ \bibinfo {pages} {L020502} (\bibinfo {year} {2021})}\BibitemShut {NoStop}%
\bibitem [{\citenamefont {Alidoust}\ \emph {et~al.}(2021)\citenamefont {Alidoust}, \citenamefont {Shen},\ and\ \citenamefont {{\v{Z}}uti{\'c}}}]{Alidoust2021}%
  \BibitemOpen
  \bibfield  {author} {\bibinfo {author} {\bibfnamefont {M.}~\bibnamefont {Alidoust}}, \bibinfo {author} {\bibfnamefont {C.}~\bibnamefont {Shen}}, \ and\ \bibinfo {author} {\bibfnamefont {I.}~\bibnamefont {{\v{Z}}uti{\'c}}},\ }\href {\doibase 10.1103/PhysRevB.103.L060503} {\bibfield  {journal} {\bibinfo  {journal} {Phys. Rev. B}\ }\textbf {\bibinfo {volume} {103}},\ \bibinfo {pages} {L060503} (\bibinfo {year} {2021})}\BibitemShut {NoStop}%
\bibitem [{\citenamefont {Yang}\ and\ \citenamefont {Wu}(2017)}]{Yang2017}%
  \BibitemOpen
  \bibfield  {author} {\bibinfo {author} {\bibfnamefont {F.}~\bibnamefont {Yang}}\ and\ \bibinfo {author} {\bibfnamefont {M.}~\bibnamefont {Wu}},\ }\href {\doibase 10.1103/PhysRevB.95.075304} {\bibfield  {journal} {\bibinfo  {journal} {Phys. Rev. B}\ }\textbf {\bibinfo {volume} {95}},\ \bibinfo {pages} {075304} (\bibinfo {year} {2017})}\BibitemShut {NoStop}%
\bibitem [{\citenamefont {Alidoust}(2020)}]{Alidoust2020}%
  \BibitemOpen
  \bibfield  {author} {\bibinfo {author} {\bibfnamefont {M.}~\bibnamefont {Alidoust}},\ }\href {\doibase 10.1103/PhysRevB.101.155123} {\bibfield  {journal} {\bibinfo  {journal} {Phys. Rev. B}\ }\textbf {\bibinfo {volume} {101}},\ \bibinfo {pages} {155123} (\bibinfo {year} {2020})}\BibitemShut {NoStop}%
\bibitem [{\citenamefont {Jacobsen}\ \emph {et~al.}(2015)\citenamefont {Jacobsen}, \citenamefont {Ouassou},\ and\ \citenamefont {Linder}}]{Jacobsen2015}%
  \BibitemOpen
  \bibfield  {author} {\bibinfo {author} {\bibfnamefont {S.~H.}\ \bibnamefont {Jacobsen}}, \bibinfo {author} {\bibfnamefont {J.~A.}\ \bibnamefont {Ouassou}}, \ and\ \bibinfo {author} {\bibfnamefont {J.}~\bibnamefont {Linder}},\ }\href {\doibase 10.1103/PhysRevB.92.024510} {\bibfield  {journal} {\bibinfo  {journal} {Phys. Rev. B}\ }\textbf {\bibinfo {volume} {92}},\ \bibinfo {pages} {024510} (\bibinfo {year} {2015})}\BibitemShut {NoStop}%
\bibitem [{\citenamefont {Lee}\ \emph {et~al.}(2021)\citenamefont {Lee}, \citenamefont {Ikegaya},\ and\ \citenamefont {Asano}}]{Lee2021}%
  \BibitemOpen
  \bibfield  {author} {\bibinfo {author} {\bibfnamefont {J.}~\bibnamefont {Lee}}, \bibinfo {author} {\bibfnamefont {S.}~\bibnamefont {Ikegaya}}, \ and\ \bibinfo {author} {\bibfnamefont {Y.}~\bibnamefont {Asano}},\ }\href {\doibase 10.1103/PhysRevB.103.104509} {\bibfield  {journal} {\bibinfo  {journal} {Phys. Rev. B}\ }\textbf {\bibinfo {volume} {103}},\ \bibinfo {pages} {104509} (\bibinfo {year} {2021})}\BibitemShut {NoStop}%
\bibitem [{\citenamefont {Liu}\ \emph {et~al.}(2014)\citenamefont {Liu}, \citenamefont {Jain},\ and\ \citenamefont {Liu}}]{Liu2014}%
  \BibitemOpen
  \bibfield  {author} {\bibinfo {author} {\bibfnamefont {X.}~\bibnamefont {Liu}}, \bibinfo {author} {\bibfnamefont {J.}~\bibnamefont {Jain}}, \ and\ \bibinfo {author} {\bibfnamefont {C.-X.}\ \bibnamefont {Liu}},\ }\href {\doibase 10.1103/PhysRevLett.113.227002} {\bibfield  {journal} {\bibinfo  {journal} {Phys. Rev. lett.}\ }\textbf {\bibinfo {volume} {113}},\ \bibinfo {pages} {227002} (\bibinfo {year} {2014})}\BibitemShut {NoStop}%
\bibitem [{\citenamefont {Bernevig}\ \emph {et~al.}(2006)\citenamefont {Bernevig}, \citenamefont {Orenstein},\ and\ \citenamefont {Zhang}}]{HelixZhang}%
  \BibitemOpen
  \bibfield  {author} {\bibinfo {author} {\bibfnamefont {B.~A.}\ \bibnamefont {Bernevig}}, \bibinfo {author} {\bibfnamefont {J.}~\bibnamefont {Orenstein}}, \ and\ \bibinfo {author} {\bibfnamefont {S.-C.}\ \bibnamefont {Zhang}},\ }\href {\doibase 10.1103/PhysRevLett.97.236601} {\bibfield  {journal} {\bibinfo  {journal} {Phys. Rev. Lett.}\ }\textbf {\bibinfo {volume} {97}},\ \bibinfo {pages} {236601} (\bibinfo {year} {2006})}\BibitemShut {NoStop}%
\bibitem [{\citenamefont {Kohda}\ \emph {et~al.}(2012)\citenamefont {Kohda}, \citenamefont {Lechner}, \citenamefont {Kunihashi}, \citenamefont {Dollinger}, \citenamefont {Olbrich}, \citenamefont {Sch\"onhuber}, \citenamefont {Caspers}, \citenamefont {Bel'kov}, \citenamefont {Golub}, \citenamefont {Weiss}, \citenamefont {Richter}, \citenamefont {Nitta},\ and\ \citenamefont {Ganichev}}]{HelixKohda}%
  \BibitemOpen
  \bibfield  {author} {\bibinfo {author} {\bibfnamefont {M.}~\bibnamefont {Kohda}}, \bibinfo {author} {\bibfnamefont {V.}~\bibnamefont {Lechner}}, \bibinfo {author} {\bibfnamefont {Y.}~\bibnamefont {Kunihashi}}, \bibinfo {author} {\bibfnamefont {T.}~\bibnamefont {Dollinger}}, \bibinfo {author} {\bibfnamefont {P.}~\bibnamefont {Olbrich}}, \bibinfo {author} {\bibfnamefont {C.}~\bibnamefont {Sch\"onhuber}}, \bibinfo {author} {\bibfnamefont {I.}~\bibnamefont {Caspers}}, \bibinfo {author} {\bibfnamefont {V.~V.}\ \bibnamefont {Bel'kov}}, \bibinfo {author} {\bibfnamefont {L.~E.}\ \bibnamefont {Golub}}, \bibinfo {author} {\bibfnamefont {D.}~\bibnamefont {Weiss}}, \bibinfo {author} {\bibfnamefont {K.}~\bibnamefont {Richter}}, \bibinfo {author} {\bibfnamefont {J.}~\bibnamefont {Nitta}}, \ and\ \bibinfo {author} {\bibfnamefont {S.~D.}\ \bibnamefont {Ganichev}},\ }\href {\doibase 10.1103/PhysRevB.86.081306} {\bibfield  {journal} {\bibinfo  {journal} {Phys. Rev. B}\ }\textbf {\bibinfo {volume} {86}},\ \bibinfo {pages}
  {081306} (\bibinfo {year} {2012})}\BibitemShut {NoStop}%
\bibitem [{\citenamefont {Ikegaya}\ \emph {et~al.}(2015)\citenamefont {Ikegaya}, \citenamefont {Asano},\ and\ \citenamefont {Tanaka}}]{HelixIkegaya}%
  \BibitemOpen
  \bibfield  {author} {\bibinfo {author} {\bibfnamefont {S.}~\bibnamefont {Ikegaya}}, \bibinfo {author} {\bibfnamefont {Y.}~\bibnamefont {Asano}}, \ and\ \bibinfo {author} {\bibfnamefont {Y.}~\bibnamefont {Tanaka}},\ }\href {\doibase 10.1103/PhysRevB.91.174511} {\bibfield  {journal} {\bibinfo  {journal} {Phys. Rev. B}\ }\textbf {\bibinfo {volume} {91}},\ \bibinfo {pages} {174511} (\bibinfo {year} {2015})}\BibitemShut {NoStop}%
\bibitem [{\citenamefont {Banerjee}\ and\ \citenamefont {Scheurer}(2024)}]{Banerjee2024altermagnetic}%
  \BibitemOpen
  \bibfield  {author} {\bibinfo {author} {\bibfnamefont {S.}~\bibnamefont {Banerjee}}\ and\ \bibinfo {author} {\bibfnamefont {M.~S.}\ \bibnamefont {Scheurer}},\ }\href@noop {} {\bibfield  {journal} {\bibinfo  {journal} {Physical Review B}\ }\textbf {\bibinfo {volume} {110}},\ \bibinfo {pages} {024503} (\bibinfo {year} {2024})}\BibitemShut {NoStop}%
\bibitem [{\citenamefont {Cheng}\ and\ \citenamefont {Sun}(2024)}]{Cheng2024}%
  \BibitemOpen
  \bibfield  {author} {\bibinfo {author} {\bibfnamefont {Q.}~\bibnamefont {Cheng}}\ and\ \bibinfo {author} {\bibfnamefont {Q.-F.}\ \bibnamefont {Sun}},\ }\href {\doibase 10.1103/PhysRevB.109.024517} {\bibfield  {journal} {\bibinfo  {journal} {Phys. Rev. B}\ }\textbf {\bibinfo {volume} {109}},\ \bibinfo {pages} {024517} (\bibinfo {year} {2024})}\BibitemShut {NoStop}%
\bibitem [{\citenamefont {Tanaka}\ and\ \citenamefont {Kashiwaya}(1996)}]{TKJosephson96}%
  \BibitemOpen
  \bibfield  {author} {\bibinfo {author} {\bibfnamefont {Y.}~\bibnamefont {Tanaka}}\ and\ \bibinfo {author} {\bibfnamefont {S.}~\bibnamefont {Kashiwaya}},\ }\href {\doibase 10.1103/PhysRevB.53.R11957} {\bibfield  {journal} {\bibinfo  {journal} {Phys. Rev. B}\ }\textbf {\bibinfo {volume} {53}},\ \bibinfo {pages} {R11957} (\bibinfo {year} {1996})}\BibitemShut {NoStop}%
\bibitem [{\citenamefont {Tanaka}\ and\ \citenamefont {Kashiwaya}(1997{\natexlab{a}})}]{TKJosephson97}%
  \BibitemOpen
  \bibfield  {author} {\bibinfo {author} {\bibfnamefont {Y.}~\bibnamefont {Tanaka}}\ and\ \bibinfo {author} {\bibfnamefont {S.}~\bibnamefont {Kashiwaya}},\ }\href {\doibase 10.1103/PhysRevB.56.892} {\bibfield  {journal} {\bibinfo  {journal} {Phys. Rev. B}\ }\textbf {\bibinfo {volume} {56}},\ \bibinfo {pages} {892} (\bibinfo {year} {1997}{\natexlab{a}})}\BibitemShut {NoStop}%
\bibitem [{\citenamefont {Tanaka}\ and\ \citenamefont {Kashiwaya}(1997{\natexlab{b}})}]{TANAKAPhyscaC1997}%
  \BibitemOpen
  \bibfield  {author} {\bibinfo {author} {\bibfnamefont {Y.}~\bibnamefont {Tanaka}}\ and\ \bibinfo {author} {\bibfnamefont {S.}~\bibnamefont {Kashiwaya}},\ }\href {\doibase https://doi.org/10.1016/S0921-4534(97)80002-0} {\bibfield  {journal} {\bibinfo  {journal} {Physica C: Superconductivity}\ }\textbf {\bibinfo {volume} {274}},\ \bibinfo {pages} {357} (\bibinfo {year} {1997}{\natexlab{b}})}\BibitemShut {NoStop}%
\bibitem [{\citenamefont {Tanaka}\ and\ \citenamefont {Kashiwaya}(2000)}]{Tanaka2000}%
  \BibitemOpen
  \bibfield  {author} {\bibinfo {author} {\bibfnamefont {Y.}~\bibnamefont {Tanaka}}\ and\ \bibinfo {author} {\bibfnamefont {S.}~\bibnamefont {Kashiwaya}},\ }\href {\doibase 10.1143/JPSJ.69.1152} {\bibfield  {journal} {\bibinfo  {journal} {Journal of the Physical Society of Japan}\ }\textbf {\bibinfo {volume} {69}},\ \bibinfo {pages} {1152} (\bibinfo {year} {2000})},\ \Eprint {http://arxiv.org/abs/https://doi.org/10.1143/JPSJ.69.1152} {https://doi.org/10.1143/JPSJ.69.1152} \BibitemShut {NoStop}%
\end{thebibliography}%
\end{document}